\documentclass{ws-p9-75x6-50}

\newcommand{\kFB}{{k_{F_B}}}
\newcommand{\kFL}{{k_{F_L}}}
\newcommand{\bfxm}{\mbox{\mbox{\boldmath$x$}}}
\newcommand{\ebnhv}{{\Lambda^{00}_{\rm Bonn}}+{\rm HV}}
\newcommand{\ehehfv}{{\Lambda^{00}_{\rm HEA}}+{\rm HFV}}

\newcommand{\eBrBhfv}{{\Lambda^{\rm BHF}_{\rm BroB}}+{\rm HFV}}
\newcommand{\efp}{{\rm FP(V}_{14}+{\rm TNI)}}

\newcommand{\ewut}{{\rm WFF(UV}_{14}+{\rm TNI)}}

\newcommand{\eapr}{{\rm APR(A}_{18}+{\rm UIX)}}
\newcommand{\eut}{{\rm UV}_{14}\!+\!{\rm TNI}}
\newcommand{\euu}{{\rm UV}_{14}\!+\!{\rm UVII}}
\newcommand{\eau}{{\rm AV}_{14}\!+\!{\rm UVII}}

\newcommand{\egth}{{\rm G}_{300}}
\newcommand{\egthpi}{{\rm G}^\pi_{300}}
\newcommand{\egdcma}{{\rm G}^{\rm DCM1}_{\rm B180}}
\newcommand{\egdcmb}{{\rm G}^{\rm DCM1}_{225}}
\newcommand{\egdcmc}{{\rm G}^{\rm DCM2}_{\rm B180}}
\newcommand{\egdcmd}{{\rm G}^{\rm DCM2}_{\rm 265}}
\newcommand{\egthK}{{\rm G}^{K^-}_{300}}

\newcommand{\msun}{M_{\odot}}
\newcommand{\gcmt}{{\rm g/cm}^3}
\newcommand{\gcmsq}{{{\rm g}\; {\rm cm}}^2}

\newcommand{\bag}{B^{1/4}}

\newcommand{\ecrusti}{\epsilon_{\rm crust}}
\newcommand{\edrip}{\epsilon_{\rm drip}}
\newcommand{\pdrip}{P_{\rm drip}}

\newcommand{\icrust}{I_{\rm crust}}
\newcommand{\itotal}{I_{\rm total}}

\newcommand{\brho}{\mbox{\mbox{\boldmath$\rho$}}}

\newcommand{\btau}{\mbox{\mbox{\boldmath$\tau$}}}
\newcommand{\bpi}{\mbox{\mbox{\boldmath$\pi$}}}

\newcommand{\bcdot}{\mbox{\mbox{\boldmath$\cdot$}}}
\newcommand{\bfG}{{\bf G}}

\newcommand{\bfpm}{\mbox{\mbox{\boldmath$p$}}}

\newcommand{\psiB}{{\psi_B}}

\newcommand{\bpsiB}{{\bar\psi_B}}

\newcommand{\KBt}{{\rm G}^{\rm K300}_{\rm B180}}
\newcommand{\KM}{{\rm G}^{\rm K240}_{\rm M78}}
\newcommand{\KB}{{\rm G}^{\rm K240}_{\rm B180}}
\newcommand{\okgr}{{\Omega_{\rm K}}}
\newcommand{\mevt}{{\rm MeV/fm}^3}
\newcommand{\gsim}{\stackrel{\textstyle >}{_\sim}}
\newcommand{\lsim}{\stackrel{\textstyle <}{_\sim}}

\newcommand{\logt}{{\rm log}_{10}}
\newcommand{\loglum}{{\rm log}_{10}\, L}
\newcommand{\logtem}{{\rm log}_{10}\, T}
\newcommand{\secm}{{\rm s}^{-1}}
\newcommand{\pkgr}{{P_{\,\rm K}}}

\newcommand{\kFo}{k_{F_0}}
\newcommand{\kFosq}{{k_{F_0}^2}}

\begin{document}

\title{Nuclear and High-Energy Astrophysics}

\author{Fridolin Weber}

\address{Joint Center for Nuclear Astrophysics, Department of Physics,
University of Notre Dame, 225 Nieuwland Science Hall, Notre Dame, IN
46556-5670, USA \\ E-mail: fweber@nd.edu}

\maketitle

\abstracts{There has never been a more exciting time in the
overlapping areas of nuclear physics, particle physics and
relativistic astrophysics than today. Orbiting observatories such as
the Hubble Space Telescope, Rossi X-ray Timing Explorer (RXTE),
Chandra X-ray satellite, and the X-ray Multi Mirror Mission (XMM) have
extended our vision tremendously, allowing us to see vistas with an
unprecedented clarity and angular resolution that previously were only
imagined, enabling astrophysicists for the first time ever to perform
detailed studies of large samples of galactic and extragalactic
objects. On the Earth, radio telescopes (e.g., Arecibo, Green Bank,
Parkes, VLA) and instruments using adaptive optics and other
revolutionary techniques have exceeded previous expectations of what
can be accomplished from the ground. The gravitational wave detectors
LIGO, LISA VIRGO, and Geo-600 are opening up a window for the
detection of gravitational waves emitted from compact stellar objects
such as neutron stars and black holes.  Together with new experimental
forefront facilities like ISAC, ORLaND and RIA, these detectors
provide direct, quantitative physical insight into nucleosynthesis,
supernova dynamics, accreting compact objects, cosmic-ray
acceleration, and pair-production in high energy sources which
reinforce the urgent need for a strong and continuous feedback from
nuclear and particle theory and theoretical astrophysics.  In my
lectures, I shall concentrate on three selected topics, which range
from the behavior of superdense stellar matter, to general
relativistic stellar models, to strange quark stars and possible
signals of quark matter in neutron stars.}

\section{Introduction}\label{sec:intro}

\begin{figure}[tb] 
\begin{center}
\leavevmode
\psfig{figure=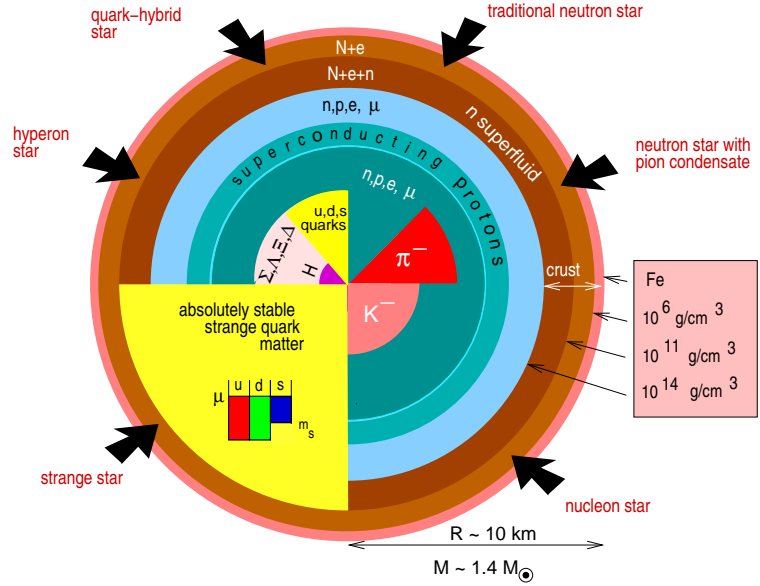,width=9cm,angle=-90}
\caption[]{Competing structures and novel phases of subatomic matter
predicted by theory to make their appearance in the cores ($R\lsim
8$~km) of neutron stars.\protect{\cite{weber99:book}}}
\label{fig:cross}
\end{center}
\end{figure}

A forefront area of research, both experimental and theoretical,
concerns the exploration of the subatomic structure of superdense
matter and the determination of the equation of state -- that is, the
relation between pressure $P$, temperature $T$ and density $\epsilon$ --
associated with such matter.\cite{greiner96:wilderness}  Knowing its
properties is of key importance for our understanding of the physics
of the early universe, its evolution in time to the present day,
compact stars, various astrophysical phenomena, and laboratory
physics.  The high-temperature domain of the phase diagram of
superdense matter is probed by relativistic heavy-ion
colliders. Complementary to this, neutron stars contain cold
superdense matter permanently in their centers (cf.\
Fig.\ \ref{fig:cross}), which make them superb astrophysical
laboratories for probing the low-density high-density domain of the
phase diagram of superdense matter.\cite{blandford92:b,weber99:book}

Neutron stars are dense, neutron-packed remnants of massive stars that
blew apart in supernova explosions. They are typically about 10
kilometers across and spin rapidly, often making several hundred
rotations per second.  The discovery rate of new rotating neutron
stars, spotted as pulsars by radio telescopes
\cite{manchester77:a,lyne90:a}, is rather high. To date about 1400
pulsars are known. Depending on star mass and rotational frequency,
gravity compresses the matter in the core regions of pulsars up to
more than ten times ($\sim 1.5$~GeV/fm$^3$) the density of
ordinary atomic nuclei, thus providing a high-pressure environment in
which numerous subatomic particle processes plausibly compete with
each other and novel phases of matter may exist. The most spectacular
ones stretch from the generation of new baryonic particles (e.g.,
$\Sigma, \Lambda, \Xi, \Delta$) to quark ($u, d, s$) deconfinement to
the formation of Boson condensates ($\pi^-, K^-$, H-matter), as
illustrated in Fig.\ \ref{fig:cross}. There are theoretical
suggestions of even more exotic processes inside pulsars, such as the
formation of absolutely stable quark matter, a configuration of matter
even more stable than the most stable atomic nucleus, $^{56}\rm{Fe}$!
In the latter event, pulsars would be largely composed of pure quark
matter, eventually enveloped in thin nuclear crusts (bottom-left
portion in Fig.\ \ref{fig:cross}). No matter which physical processes
are actually realized inside neutron stars, each one leads to
fingerprints, some more pronounced than others though, in the
observable stellar quantities. Paired with the unprecedented wealth of
new observational pulsar data, it seems to be within reach for the
first time ever to seriously explore the subatomic structure of matter
in the high-density low-temperature portion of its phase diagram from
observed pulsar data. To this aim Einstein's field equation of
relativistic gravity,
\begin{eqnarray}
  G^{\mu\nu} \equiv R^{\mu\nu} - {{1}\over{2}} g^{\mu\nu} R = 8 \, \pi
  \, T^{\mu\nu}(\epsilon,P(\epsilon)) \, ,
\label{eq:intro.1}
\end{eqnarray} 
is to be solved in combination with the latest theories of the
subatomic structure of matter.\cite{weber99:book} The latter follow
according to the scheme
\begin{eqnarray}
  { {\partial {\cal L}(\{\phi\}) }\over{\partial \phi} } -
    \partial_\mu \; { {\partial {\cal L}(\{\phi\})}\over{\partial
    (\partial_\mu \phi)} } = 0 \quad \Rightarrow \quad P(\epsilon) \, ,
\label{eq:intro.2}
\end{eqnarray} 
where ${\cal L}(\{\phi\})$ denotes a given stellar matter
lagrangian.\cite{weber99:book} In general, ${\cal L}$ is a complicated
functional of the numerous hadron and quark fields, collectively
written as $\{ \phi \}$, that acquire finite amplitudes up to the
highest densities reached in the cores of neutron stars.  According to
what has been said just above, plausible candidates for $\phi$ are the
charged states of the SU(3) baryon octet, $p, n, \Sigma, \Lambda, \Xi$
\cite{glen85:b}, the charged states of the $\Delta$
\cite{weber89:e,huber98:a}, $\pi^-$ \cite{umeda92:a} and $K^-$
\cite{kaplan86:a,li97:a,li97:b,brown96:a,brown97:a} mesons, as well as
$u, d, s$ quarks.\cite{fritzsch73:a,baym78:a,kettner94:b} The
conditions of chemical equilibrium and electric charge neutrality of
stellar matter require the presence of leptons too, in which case
$\phi=e^-, \mu^-$.  Theories of superdense matter enter Einstein's
field equation (\ref{eq:intro.1}) via the energy-momentum tensor
$T^{\mu\nu}$, which contains the equation of state of the stellar
matter, $P(\epsilon)$.  Because of the rather uncertain behavior of
the matter at supernuclear densities, the models derived for the
equation of state differ considerably with from each other. This has
its origin in various sources such as: (1) the many-body technique
used for the determination of the equation of state, (2) the model
adopted for the nucleon--nucleon force, (3) assumptions about the
fundamental building blocks of neutron star matter, (4) the inclusion
of boson condensates, and (5) the considerations of a possible phase
transition of confined hadronic matter into deconfined quark
matter.\cite{weber99:book,heiselberg00:a}

In general Eqs.\ (\ref{eq:intro.1}) and (\ref{eq:intro.2}) were to be
solved simultaneously since the particles move in curved spacetime
whose geometry, determined by Einstein's field equations, is coupled
to the total mass energy $\epsilon$ of the matter.  In the case of
neutron stars, however, the long-range gravitational force can be
cleanly separated from the short-range nuclear force so that Eqs.\
(\ref{eq:intro.1}) and (\ref{eq:intro.2}) decouple from each other.
This simplifies the study of compact stellar objects considerably.

\section{Models for the Equation of State of Superdense Neutron Star Matter}
\label{sec:eos}

\subsection{Nonrelativistic Models}\label{ssec:nonr.eoss}

For non-relativistic models, the starting point are phenomenological
nucleon-nucleon interactions, denoted $V_{ij}$, which fit the
nucleon-nucleon scattering data and the deuteron properties. In order
to achieve the correct binding energy of nuclear matter at the
empirical saturation density $\rho_0$ ($\rho_0=0.15$~nucleons/fm$^3$
which corresponds to a mass density of $\epsilon = 2.5\times 10^{14}$
g/cm$^3$), two-nucleon potentials are supplemented with three-nucleon
interactions $V_{ijk}$. The hamiltonian is then of the form
\begin{equation}
 {\cal H} \; = \; \sum_i \left( {{-\,\hbar^2}\over{2\, m}} \right) \,
              \nabla_i^2 \; + \; \sum_{i<j} V_{ij} \; + \;
              \sum_{i<j<k} V_{ijk} \, .
\label{eq:hamil}
\end{equation}
A proven many-body method frequently adopted to solve the associated
many-body Schroedinger equation ${\cal H} |\Psi> = E |\Psi>$ is the
variational approach \cite{pandharipande74:a,pandharipande79:a} where
a variational trial function $|\Psi_v>$ is constructed from a
symmetrized product of two-body correlation operators ($F_{ij}$)
acting on an unperturbed ground-state, i.e.,
\begin{equation}
      |\Psi_v> \; = \; \Biggl[ \hat S \; \prod_{i<j} \; F_{ij} \Biggr]
      |\Phi> \; ,
\label{eq:psiv}
\end{equation}
where $|\Phi>$ denotes the antisymmetrized Fermi-gas wave function,
\begin{equation}
       |\Phi> \; = \; \hat A \; \prod_j \; {\rm exp}(i\, \bfpm_j
     \bcdot \bfxm_j) \, .
\label{eq:phi} 
\end{equation}
The correlation operator contains variational parameters which are
varied to minimize the energy per baryon for a given density
$\rho$:\cite{pandharipande74:a,pandharipande79:a}
\begin{equation}
      E_v(\rho) \; = \; {\rm min} \;\;\; \left\{ { {<\, \Psi_v\, |\,
{\cal H}\, |\, \Psi_v\, >} \over{<\, \Psi_v\, |\, \Psi_v\, >} }
\right\} \; \; \geq \; \; E_0 \, .
\label{eq:evar}
\end{equation}
As indicated, $E_v$ constitutes an upper bound to the ground-state
energy $E_0$. The energy density $\epsilon(\rho)$ and pressure
$P(\rho)$ are obtained from Eq.\ (\ref{eq:evar}) through the following
manipulations:
\begin{equation}
         \epsilon(\rho) \; = \; \rho \; \left( E_v(\rho)\, + \, m
\right) \; , \qquad \quad P(\rho)\; = \; \rho^2 \; { {\partial\;\;\,
}\over {\partial \rho} }\; E_v(\rho)\; ,
\label{eq:f52}
\end{equation}
which can be combined to an equation of state of the form
$P(\epsilon)$ applicable to the stellar structure calculations.

A representative collection of non-relativistic nuclear equations of
\begin{figure}[tb]
\begin{center}
\leavevmode \psfig{figure=eos2.ps.bb,width=7.0cm,angle=90}
\caption[]{Graphical illustration of $\eau$, $\euu$, $\eut$, $\egth$,
  and $\egthpi$.\protect{\cite{weber99:book}}}
\label{fig:eos2}
\end{center}
\end{figure} 
state is listed in Tables \ref{tab:nonrel.eos} and
\ref{tab:bulk.nonrel.eos}, where the following abbreviations are used:
N = pure neutron matter; NP = $n,\,p$, and leptons in chemical
equilibrium; and $K$ denotes the incompressibility (in MeV) of symmetric
nuclear matter at saturation density.  Several of these equations of
state are shown in Figs.\ \ref{fig:eos2} and \ref{fig:eosbary} where
the pressure is plotted as a function
\begin{figure}[tb]
\begin{center}
\leavevmode
\psfig{figure=eoss.bb,width=7.0cm,angle=90}
\caption[]{Graphical illustration of equations of state HV, HFV,
$\efp$, $\egth$, and $\egdcma$.}
\label{fig:eosbary}
\end{center}
\end{figure}
of total energy density (in units of the density of normal nuclear
matter, $\epsilon_0=140~\mevt$).  The Thomas-Fermi equation of state
TF96 of Table \ref{tab:nonrel.eos} is based on the new Thomas-Fermi
approach of Myers and Swiatecki.\cite{myers95:a}  The effective
interaction $v_{12}$ of this new
\begin{table}[tb]
\caption[]{Non-relativistic models for the equation of state of
neutron star matter.}
\label{tab:nonrel.eos}
\begin{center}
\begin{tabular}{|lll|} \hline
     EOS   &Properties (see text)         &References      \\  \hline 
{\small WFF(UV$_{14}$+TNI)}       &NP, $K$=261 &\cite{wiringa88:a}\\
{\small WFF(UV$_{14}$+UVII)}      &NP, $K$=202 &\cite{wiringa88:a}\\
{\small WFF(AV$_{14}$+UVII)}      &NP, $K$=209 &\cite{wiringa88:a}\\
{\small FP(V$_{14}$+TNI)}         &N,  $K$=240 &\cite{friedman81:a}\\
{\small APR(A$_{18}$+UIX)}        &NP          &\cite{akmal98:a}  \\
{\small TF96}                     &N,  $K$=234 
          &\cite{myers90:a,myers91:a,strobel97:a,strobel99:a}  \\ \hline
\end{tabular}
\end{center}
\end{table}
approach consists of the Seyler-Blanchard potential
\cite{seyler61:a}, generalized by the addition of one momentum
dependent and one density dependent term,\cite{myers95:a}
\begin{eqnarray}
  v_{12} &=& - \frac{2\, T_{0}}{\rho_0} \, Y\bigl(r_{12}\bigr)
\nonumber  \\ 
&\times& \Bigl\{ \frac{1}{2} (1 \mp \xi) \, \alpha - \frac{1}{2}(1
\mp \zeta) \Bigl(\beta \Bigl( \frac{p_{12}}{\kFo} \Bigr)^2 - \gamma \,
\frac{\kFo}{p_{12}} + \sigma \Bigl( \frac{2 \bar \rho}{\rho_0}
\Bigr)^{\frac{2}{3}} \Bigr) \Bigr\} \, . \ ~~~~~~ \ 
\label{eq:v.tf96}
\end{eqnarray} The upper (lower) sign in (\ref{eq:v.tf96}) corresponds
to nucleons with equal (unequal) isospin.  The quantities $\kFo$,
$T_{0}$ $(= \kFosq / 2m)$, and $\rho_0$ are the Fermi momentum, the
Fermi energy and the particle density of symmetric nuclear matter. The
potential's radial dependence is describe by the normalized Yukawa
interaction
\begin{equation}
  Y\bigl(r_{12}\bigr) = {{1}\over{4\, \pi\, a^3}} \; { {e^{-\,
        r_{12}/a}}\over{r_{12}/a} } \, .
\label{eq:2.ms96}
\end{equation} Its strength depends both on the magnitude of the
particles' relative momentum, $p_{12}$, and on an average of the
densities at the locations of the particles.  The parameters $\xi$ and
$\zeta$ were introduced in order to achieve better agreement with
asymmetric nuclear systems, and the behavior of the optical potential
is improved by the term $\sigma (2\bar\rho/\rho_0)^{2/3}$. Here the
average density is defined by $\overline{\rho}^{2/3} = (\rho_{1}^{2/3}
+ \rho_{2}^{2/3})/2$, where $\rho_1$ and $\rho_2$ are the relevant
densities of the interacting particle (neutron or protons) at points 1
\begin{table}[tb]
\caption[]{Nuclear matter properties of the equations of state
compiled in Table \ref{tab:nonrel.eos}.}
\label{tab:bulk.nonrel.eos} 
\begin{center}
\begin{tabular}{|llllll|} \hline
{\small EOS}  &$E/A$  &$\rho_0$     &$K$    &$M^*$  &$a_{\rm sy}$   \\
              &(MeV)  &(fm$^{-3})$  &(MeV)  &(MeV)  &(MeV)   \\ \hline 
{\small WFF(UV$_{14}$+TNI)} &$-16.6$    &0.157  &261   &0.65 &30.8  \\
{\small WFF(UV$_{14}$+UVII)}&$-11.5$    &0.175  &202   &0.79 &29.3  \\
{\small WFF(AV$_{14}$+UVII)}&$-12.4$    &0.194  &209   &0.66 &27.6  \\
{\small FP(V$_{14}$+TNI)}   &$-16.00$   &0.159  &240   &0.64 &$-$    \\
{\small APR(A$_{18}$+UIX)}  &$-16.00$   &0.16   &-     &-    &$-$    \\
{\small TF96}               &$-16.04$   &0.161  &234   &$-$  &32.0  \\ \hline
\end{tabular}
\end{center}
\end{table}
and 2. The potential's seven free parameters $\alpha,~ \beta,~
\gamma,~ \sigma,~ \xi,~ \zeta,~ a$ are adjusted to the properties of
finite nuclei, the parameters of the mass formula, and the behavior of
the optical potential.\cite{strobel97:a} The nuclear matter properties
at saturation obtained for TF96 are summarized in Table
\ref{tab:bulk.nonrel.eos}, where the listed quantities are: binding
energy of normal nuclear matter at saturation density, $E/A$;
compression modulus, $K$; effective nucleon mass, $M^*$ $(\equiv
m^*/m$ where $m$ denotes the free nucleon mass); and the asymmetry
energy, $a_{\rm sy}$.  The new Thomas-Fermi force has the advantage
over the standard Seyler-Blanchard interaction to not only reproduce
the ground-state properties of finite nuclei and infinite symmetric
nuclear matter, but also the optical potential and, as revealed by
stellar structure calculations \cite{strobel97:a}, the properties of
neutron stars as well. These features render the new Thomas-Fermi
model very attractive for investigations of the properties of dense
nuclear matter.

\subsection{Relativistic Fieldtheoretical Models}\label{ssec:r.eoss}

\begin{figure}[tb]
\begin{center}
\leavevmode
\psfig{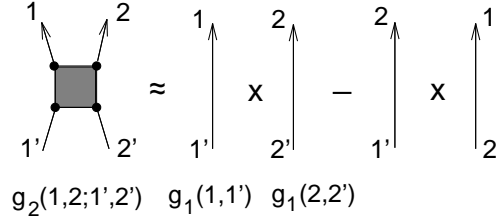}
\caption[]{Factorization of four-point baryon Green function $g_2$
  into antisymmetrized products of two-point baryon Green functions,
  $g_1 \times g_1$. Direct (Hartree) and exchange (Fock) contribution
  are shown. This factorization scheme truncates the many-body
  equations at the Hartree-Fock level.}
\label{fig:g2hf}
\end{center}
\end{figure}

Relativistic, field-theoretical theories of dense nuclear matter and
finite nuclei have enjoyed a renaissance in recent years, and they
have the virtue of describing nuclear matter at saturation, many
features of finite nuclei, both spherical and deformed, and they
extrapolate causally to high density.\cite{serot86:a} The starting
point of relativistic theories of superdense neutron star matter is a
lagrangian of the following type,\cite{weber99:book}
\begin{eqnarray}
{\cal L}(x) \, &=&  \, \sum_{B=p,n,\Sigma^{\pm,0},\Lambda,\Xi^{0,-},
        \Delta^{++,+,0,-} }  {\cal L}^0_B(x) \label{eq:f31} \\
  &+& \sum_{M=\sigma,\omega,\pi,\varrho,\eta,\delta,\phi}
     \left\{ {\cal L}^0_M(x) \;+\; \sum_{B=p,n,...,\Delta^{++,+,0,-}}
  {\cal L}^{\rm Int}_{B,M} (x) \right\} 
  + \sum_{L=e^-,\mu^-} \;{\cal L}_L (x)\;. \nonumber
\end{eqnarray} The subscript $B$ runs over all baryon species that may become
populated in dense neutron star matter (Table \ref{tab:bary}). The
nuclear forces are mediated by a collection of scalar, vector, and
isovector mesons \cite{machleidt87:a} which is compiled in
Table \ref{tab:mesons}.
\begin{table}
\caption[]{Masses, quantum numbers (spin $J_B$, isospin $I_B$,
strangeness $S_B$, hypercharge $Y_B$, third component of isospin
$I_{3B}$), and electric charges ($q_B$) of baryons.}
\label{tab:bary}
\begin{center}
\begin{tabular}{|cccrrrrr|}\hline
Baryon ($B$) &$m_B\;($MeV$)$ &$J_B$ &$I_B$ &$S_B$ &$Y_B$ &$I_{3B}$ &$q_B$\\
                                                        \hline 
$n$ &939.6    &$1/2$ &$1/2$ &0     &1    &$-1/2$ &0            \\
$p$ &938.3    &$1/2$ &$1/2$ &0     &1    &$1/2 $ &1            \\
$\Sigma^+$    &1189  &$1/2$ &1     &$-1$ &0      &1      &1    \\
$\Sigma^0$    &1193  &$1/2$ &1     &$-1$ &0      &0      &0    \\
$\Sigma^-$    &1197  &$1/2$ &1     &$-1$ &0      &$-1$   &$-1$ \\
$\Lambda $    &1116  &$1/2$ &0     &$-1$ &0      &0      &0    \\
$\Xi^0$       &1315  &$1/2$ &$1/2$ &$-2$ &$-1$   &$1/2$  &0    \\
$\Xi^-$       &1321  &$1/2$ &$1/2$ &$-2$ &$-1$   &$-1/2$ &$-1$ \\
$\Delta^{++}$ &1232  &$3/2$ &$3/2$ &0    &1      &$3/2$  &2    \\
$\Delta^{+}$  &1232  &$3/2$ &$3/2$ &0    &1      &$1/2$  &1    \\
$\Delta^{0}$  &1232  &$3/2$ &$3/2$ &0    &1      &$-1/2$ &0    \\
$\Delta^{-}$  &1232  &$3/2$ &$3/2$ &0    &1      &$-3/2$ &$-1$ \\  \hline
\end{tabular}
\end{center}
\end{table} The equations of motion that follow from Eq.\ (\ref{eq:f31}) 
for the baryon fields are given by\cite{weber99:book}
\begin{eqnarray}
  && \left( i \gamma^\mu\partial_\mu-m_B \right)\, \psi_B(x) = 
  g_{\sigma B}\, \sigma(x) \, \psi_B(x) \nonumber \\ 
  &&\qquad\qquad~~~+ \Bigl\{ g_{\omega B}\gamma^\mu\omega_\mu(x)
  +{f_{\omega B}\over{4m_B}} \sigma^{\mu\nu} F_{\mu\nu}(x)\Bigr\} \,
  \psi_B(x) \nonumber \\ &&\qquad\qquad~~~+ \Bigl\{ g_{\rho B}\;
  \gamma^\mu \; {\btau}
\bcdot {\brho}_\mu(x) + {f_{\rho B}\over{4\,m_B}}
  \; \sigma^{\mu\nu} \; {\btau}\bcdot{\bfG}_{\mu\nu}(x) \Bigr\}\,
  \psi_B(x) \nonumber \\ &&\qquad\qquad~~~ + {f_{\pi B}\over{m_\pi}}\;
  \gamma^\mu \gamma^5 \; \bigl(\partial_\mu\;
  {\btau}\bcdot{\bpi}(x)\bigr) \; \psi_B(x) \, ,
\label{eq:eompsi}
\end{eqnarray} where the standard field-theoretical notation of 
Ref.\ \cite{bjorken65:a} is used.  The meson fields in
(\ref{eq:eompsi}) obey the following field equations,\cite{weber99:book}
\begin{eqnarray}
  \bigl( \partial^\mu\partial_\mu+m^2_\sigma)\, \sigma(x) = &-& \sum_B
  g_{\sigma B}\, \bar\psi_B(x) \psi_B(x) 
-  m_N \, b_N
  \, g_{\sigma N} \left( g_{\sigma N} \sigma(x) \right)^2 \nonumber \\
  &-& c_N\, g_{\sigma N} \left( g_{\sigma N} \sigma(x) \right)^3 \, ,
\label{eq:eomsigma} 
\end{eqnarray}
\begin{eqnarray}
\partial^\mu F_{\mu\nu}(x) + m_\omega^2 \, \omega_\nu(x) &=& \sum_B 
\Bigl\{ g_{\omega B}\; \bpsiB(x) \gamma_\nu \psiB(x) 
\nonumber \\ &&~~~~~~~ 
- {{f_{\omega B}}\over{2m_B}}\, \partial^\mu \left(
\bpsiB(x) \sigma_{\mu\nu} \psiB(x) \right) \Bigr\} \, , ~~~~~ \ ~
\label{eq:eomom5} \\
  \left( \partial^\mu \partial_\mu + m^2_\pi \right) \, {\bpi}(x) &=&
  \sum_B {f_{\pi B}\over{m_\pi}} \; \partial^\mu \left( \bar \psi_B(x)
  \;\gamma_5 \,\gamma_\mu \, {\btau} \; \psi_B(x) \right)\, ,
\label{eq:eompi4} \\
\partial^\mu {\bfG}_{\mu\nu}(x) + m_\rho^2 \, \brho_\nu(x)
&=& \sum_B \Bigl\{ g_{\rho B}\; \bpsiB(x) \btau \gamma_\nu \psiB(x)
\nonumber \\ &&~~~~~~~ - {{f_{\rho B}}\over{2m_B}} \; \partial^\lambda
\left( \bpsiB(x) \btau \sigma_{\mu\nu} \psiB(x) \right) \Bigr\} \, ,
\label{eq:eomrho5}
\end{eqnarray} with the field tensors $F^{\mu\nu}$ and ${\bfG}^{\mu\nu}$ 
defined as $F^{\mu\nu} = \partial^\mu\omega^\nu -
\partial^\nu\omega^\mu$ and ${\bfG}^{\mu\nu} = \partial^\mu{\brho}^\nu
- \partial^\nu{\brho}^\mu$.  Equations~(\ref{eq:eompsi}) through
(\ref{eq:eomrho5}) are to be solved simultaneously in combination with
the neutron star matter constraints of electric charge neutrality and
chemical equilibrium. The former can be written as
\begin{eqnarray}
  \sum_B q_B \; (2J_B+1)\; {{p_{F,B}^3}\over{6\pi^2}}\; - \;
  \sum_{\lambda=e,\mu} { {p_{F,\lambda}^3}\over{3\pi^2}} \, - \,
  \varrho_M \; \Theta(\mu^M - m_M) \; = \; 0 \, ,
\label{eq:chargeneut}
\end{eqnarray} while chemical ($\beta$) equilibrium is expressed as 
\begin{equation}
\mu^B = \mu^n - q_B \, \mu^e \, ,
\label{eq:mub}
\end{equation} 
where $\mu^n$ and $\mu^e$ denote the chemical potentials of neutrons
and electrons whose knowledge is sufficient to determine the chemical
\begin{table}[tb]
\caption[]{Mesons and their quantum numbers.\protect{\cite{pdg84:a}}
  The entries are: spin $J_M$, parity $\pi$, isospin $I_M$, and mass
  $m_M$ of meson $M$.}
\label{tab:mesons}
\begin{center}
\begin{tabular}{|ccclc|}\hline
Meson ($M$) &$J^\pi_M$  &$I_M$  &Coupling          &Mass (MeV)   \\ \hline
$\sigma$    &$0^+$      &0      &scalar         &550        \\
$\omega$    &$1^-$      &0      &vector         &783        \\
$\pi^\pm$   &$0^-$      &1      &pseudovector   &140         \\
$\pi^0$     &$0^-$      &1      &pseudovector   &135         \\
$\rho$      &$1^-$      &1      &vector         &769         \\
$\eta$      &$0^-$      &0      &pseudovector   &549         \\
$\delta$    &$0^+$      &1      &scalar         &983         \\
$\phi$      &$1^-$      &0      &vector         &1020        \\  
$K^+$       &$0^-$     &1/2     &pseudovector   &494          \\
$K^-$       &$0^-$     &1/2     &pseudovector   &494         \\
\hline
\end{tabular}
\end{center}
\end{table}
potential of any other baryon present in the system.  The only mesons
that may plausibly condense (density $\rho_M$) in neutron star matter
are the $\pi^-$ \cite{umeda92:a,tatsumi91:a} or the $K^-$
\cite{kaplan86:a,brown87:a,lee95:a,pons01:b}.
Equations~(\ref{eq:chargeneut}) and (\ref{eq:mub}) leads to
constraints on the Fermi momenta of baryons and leptons, $\kFB$ and
$\kFL$ respectively.  The leptons in neutron star matter can be
treated as free particles, which obey the following Dirac equation,
\begin{equation}
\left( \i \gamma^\mu\partial_\mu - m_L \right)\; \psi_L(x) = 0 \, ,
\label{eq:ffermiions}
\end{equation} 
with $L=e^-, \mu^-$. An elegant mathematical framework that enables
one to solve the equations of motion (\ref{eq:eompsi}) through
(\ref{eq:ffermiions}) is the relativistic Greens function method, which
is outlined next.  The starting point is the Martin-Schwinger
hierarchy of coupled Greens functions.\cite{wilets79:a} In the lowest
order, the Martin-Schwinger hierarchy can be truncated by factorizing
the four-point Greens function $g_2(1,2;1'2')$, where primed
(unprimed) arguments refer to ingoing (outgoing) fermions, into a
product of two-point Greens functions $g_1(1;1')$, as schematically
shown in Fig.\ \ref{fig:g2hf}.  This leads to the well-known
relativistic Hartree (i.e., mean-field) and Hartree-Fock
approximations.  The $T$-matrix approximation, also known as ladder
($\Lambda$) approximation, goes beyond Hartree and Hartree-Fock by
truncating the Martin-Schwinger hierarchy by factorizing the six-point
Greens function $g_3(123;1'2'3')$ into products of four-point and
two-point functions by means of which dynamical two-particle
correlations in matter, which are connected to a two-body
boson-exchange potential $v$, are taken into account. The $T$ matrix
(effective two-particle interaction in matter) is computed from an
integral equation of the following type, \cite{weber99:book}
\begin{equation}
T\; =\; v - v^{\rm ex} + \int v \; \Lambda \; T \, ,
\label{eq:tmatrix}
\end{equation} with $\Lambda$ the baryon-baryon propagator
in intermediate scattering states. In the framework of the so-called
$\Lambda^{00}$ approximation, the baryons in intermediate states
propagate as free particles, leading to a $\Lambda \approx
\Lambda^{00}\equiv i g^0_1 g^0_1$. This is in sharp contrast to the
relativistic Brueckner-Hartree-Fock (RBHF) approximation where the
intermediate baryons are coupled to the nuclear background and,
therefore, are constrained to scattering states outside their
respective Fermi seas.\cite{huber93:a,huber94:a,huber95:a} The basic
input quantity in Eq.\ (\ref{eq:tmatrix}) is the nucleon-nucleon
interaction in free space. Representative models are, for instance,
the Bonn meson-exchange model \cite{machleidt87:a} and the
boson-exchange potentials of Brockmann and Machleidt.\cite{brock90:a}
A characteristic feature of these potentials is that
their parameters are adjusted to the relativistic two-nucleon
scattering data and the properties of the deuteron. In this respect a
parameter-free treatment of the many-body problem is achieved.  The
Born approximation to the $T$-matrix, which merely sums numerous
boson-exchange potentials,
\begin{equation} <1 2\mid v \mid 1' 2'>\; = \;
\sum_{M=\sigma,\omega,\pi,\rho,\eta,\delta, \phi} \; \delta^4_{11'}
\;\; \Gamma^M_{11'} \;\; \Delta^M_{12} \;\; \Gamma^M_{22'} \;\;
\delta^4_{22'} \, ,
\label{eq:vborn}
\end{equation} neglects dynamical correlations among baryons. It is
this approximation which reduces the ladder approximation to the
relativistic Hartree and Hartree-Fock
approximations.\cite{weber99:book} The symbols $\Gamma^M$ in Eq.\
(\ref{eq:vborn}) stands for the various meson-nucleon vertices, and
$\Delta^M$ denotes the free meson propagator of a meson of type $M$
whose explicit forms are given Ref.\ \cite{weber99:book}.  The baryon
self-energy, or mass operator, is obtained from the $T$ matrix
according to\cite{weber99:book}
\begin{equation}  
\Sigma^B \; = \; i \; \sum_{B'=p,n,\Sigma^{\pm,0},\Lambda,\Xi^{0,-},
\Delta^{++,+,0,-}} \int \; \left[ \; {\rm tr}\; \left(
T^{BB'} \; g_1^{B'} \right) - T^{BB'} \; g_1^{B'}\; \right] \, ,
\label{eq:sigmal}
\end{equation}
where $B'$ sums all the charged baryon states whose thresholds are
reached in neutron star matter treated within this framework. 
\begin{figure}[tb]
\begin{center}
\leavevmode 
\psfig{figure=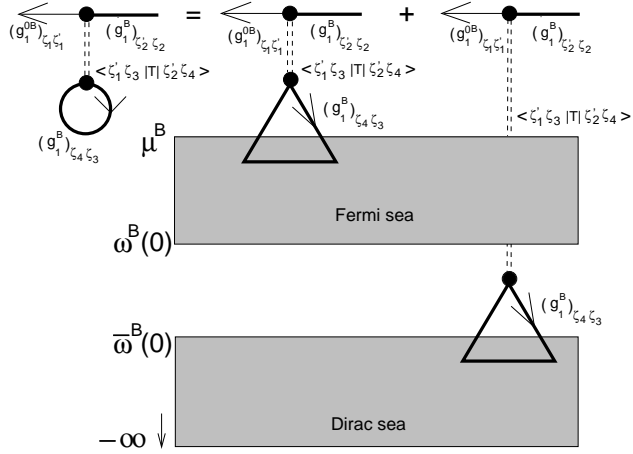,width=6.0cm,angle=-90}
\caption[]{Graphical representation of $g^{0B}_1 \Sigma^B g^B_1$ of
Dyson's equation (\protect{\ref{eq:dyson}}).  The Fermi sea of each
baryon is filled up to the highest single-particle energy state,
$\mu^B$. The infinite Dirac sea is filled with antibaryons, which too
modify the motion of the baryons.  The Fermi and Dirac-sea graphs are
referred to as medium and vacuum polarization contributions,
respectively.\cite{serot86:a,horowitz87:a}}
\label{fig:fermisea}
\end{center}
\end{figure}
The baryon Greens functions $g^B_1$ in Eq.\ (\ref{eq:sigmal}) are
given as solutions of Dyson's equation,
\begin{equation}
      g^B_1 \; = \; g^{0\,B}_1 \; + \; g^{0\,B}_1 \; \Sigma^B(\{
      g^{B'}_1 \}) \; g^B \, ,
\label{eq:dyson} 
\end{equation} which terminates the set of equations to be solved
self-consistently.  The diagrammatic representation of the second term
\begin{table}[tb]
\caption[]{Relativistic, field-theoretical models for the equation of
state of neutron star matter.}
\label{tab:rel.eos}
\begin{center}
\begin{tabular}{|lll|} \hline
     EOS   &Properties (see text)         &References      \\  \hline 
G$^{\rm DCM1}_{225}$        &H, $K$=225   &\cite{glen91:b} \\
G$^{\rm DCM2}_{265}$        &H, $K$=265   &\cite{glen91:b} \\
$\KM$                       &H, $K$=240   &\cite{glen97:book} \\
G$_{300}$                   &H, $K$=300   &\cite{glen89:a493} \\
$\egthK$                    &H,$K^-$, $K$=300   &\cite{glen89:a493}    \\
G$^\pi_{200}$               &H, $\pi$, $K$=200 &\cite{glen86:b} \\
G$^\pi_{300}$               &H, $\pi$, $K$=300 &\cite{glen89:a493} \\
HV                          &H, $K$=285   &\cite{glen85:b,weber89:e} \\
$\ebnhv$                    &H, $K$=186        &\cite{weber90:a} \\
HFV                         &H,\,$\Delta$, $K$=376  &\cite{weber89:e} \\
$\ehehfv$                   &H,\,$\Delta$, $K$=115 
                                        &\protect{\cite{weber90:a}} \\
$\eBrBhfv$                  &H,\,$\Delta$, $K$=249
                             &\cite{huber93:a,huber94:a,huber95:a} \\ \hline
\end{tabular}
\end{center}
\end{table}  
of Dyson's equation is shown in Fig.\ \ref{fig:fermisea}. It is this
term which corrects the free Greens function for medium effects
originating from the presence of Fermi seas of filled baryons.  The
equation of state follows from the stress-energy density tensor,
${T}_{\mu\nu}$, of the system as\cite{weber99:book}
\begin{eqnarray} E(\rho) \; &=& \; <\,{T}_{00}\,> / \rho \; - \; m \; ,
 \qquad\qquad {\rm where} \\
\label{eq:engpn}
{T}_{\mu\nu}(x)\; &=&\; \sum_{\chi=B, L} \partial_\nu\psi_\chi(x)
\; {{\partial {\cal L}(x)} \over{\partial
\;(\partial^\mu\,\psi_\chi(x))}} \;-\;g_{\mu\nu} \; {\cal L}(x)\, .
\label{eq:emtensor}
\end{eqnarray} 
The pressure is obtained from $E(\rho)$ as in Eq.\ (\ref{eq:f52}).

A broad collection of relativistic models for the equation of state of
superdense neutron star matter is compiled in Table \ref{tab:rel.eos}.
The specific properties of these equations of state are described in
Table \ref{tab:rel.eos.bulk}, where the
\begin{table}[tb]
\caption[]{Nuclear matter properties of the equations of state
compiled in Table \ref{tab:rel.eos}.}
\label{tab:rel.eos.bulk}
\begin{center}
\begin{tabular}{|llllll|} \hline
{\small EOS}  &$E/A$  &$\rho_0$     &$K$    &$M^*$  &$a_{\rm sy}$   \\
              &(MeV)  &(fm$^{-3})$  &(MeV)  &(MeV)  &(MeV)   \\ \hline 
$\egdcmb$                &$-16.0$   &0.16   &225    &0.796  &32.5   \\
$\egdcmd$                &$-16.0$   &0.16   &265    &0.796  &32.5  \\
$\KM$                    &$-16.3$   &0.153  &240    &0.78   &32.5  \\
G$_{300}$                &$-16.3$   &0.153  &300    &0.78   &32.5  \\
$\egthK$                 &$-16.3$   &0.153  &300    &0.78   &32.5  \\
G$^\pi_{200}$            &$-15.95$  &0.145  &200    &0.8    &36.8  \\
G$^\pi_{300}$            &$-16.3$   &0.153  &300    &0.78   &32.5  \\
HV                       &$-15.98$  &0.145  &285    &0.77   &36.8  \\
$\ebnhv$                 &$-11.9$   &0.134  &186    &0.79   &34    \\
HFV                      &$-15.54$  &0.159  &376    &0.62   &30    \\
$\ehehfv$                &$-8.7$    &0.132  &115    &0.82   &29   \\
$\eBrBhfv$               &$-15.73$  &0.172  &249    &0.73   &34.3   \\ \hline
\end{tabular}
\end{center}
\end{table}
following abbreviations are used: N = pure neutron; NP = $n,\,p$,
leptons; $\pi$ = pion condensation; H = composed of $n,\,p$, hyperons
($\Sigma^{\pm,0},\,\Lambda,\,\Xi^{0,-}$), and leptons; $\Delta$ =
$\Delta_{1232}$-resonance; $K^-$ = condensate of negatively charged
kaons; and $K$ = incompressibility (in MeV). All equations of state of
this collection account for neutron star matter in full baryonic
equilibrium. A few selected equations of state of this collection are
shown in Fig.\ \ref{fig:eosbary}.  An inherent feature of relativistic
equations of state is that they do not violate causality, i.e., the
velocity of sound given by $v_s = \sqrt{dP/d\epsilon}$ is smaller than
the velocity of light at all densities. This is not the case for most
non-relativistic models for the equation of state.  Among the latter,
$\ewut$, $\eapr$, and the TF96 models for the equation of state do
not violate causality up to densities relevant for the construction of
models of neutron stars from them.

\subsection{Baryon-Lepton Composition of Neutron Star Matter}\label{sec:comp}

At densities lower than the density of normal nuclear matter,
$\epsilon_0$ ($= 140~\mevt$), neutron star matter consists of only $p,
n, e^-,$ and $\mu^-$ whose densities are determined by the matter
equations (\ref{eq:chargeneut}) through (\ref{eq:dyson}). At densities
larger than $\epsilon_0$ the more massive baryon states
$\Sigma^{\pm,0},\Lambda,\Xi^{0,-},\Delta^{++,\pm,0}$ become populated
since their chemical potentials, $\mu^B$, become larger than the
lowest-lying energy eigenstates $\omega^B({\bfpm}=0)$ of these
particles,
\begin{eqnarray}
\mu^B \equiv \mu^n - q_B\, \mu^e &\geq& \omega^B({\bfpm}=0) \, .
\label{eq:thresh}
\end{eqnarray}
In relativistic field theory, the energy eigenstates are given
by\cite{weber99:book}
\begin{equation}
\omega^B({\bfpm}) = \Sigma^B_0(\omega^B({{\bfpm}}),{{\bfpm}}) + I_{3
B} \; \Sigma^B_{03}(\omega^B({{\bfpm}}),{{\bfpm}}) + [ m_B +
\Sigma^B_S(\omega^B({{\bfpm}}),{{\bfpm}})] \, ,
\end{equation}
while in the non-relativistic case one has
\begin{equation}
  \omega^B({\bfpm}) = {1 \over {2\, m_B}} \; {\bfpm}^2 +
  \Sigma^B({{\bfpm}}) \, ,
\end{equation} 
where $\Sigma^B$ denotes the non-relativistic one-particle potential
felt by a baryon in matter.  Since the chemical potentials $\mu^n$ and
$\mu^e$ are positive and the self-energies $\Sigma^B_{03}$ are
negative ($\Sigma^B_{03}$ is proportional to the difference of proton
and neutron densities, which is negative for neutron star matter
because of its neutron excess), it follows from Eq.\ (\ref{eq:thresh})
that (i) negatively charged baryons are charge-favored, and (ii)
baryons having the opposite isospin orientation as the neutron (i.e.,
$I_{3B} > 0$) are isospin-favored.\cite{glen85:b} The physical reason
behind item (i) is that
\begin{figure}[tb]
\begin{center}
\parbox[t]{5.9 cm}
{\leavevmode \psfig{figure=popbl2.ps.bb,width=6.5cm,angle=90}
  {\caption[]{Baryon-lepton composition, normalized to the total baryon
      density $\rho$, of neutron star matter computed for HV.}
\label{fig:f3twn}}}
\ \hskip 0.5cm   \
\parbox[t]{5.9 cm}
{\leavevmode \psfig{figure=popbl3.ps.bb,width=6.5cm,angle=90}
  {\caption[]{Same as figure \protect{\ref{fig:f3twn}}, but computed for HFV.}
\label{fig:f4twn}}}
\end{center}
\end{figure} 
negatively charged baryons replace electrons with high Fermi momenta,
while (ii) is making the matter more isopin symmetric. Both items
enable neutron star matter to settle down in an energetically more
favorable state.  Figures \ref{fig:f3twn} through \ref{fig:10.Hub97a}
illustrate possible particle compositions of neutron star matter
computed for different many-body techniques as well as different
nuclear forces.  One sees that in relativistic Hartree (HV) the
$\Lambda$ hyperon, which remains unaffected by the effects (i) and
(ii) since its electric charge $q_\Lambda=0$ and the third component
of isospin $I_{3\Lambda}=0$, has the lowest threshold of all hyperons.
In the case of relativistic Hartree-Fock (HFV), the charge favored but
\begin{figure}[tb]
\begin{center}
\parbox[t]{5.9 cm} {\leavevmode
\psfig{figure=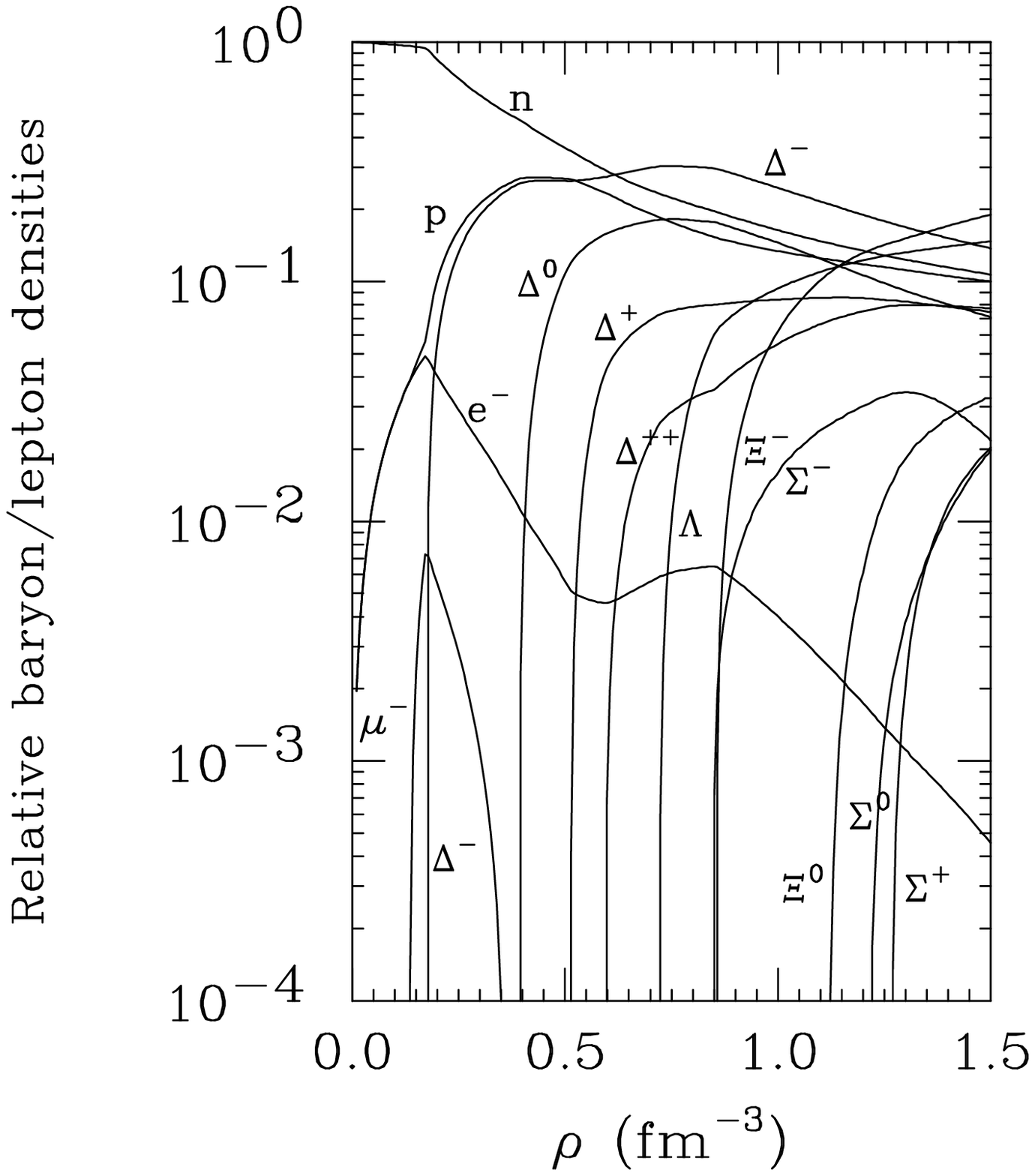,width=6.5cm,angle=90} {\caption[]{Same as
figure \protect{\ref{fig:f3twn}}, but for RBHF\,1 parameter set (see
Ref.\ \protect{\cite{weber99:book}}). The hyperons are non-universally
coupled, the $\Delta$'s universally.}
\label{fig:9.Hub97a}}}
\ \hskip 0.5cm   \
\parbox[t]{5.9 cm} {\leavevmode
\psfig{figure=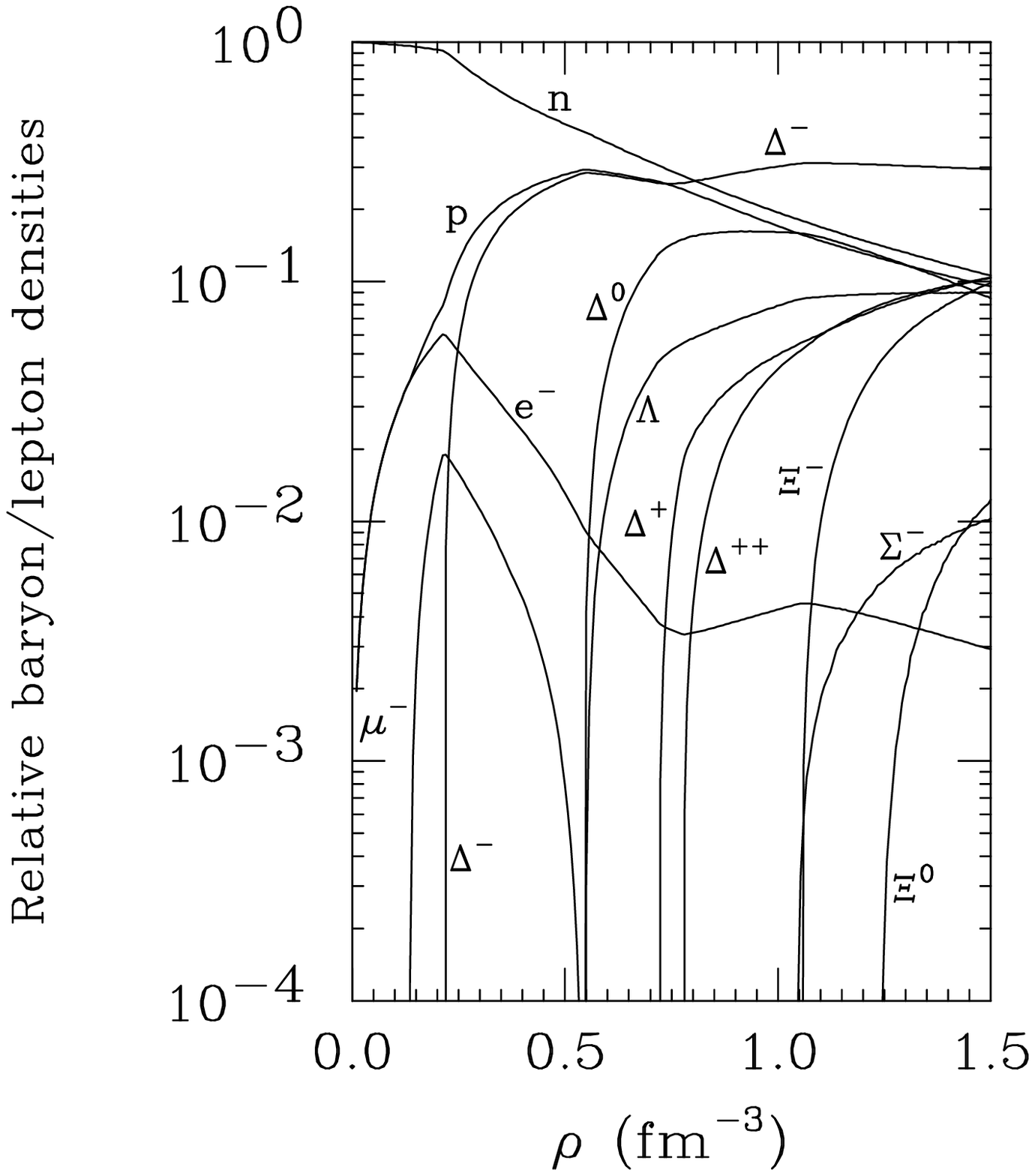,width=6.5cm,angle=90} {\caption[]{Same
as Fig.\ \protect{\ref{fig:9.Hub97a}}, but for non-universally
coupled hyperons and $\Delta$'s.\protect{\cite{huber98:a}}}
\label{fig:10.Hub97a}}}
\end{center}
\end{figure}
isospin unfavored $\Sigma^-$ possesses the lowest threshold, and the
$\Delta$ makes its appearance too. The latter is also the case for
relativistic Brueckner-Hartree-Fock type calculations.\cite{huber98:a}
The important net effect of all these new degrees of freedom is a
softening of the equation of state at supernuclear densities, which
reduces the maximum mass of a neutron star.

\subsection{H-dibaryons}

A novel particle that may make its appearance in the center of a
neutron star is the H-dibaryon, a doubly strange six-quark composite
with spin and isospin zero, and baryon number two.\cite{jaffe77:a}
Since its first prediction in 1977, the H-dibaryon has been the
subject of many theoretical and experimental studies as a
\begin{figure}[tb]
\begin{center}
\leavevmode 
\psfig{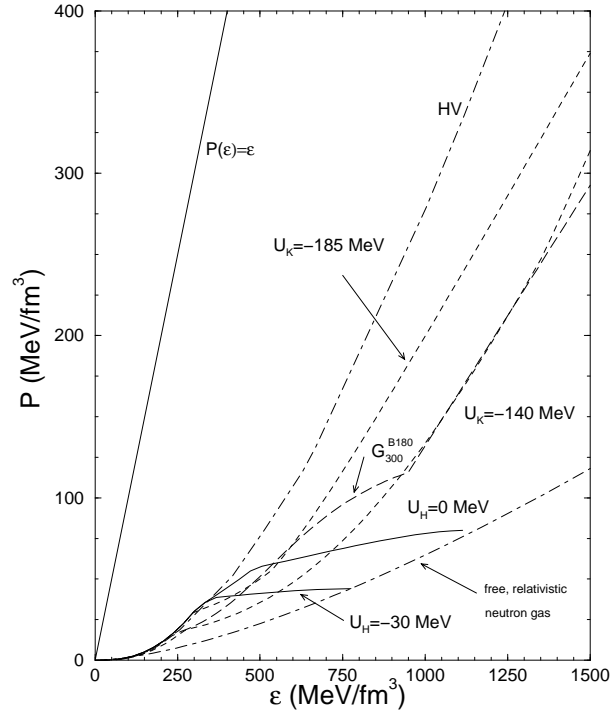} 
\caption[]{Models for the equation of state of neutron star
  matter.\protect{\cite{weber99:book}}}
\label{fig:eos_coll}
\end{center}
\end{figure} 
possible candidate for a strongly bound exotic state.  In neutron
stars, which may contain a significant fraction of $\Lambda$ hyperons
(Figs.\ \ref{fig:f3twn} through \ref{fig:10.Hub97a}), the $\Lambda$'s
could combine to form H-dibaryons, which could give way to the
formation of H-matter at densities somewhere between $3\, \epsilon_0$
and $6\, \epsilon_0$ depending on the in-medium properties of the
H-dibaryon.\cite{glen98:a,tamagaki91:a,sakai97:a} This could cause a
significant softening of the equation of state as shown in Fig.\
\ref{fig:eos_coll} (curves labeled $U_H$).  In Ref.\ \cite{glen98:a}
it was shown that H-dibaryons with a vacuum mass of about 2.2~GeV and
a moderately attractive potential in the medium of about $U_H = -
30$~MeV could go into a Bose condensate in the cores of neutron stars
if the limiting star mass is about that of the Hulse-Taylor pulsar
PSR~1913+16, $M=1.444\, \msun$.  Conversely, if the medium potential
were moderately repulsive, around $U_H = + 30$~MeV, the formation of
H-dibaryons may only take place in heavier neutron stars of mass
$M\gsim 1.6\, \msun$. If indeed formed, however, H-matter may not
remain dormant in neutron stars but, because of its instability
against compression could trigger the conversion of neutron stars into
hypothetical strange stars.\cite{sakai97:a,faessler97:a,faessler97:b}

\subsection{Condensation of $K^-$ Mesons}

\begin{figure}[tb] 
\begin{center}
\leavevmode
\psfig{figure=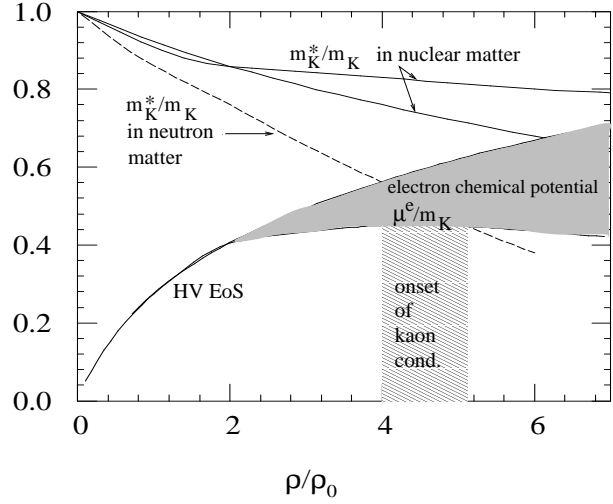,width=8.0cm,angle=0}
\caption[]{Effective kaon mass in nuclear and neutron star
  matter.\protect{\cite{mao99:a,waas97:a,gal01:a,cieply01:a,kaiser01:a}}}
\label{fig:Kmass}
\end{center}
\end{figure}

Once the reaction 
\begin{equation}
  e^- \rightarrow K^- + \nu
\label{eq:kaon.1}
\end{equation} becomes possible in a neutron star, it becomes energetically
advantageous for the star to replace the fermionic electrons with the
bosonic $K^-$ mesons. Whether or not this actually happens depends on
how quickly the $K^-$ drops with density in dense matter (Fig.\
\ref{fig:Kmass}).
An analysis of the KaoS data \cite{barth97:a} shows that the
attraction from nuclear matter may bring the $K^-$ mass down to
$m^*_{K^-}\simeq 200$~MeV at $\rho\sim 3\, \rho_0$. For neutron-rich
matter, the $K^-$ mass may drop as\cite{li97:a,li97:b,brown96:a,brown97:a} 
\begin{equation}
  m^*_{K^-}(\rho) \simeq m_{K^-} \left( 1 - 0.2 \,
  {{\rho}\over{\rho_0}} \right) \, ,
\label{eq:meff02}
\end{equation} 
with $m_K = 495$~MeV the $K^-$ vacuum mass.  Combining this relation
with the outcome for $\mu^e$ shown in Fig.\ \ref{fig:Kmass} shows that
the threshold condition for the onset of $K^-$ condensation, $\mu^e =
m^*_K$ would be fulfilled at densities $\rho \gsim 3 \rho_0$, which
are easily reached in neutron stars.

\subsection{Quark Deconfinement}\label{ssec:qdeconf}

\begin{figure}[tb] 
\begin{center}
\leavevmode
\psfig{figure=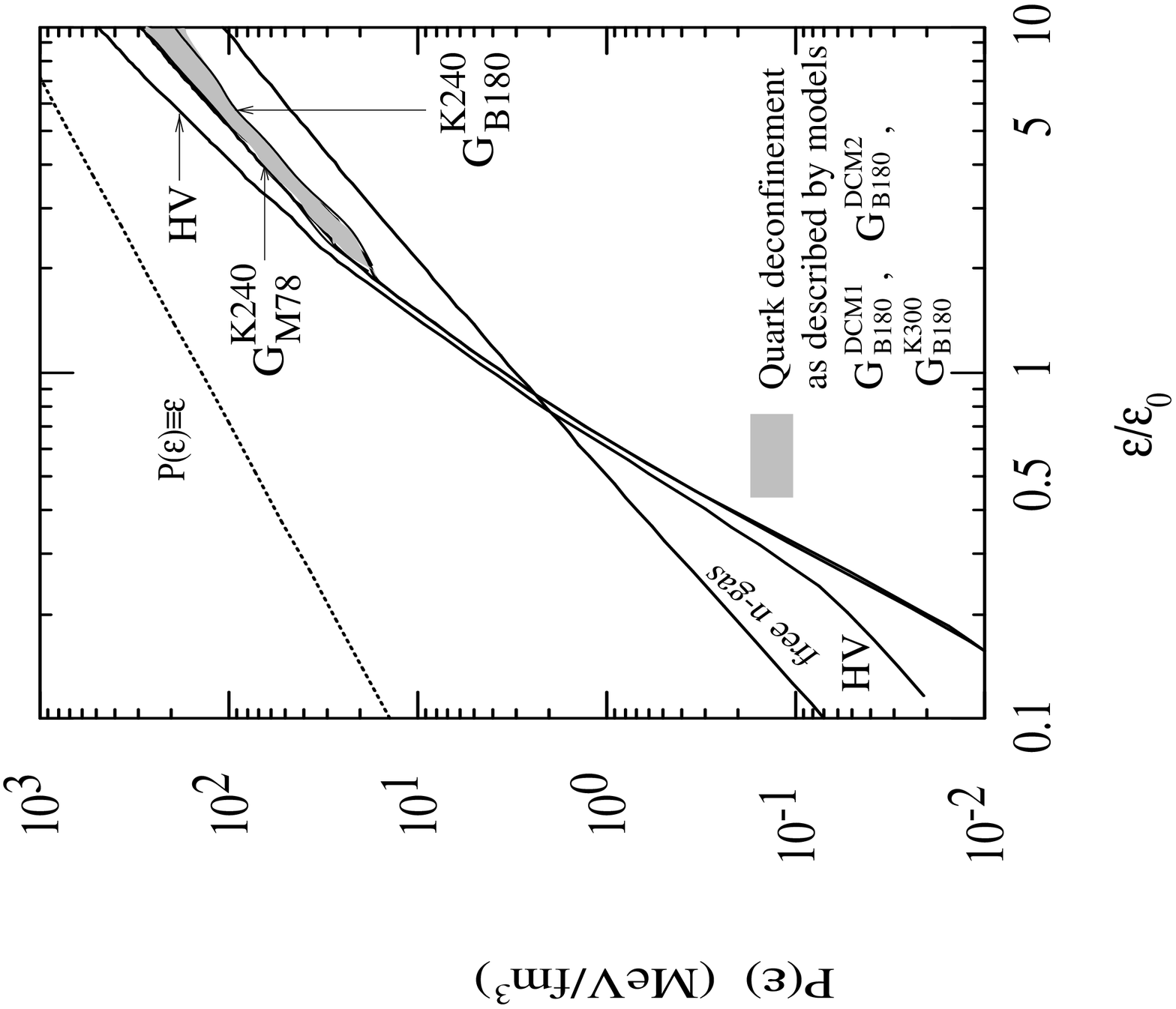,width=8.0cm,angle=-90}
\caption[]{Models for the equations of state of neutron star
matter. The shaded area marks quark deconfinement taken into account
in models $\KBt$, $\egdcmc$, $\KB$ $\egdcma$.\cite{weber99:topr}}
\label{fig:eos.qm}
\end{center}
\end{figure}

The phase transition between confined hadronic matter and quark matter
is characterized by the conservation of baryon charge and electric
charge. The Gibbs condition for phase equilibrium then is that the two
associated chemical potentials $\mu^n$ and $\mu^e$, corresponding to
baryon and electric charge conservation, and the pressure in the two
phases be equal,
\begin{eqnarray}
  P_{\rm H}(\mu^n,\mu^e, \{ \phi \}, T) = P_{\rm Q}(\mu^n,\mu^e,T) \,
  .
\label{eq:gibbs1}
\end{eqnarray} Here $P_{\rm H}$ denoted the pressure of hadronic
matter computed for a hadronic Lagrangian ${\cal L}(\{\phi\})$ as
given in Eq.\ (\ref{eq:f31}). The pressure of quark matter, $P_{\rm
Q}$, is obtainable from the bag model.\cite{kettner94:b}
The quark chemical potentials $\mu^u, \mu^d, \mu^s$ are related to the
baryon and charge chemical potentials as
\begin{eqnarray}
  \mu^u = {{1}\over{3}} \, \mu^n - {{2}\over{3}} \, \mu^e\, ,\qquad
  \mu^d = \mu^s = {{1}\over{3}} \, \mu^n + {{1}\over{3}} \, \mu^e \, .
\label{eq:cp.ChT}
\end{eqnarray} Equation~(\ref{eq:gibbs1}) is to be supplemented with
the two global relations for conservation of baryon charge and
electric charge within an unknown volume $V$ containing $A$
baryons.\cite{glen91:pt} The first one is given by
\begin{equation}
  \rho \equiv {A\over V} = (1-\chi) \, \rho_{\rm H}(\mu^n,\mu^e,T) +
  \chi \, \rho_{\rm Q}(\mu^n,\mu^e,T) \, ,
\label{eq:bcharge}
\end{equation} where $\chi\equiv V_{\rm Q}/V$ denotes the volume
proportion of quark matter, $V_{\rm Q}$, in the unknown volume $V$,
\begin{figure}[tb]
\begin{center}
\leavevmode
\psfig{figure=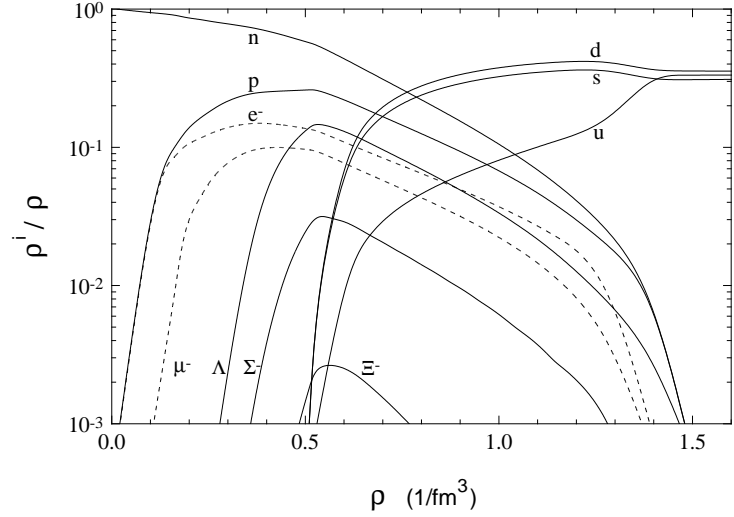,width=7.0cm,angle=-90}
\caption[]{Sample composition of chemically equilibrated neutron
(quark-hybrid) star matter.\protect{\cite{weber99:book,weber99:topr}}}
\label{fig:3.13.1}
\end{center}
\end{figure}
and $\rho_i$ $(i = {\rm H}, {\rm Q}, {\rm L})$ is the particle  number
density.  Global neutrality of electric charge within the volume $V$
can be written as
\begin{equation}
  0 = {Q\over V} = (1-\chi) \, q_{\rm H}(\mu^n,\mu^e,T) + \chi \,
  q_{\rm Q}(\mu^n,\mu^e,T)  +  q_{\rm L} \, ,
\label{eq:echarge}
\end{equation} with $q_i$ the electric charge densities of hadrons, quarks, 
and leptons.  For a given temperature $T$, Eqs.~(\ref{eq:gibbs1})
through (\ref{eq:echarge}) serve to determine the two independent
chemical potentials and the volume $V$ for a specified volume fraction
$\chi$ of the quark phase in equilibrium with the hadronic phase.
After completion $V_{\rm Q}$ is obtained as $V_{\rm Q}=\chi V$.
Through Eqs.~(\ref{eq:gibbs1}) to (\ref{eq:echarge}), the chemical
\begin{table}[b]
\caption[]{Models for the equation of state of neutron star matter
accounting for quark deconfinement.}
\label{tab:eos.quark}
\begin{center}
\begin{tabular}{|cll|} \hline
     EOS   &Properties (see text) &Reference   \\  \hline 
$\egdcma$  &H,Q, $K$=225, $B^{1/4}=180$  &\cite{glen91:pt,glen91:c}  \\
$\egdcmc$  &H,Q, $K$=265, $B^{1/4}=180$  &\cite{glen91:pt,glen91:c}  \\
$\KB$      &H,Q, $K$=240, $B^{1/4}=180$  &\cite{glen97:book} \\
$\KBt$     &H,Q, $K$=300, $B^{1/4}=180$  &\cite{glen97:book} \\ \hline
\end{tabular}
\end{center}
\end{table}
potentials obviously depend on the proportion $\chi$ of the phases in
equilibrium, and hence so also all properties that depend on them, the
energy densities, baryon and charge densities of each phase, and the
common pressure.  For the mixed phase, the volume proportion of quark
matter varies from $0 \leq \chi \leq 1$, and the energy density is the
linear combination of the two phases,\cite{glen91:pt}
\begin{equation}
  \epsilon = (1-\chi) \, \epsilon_{\rm H}(\mu^n,\mu^e, \{\phi\}, T) +
  \chi \, \epsilon_{\rm Q}(\mu^n,\mu^e,T) \, .
\label{eq:eps.chi}
\end{equation} 
By solving the models of confined and deconfined phases in both pure
phases and in the mixed phase, we can compute the baryon, lepton and
quark populations in neutron star matter from Eqs.~(\ref{eq:gibbs1})
through (\ref{eq:echarge}). One such sample outcome is shown in Fig.\
\ref{fig:3.13.1}.  Three features emerge immediately from this
population. Firstly, one sees that the transition from pure hadronic
matter to the mixed phase occurs at rather low density of about $3\,
\rho_0$. Depending on the bag constant and the underlying nuclear
many-body approximation, threshold values even as small as about $2\,
\rho_0$ can be obtained.\cite{glen91:pt,weber99:topr} Secondly, the
lepton density saturates as soon as deconfined quark matter is
generated, since charge neutrality can be achieved more economically
among the baryon-charge carrying particles themselves. Thirdly, the
presence of quark matter enables the hadronic regions of the mixed
phase to arrange themselves to be more isospin symmetric than in the
pure
\begin{table}[tb]
\caption[]{Nuclear matter properties of the equations of state
compiled in Table \ref{tab:eos.quark}.}
\label{tab:eos.quark.bulk}
\begin{center}
\begin{tabular}{|llllll|} \hline
{\small EOS}  &$E/A$  &$\rho_0$     &$K$    &$M^*$  &$a_{\rm sy}$   \\
              &(MeV)  &(fm$^{-3})$  &(MeV)  &(MeV)  &(MeV)   \\ \hline 
$\egdcma$                &$-16.0$   &0.16   &225    &0.796  &32.5  \\
$\egdcmc$                &$-16.0$   &0.16   &265    &0.796  &32.5  \\
$\KB$                    &$-16.3$   &0.153  &240    &0.78   &32.5  \\
$\KBt$                   &$-16.3$   &0.153  &300    &0.70   &32.5  \\
\hline
\end{tabular}
\end{center}
\end{table}
phase by transferring charge to the quark phase in equilibrium with
it. Symmetry energy will be lowered thereby at only a small cost in
rearranging the quark Fermi surfaces. Thus the mixed phase region
consists of positively charged nuclear matter and negatively charged
quark matter.  Models for the equation of state of neutron star matter
which account for quark deconfinement are listed in Tables
\ref{tab:eos.quark} and \ref{tab:eos.quark.bulk}.  A graphical
illustration of these equations of state is shown in Fig.\
\ref{fig:eos.qm}.

\section{Observed Neutron Star Properties}\label{sec:obsd}

Orbiting observatories such as the Hubble Space Telescope, the Chandra
X-ray satellite, and the X-ray Multi Mirror Mission have extended our
vision tremendously, allowing us to look at neutron stars with an
unprecedented clarity and angular resolution that previously were only
imagined, The properties of such objects such as masses, rotational
frequencies, radii, moments of inertia, redshifts, and temperatures
are known to be sensitive to the adopted microscopic model for the
nucleon-nucleon interaction and thus to the nuclear equation of
state.\cite{weber99:book,blaschke01:trento} Hence by means of
comparing theoretically determined values for these quantities with
observed ones one may be able to constrain the behavior of superdense
matter from the (unprecedented wealth of new) observational pulsar
data. In the following we briefly summarize several important star
properties.

\subsection{Masses}\label{ssec:masses}

The gravitational mass is of special importance since it can be
inferred directly from observations of X-ray binaries and binary radio
pulsar systems. The masses of the latter appear to be centered about
$1.35\, \msun$.\cite{thorsett99:a} The mass of the Hulse-Taylor radio
pulsar PSR~1913+16~\cite{taylor89:a}, given by $1.444\pm 0.003) \,
\msun$, is somewhat higher than this value. Examples of neutron stars
in X-ray binaries are Vela X-1 and the burster Cygnus X-2. Their
respective masses are $M=1.87^{+0.23}_{-0.17} \,
\msun$~\cite{kerkwijk00:a,barziv01:a} and $M=(1.8\pm 0.4) \,
\msun$~\cite{orosz00:a}. Indications for the possible existence of
very heavy neutron stars, with masses around $2\,\msun$ (e.g., neutron
star 4U~1636--536), may also come from the
observation of quasi-periodic oscillations in luminosity in low-mass
X-ray binaries.\cite{klis00:a} These mass values, which constitute
upper boundaries on the star mass, are larger than the typical $1.35\,
\msun$ masses found in neutron star binaries, presumably due to
accreted matter.

\subsection{Rotational Frequencies of Fast Pulsars}\label{ssec:freq}

The rotational periods of fast pulsars provide constraints on the
equation of state when combined with the mass constraint. The fastest
pulsars observed so far have rotational periods of 1.6~ms, which
corresponds to a rotational frequency of 620 Hz. The successful model
for the nuclear equation of state must therefore account for
rotational neutron star periods of at least $P = 1.6$~ms as well as
for the masses discussed in Sect.\ \ref{ssec:masses}.

Observationally, the searches for rapidly rotating radio pulsars are
biased, being least sensitive to short periods. Therefore the world's
data of radio pulsars are unlikely to represent the true underlying
population of fast pulsars.\cite{burderi01:a} Most of the large
surveys have had very poor sensitivity to millisecond pulsars with
rotational periods below about 4~ms, thus presumably distorting the
statistics on pulsar periods. The present cutoff in short periods at
about 1~ms is therefore possibly only an artifact of the search
sensitivity.

\subsection{Radii}\label{ssec:radii}

Direct radius determinations for neutron stars do not exist. However,
combinations of data of 10 well-observed X-ray bursters with special
theoretical assumptions lead Van Paradijs \cite{paradijs78:a} to the
conclusion that the emitting surface has a radius of about 8.5
km. This value, however, may be underestimated by a factor of
two.\cite{shapiro83:a} Fujimoto and Taam derived from the
observational data of the X-ray burst source MXB 1636--536, under
somewhat uncertain theoretical assumptions, a neutron star mass and
radius of $1.45\,\msun$ and 10.3~km.\cite{fujimoto86:a} An error
analysis lead them to predicting mass and radius ranges of 1.28 to
$1.65\,\msun$ and 9.1 to 11.3 km, respectively. Very recently the
nearby, isolated neutron star
RXJ~185635--3754~\cite{walter96:a,walter97:a} has attracted particular
attention because of its interpretation as a possible strange quark
star.\cite{drake02:a} (The physics of strange quark stars will be
discussed in Sect.\ \ref{sec:strange}.) The observed X-ray flux and
temperature from this object, if originating from a uniform blackbody,
appears to correspond to an effective stellar radius significantly
smaller than $\sim 10$~km, the canonical neutron star radius. This
follows from the relation
\begin{equation}
  R_\infty = R \; \biggl( 1 - { {2\, M}\over{R} } \biggr)^{-1/2}
  \approx 7 \; {{D}\over{120~{\rm pc}}}~{\rm km} \, ,
\label{eq:rinfty}
\end{equation} where $R$ is the neutron star radius, $M$ its mass, and
$D$ the distance to the star. One major uncertainty in
(\ref{eq:rinfty}) is buried in the numerical coefficient, which could
increase by 50\% or more if the star's surface temperature is not
uniform. Moreover, the existence of an atmosphere on this neutron star
could alter the coefficient probably even more
strongly.\cite{pons02:a} The other major uncertainty in obtaining a
radius is the star's distance which, however, has now been determined
to be $D = 117\pm 12$~pc~\cite{walter02:a}, nearly double the
originally published distance. Taking these considerations into
account, the interpretation of RXJ~185635--3754 as a quark star
appears questionable.\cite{walter02:a}

\subsection{Moment of Inertia}\label{ssec:moi}

Another global neutron star property is the moment of inertia, $I$.
Early estimates of the energy-loss rate from pulsars spanned a wide
range of $I$, i.e., $7\times 10^{43} < I < 7\times
10^{44}~\gcmt$.\cite{ruderman72:a} From the luminosity of the Crab
nebula ($\sim 2-4 \times 10^{38}~ {\rm erg/sec}$), several authors
have found a lower bound on the moment of inertia of the pulsar given
by $I \gsim 4-8\times 10^{44}~
\gcmsq$.\cite{baym75:a,trimble70:a,borner73:a}

\subsection{Redshift}\label{ssec:redshift}

Next we mention the neutron star redshift, $z$, which is given by
\begin{equation}
  z = \Bigl( 1 - { {2\, M}\over{R} } \Bigr)^{-1/2} - 1 \, ,
\label{eq:redshift}
\end{equation} 
with $M$ and $R$ the star's mass and radius.  Liang has considered the
neutron star redshift data base provided by measurements of
$\gamma$-ray burst redshifted annihilation lines in the range
$300-511$ keV.\cite{liang86:a} These bursts have widely been
interpreted as gravitationally redshifted 511~keV $e^\pm$ pair
annihilation lines from the surfaces of neutron stars. From this he
showed that there is tentative evidence to support a neutron star
redshift range of $0.2 \leq z \leq 0.5$, with the highest
concentration in the narrower range $0.25 \leq z \leq 0.35$.  A
particular role plays the source of the 1979 March 5 $\gamma$-ray
burst source, which has been identified with SNR N49 by its
position. From the interpretation of its emission, which has a peak at
$\sim 430$ keV, as the 511~keV $e^\pm$ annihilation line the resulting
gravitational redshift has a value of $z=0.23 \pm 0.05$.

\subsection{Magnetic fields}\label{sec:mfields}

Neutron stars are highly magnetized objects.\cite{michel82:a} There is
a general plausibility argument to the effect that, if internal
magnetic flux in a star is conserved owing to high conductivity, then
the surface magnetic field increases as $R^{-2}$ where $R$ is the
\begin{table}[tb]
\caption[]{Overview of neutrino emitting processes relevant for
  neutron star cooling.\cite{weber99:book}}
\label{tab:emis.over}
\begin{center}
\begin{tabular}{|llc|} \hline
Name     &{Particle processes}   &Efficiency            \\ \hline
Modified Urca       &$n+n$       $\rightarrow n+p+e^-+\bar\nu_e$ &slow \\
                    &$n+p+e^-$   $\rightarrow n+n+\nu_e$        & \\
Direct Urca         &$n$         $\rightarrow p+e^-+\bar\nu_e$  &fast \\
                    &$p+e^-$     $\rightarrow n+\nu_e$          & \\
Quark modified Urca &$d+u+e^-$  $\rightarrow d+d+\nu_e$        &slow \\
                    &$u+u+e^-$  $\rightarrow u+d+\nu_e$        & \\
                    &$d+u+e^-$  $\rightarrow d+s+\nu_e$       &  \\
                    &$u+u+e^-$  $\rightarrow u+s+\nu_e$        & \\
Quark direct Urca  &$d$         $\rightarrow u+e^-+\bar\nu_e$  &fast \\
                   &$u+e^-$     $\rightarrow d+\nu_e$          & \\
                   &$s$         $\rightarrow u+e^-+\bar\nu_e$  & \\
                   &$u+e^-$     $\rightarrow s+\nu_e$          & \\
$\pi^-$ condensate &$n+<\pi^->$  $\rightarrow n+e^-+\bar\nu_e$  &fast \\
$K^-$ condensate   &$n+<K^->$    $\rightarrow n+e^-+\bar\nu_e$  &fast \\
Quark bremsstrahlung &$Q_1+Q_2$  $\rightarrow Q_1+Q_2+\nu+\bar\nu$ &slow  \\
Core  bremsstrahlung &$n+n$     $\rightarrow n+n+\nu_e+\bar\nu_e$  &slow \\
                     &$n+p$     $\rightarrow n+p+\nu_e+\bar\nu_e$  & \\
                     &$e^-+p$   $\rightarrow e^-+p+\nu_e+\bar\nu_e$ &\\
Crust bremsstrahlung &$e^-+(A,Z)$  $\rightarrow  
               e^-+(A,Z)+\nu_e+\bar\nu_e$  &slow \\ \hline
\end{tabular}
\end{center}
\end{table}
stellar radius. Thus, if the sun were to collapse to neutron star
dimensions (roughly a factor of $7\times 10^4$ in radius), one would
magnify the general 1~G solar field to about $5\times 10^9$~G. This
value, while being in good agreement with the magnetic field strength
of $B \sim 4-6\times 10^{11}$~G estimated for the X-ray pulsar
1\,E\,2259+586 \cite{iwasawa92:a}, the central source in the supernova
remnant G\,109.1--1.0, is a bit shy of $10^{12}$~G inferred from line
features in the pulsed hard X-ray spectrum of Her~X--1 and
4U~0115--63.\cite{truemper78:a,voges82:a,wheaton79:a} Which physical
mechanisms might be important to create the extraordinarily strong
magnetic fields of neutron stars is an open issue. One possible
mechanism may be differential rotation of the stellar core, as
proposed for planetary magnetic fields. Besides that, ferromagnetism
has been examined as a possible magnetic field source, while
differential rotation between say superfluid protons and normal
electrons has been another suggestion.\cite{michel82:a} Effects of
strong magnetic fields on the structure of compact stars were studied,
for instance, in Refs.\ \cite{chakrabarty97:a,cardall01:a}.
Particularly striking are so-called magnetars, which are interpreted
as neutron stars with superstrong surface magnetic fields on the order
of $\sim 10^{15}$~G.\cite{duncan92:a} This translates to interior
magnetic fields of up to $\sim 10^{18}$~G, which is about the highest
possible value a star can sustain.

\goodbreak
\subsection{Cooling Data}\label{ssec:cool}

The detection of thermal photons from a stellar surface via X-ray
observatories like ASCA and ROSAT serves as the principal window on
the properties of a star.  The surface temperatures of stars are
derivable from the measured flux and spectrum.  The cooling rate of a
hot, young neutron star is primarily dependent at early times (first
several thousand years) on the neutrino emissivity of the core's
composition. The possible existence of Bose-Einstein condensates or
quark matter enhance the neutrino emissivity from the core, leading to
a more rapid early cooling (Table~\ref{tab:emis.over}). Superfluidity,
on the other hand, has the opposite effect on cooling. Quantitative
constraints have been hampered by the relatively small number of young
pulsars known, the complication that several of them also display
non-thermal, beamed X-ray emission from their magnetospheres, and
uncertainties in distance and interstellar
absorption. Table~\ref{tab:observations1} gives an overview of
\begin{table}[tb]
\caption[]{Ages $\tau$ (in years) and luminosities $L$ (in
erg~$\secm$) / temperatures $T$ (in $K$) of
pulsars.\cite{weber99:book}}
\label{tab:observations1}
\begin{center}
\begin{tabular}{|lccll|} \hline
 Pulsar &SNR            &$\logt~\tau$  &$L$ or $T$    &Refs.  \\ \hline
   1706--44 & & 4.25     &$\loglum=32.8\pm 0.7$ & \cite{becker92:a} \\
  1823--13 & & 4.50     &$\loglum=33.2\pm 0.6$ & \cite{finley93:b} \\
  2334+61 & & 4.61      &$\loglum=33.1\pm 0.4$ & \cite{becker93:b} \\
  0531+21 & Crab & 3.09 &$\loglum=35.5\pm 0.3$ & \cite{becker95:a}\\
  1509--58 &MSH 15-52 & 3.19 & $\loglum=33.6\pm 0.4$ 
        & \cite{seward83:a,trussoni90:a} \\
  0540--69 & & 3.22     &$\loglum=36.2\pm 0.2$ & \cite{finley93:a} \\
  1951+32 &CTB 80 &5.02 & $\loglum=33.8\pm 0.5$ & \cite{safiharb95:a} \\
  1929+10 & & 6.49 &    $\loglum=28.9\pm 0.5$ 
        & \cite{oegelman95:a,yancopoulos93:a} \\
  0950+08 & & 7.24 &    $\loglum=29.6\pm 1.0$ & \cite{seward88:a} \\
  J0437--47 & & 8.88 &  $\loglum=30.6\pm 0.4$ & \cite{becker93:c} \\ 
  0833--45 &Vela & 4.05 & $\loglum=32.9\pm 0.2$ & \cite{oegelman93:a} \\
  0656+14 & & 5.04      &$\loglum=32.6\pm 0.3$ & \cite{finley92:a} \\ 
  0630+18 &Geminga & 5.51 &$\loglum=31.8\pm 0.4$ & \cite{halpern93:a} \\
  1055--52 & & 5.73 &   $\loglum=33.0\pm 0.6$ & \cite{oegelman93:b} \\ 
  J0205+6449 &3C58 & 2.91 &$\logtem=6.0334$ & \cite{slane02:a} \\ 
\hline
\end{tabular}
\end{center}
\end{table}
observed neutron star temperatures.  Because of its unusually low
temperature which appears to falls well below predictions from
standard cooling models\cite{slane02:a}, the neutron star in supernova
remnant (SNR) 3C58 received much attention recently (see Sect.\
\ref{sec:thermal}).

\section{Properties of Neutron Star Models}\label{sec:properties}

\subsection{Nonrotating Star Models}\label{sec:nonrotst}

For many studies of neutron star properties it is sufficient to treat
neutron star matter as a perfect fluid. The energy-momentum tensor of
such a fluid is given by
\begin{eqnarray}
  T^{\mu\nu} = u^\mu \, u^\nu \, \bigl(\, \epsilon + P \, \bigr) \, +
  \, g^{\mu\nu} \, P \, ,
\label{eq:85.7}
\end{eqnarray} where $u^\mu$ and $u^\nu$ are four-velocities defined as
\begin{eqnarray}
  u^\mu \equiv {{dx^\mu}\over{d\tau}} \, , \qquad u^\nu \equiv
  {{dx^\nu}\over{d\tau}} \, .
\label{eq:4vel}
\end{eqnarray} They are the components of the macroscopic velocity of
the stellar matter with respect to the actual coordinate system that
is being used to derive the stellar equilibrium equations.  The
production of curvature by the star's mass is specified by Einstein's
field equations,
\begin{eqnarray}
  G_{\mu\nu} = 8\, \pi\, T_{\,\mu\nu}\, , \quad \makebox{where} \quad
  G_{\mu\nu} \equiv R_{\,\mu\nu} - {{1}\over{2}}\, g_{\mu\nu}\, R
\label{eq:14.26}
\end{eqnarray} is the Einstein tensor.
The Ricci tensor $R_{\,\mu\nu}$ is obtained from the Riemann tensor
$R^{\,\tau}{_{\mu\sigma\nu}}$ by contraction, that is, $R_{\,\mu\nu} =
R^{\,\tau}{_{\mu\sigma\nu}}\; g^{\sigma}{_{\tau}}$ which leads to
\begin{eqnarray}
  R_{\,\mu\nu} &=& {{\partial}\over{\partial x^\nu}}\,
  \Gamma_{\mu\sigma}^{\sigma} - {{\partial}\over{\partial x^\sigma}}\,
  \Gamma_{\mu\nu}^{\sigma} + \Gamma_{\mu\sigma}^{\kappa}\,
  \Gamma_{\kappa\nu}^{\sigma} - \Gamma_{\mu\nu}^{\kappa}\,
  \Gamma_{\kappa\sigma}^{\sigma} \, ,
\label{eq:14.27} 
\end{eqnarray} 
with $\Gamma_{\mu\nu}^{\sigma}$ the Christoffel defined as
\begin{eqnarray}
  \Gamma_{\mu\nu}^{\sigma} \equiv {{1}\over{2}}\, g^{\sigma\lambda}\,
  \left( {{\partial}\over{\partial x^\nu}}\, g_{\mu\lambda} +
  {{\partial}\over{\partial x^\mu}}\, g_{\nu\lambda} -
  {{\partial}\over{\partial x^\lambda}}\, g_{\mu\nu} \right)\, .
\label{eq:14.17}
\end{eqnarray} 
The scalar curvature of spacetime $R$ in Eq.\ (\ref{eq:14.26}), also
known as Ricci scalar, follows from Eq.\ (\ref{eq:14.27}) as
\begin{equation}
  R = R_{\,\mu\nu}\, g^{\mu\nu} \, .
\label{eq:14.27a}
\end{equation}
Finally, we need to specify the metric of a non-rotating body in
general relativity theory. Assuming spherical symmetry, the metric
has the form
\begin{eqnarray}
  ds^2 = - e^{2\,\Phi(r)} \, dt^2 + e^{2\,\Lambda(r)} \, dr^2 +
  r^2 \, d\theta^2 + r^2 \, {\rm sin}^2\theta \, d\phi^2\, ,
\label{eq:15.20}
\end{eqnarray} where $\Phi(r)$ and $\Lambda(r)$ are radially
varying metric functions. Introducing the covariant components of the
metric tensor, 
\begin{equation}
g_{t t} = - \, {e^{2\,\Phi}} \, , ~
g_{r r} = {e^{2\,\Lambda}} \, , ~
g_{\theta \theta} =  {r}^{2} \, , ~
g_{\phi \phi} =  {r}^{2} \sin^2\!\theta \, ,
\label{eq:15.34}
\end{equation} 
the non-vanishing Christoffel symbols of a spherically symmetric body
are
\begin{eqnarray}
\Gamma_{t t}^r =  
 {e^{2\,\Phi-2\,\Lambda}} \, \Phi' \, , ~
\Gamma_{t r}^t =  \Phi' \, , ~
\Gamma_{r r}^r = \Lambda'  \, , 
\Gamma_{r \theta}^\theta = {r}^{-1}  \, , ~
\Gamma_{r \phi}^\phi =  {r}^{-1}  \, , ~
\Gamma_{\theta \theta}^r = - \, r \; e^{-2\,\Lambda}  \, , 
\nonumber
\end{eqnarray}
\begin{eqnarray}
\Gamma_{\theta \phi}^\phi =  {\frac {\cos\theta}{\sin\theta}} \, , ~~
\Gamma_{\phi \phi}^r = - \, r \, \sin^2\!\theta \;
e^{-2\,\Lambda}  \, , ~~
\Gamma_{\phi \phi}^\theta = -\sin\theta \, \cos\theta \, ,
\label{eq:15.54c}
\end{eqnarray} 
where accents denote differentiation with respect to the radial
coordinate.  From Eqs.\ (\ref{eq:85.7}), (\ref{eq:14.26}) and
(\ref{eq:15.54c}) one derives the structure equations of spherically
symmetric neutron stars known as Tolman-Oppenheimer-Volkoff
equations,\cite{oppenheimer39,tolman39:a}
\begin{eqnarray}
{{dP}\over{dr}} = - \, \frac{\epsilon(r)\, m(r)}{r^2} \; \frac{\left[1
+ P(r)/\epsilon(r) \right] \, \left[ 1 + 4 \pi r^3 P(r)/m(r) \right]}
{1 - 2 m(r)/r} \, .
\label{eq:f28}
\end{eqnarray}
Note that we use units for which the gravitational constant and
velocity of light are $G=c=1$ so that $\msun = 1.5$~km.  The boundary
condition to (\ref{eq:f28}) is $P(r=0) \equiv P_c = P(\epsilon_c)$,
where $\epsilon_c$ denotes the energy density at the star's center,
which constitutes an input parameter.  The pressure is to be computed
out to that radial distance where $P(r=R)=0$ which determines the
star's radius $R$.  The mass contained in a sphere of radius $r~(\leq
R)$, denoted by $m(r)$, follows as
\begin{eqnarray}
m(r) = 4 \pi \int\limits^{r}_0 dr'\; r'^2 \; \epsilon(r') \, .
\label{eq:f29}
\end{eqnarray}
The star's total gravitational mass is thus given by $M\equiv m(R)$. 

Figure \ref{fig:ovme1} exhibits the gravitational masses of
non-rotating neutron stars as a function of central energy density for
a sample of equations of state of Tables \ref{tab:nonrel.eos} and
\ref{tab:rel.eos}.  Each star sequence is shown up to densities that
are slightly larger than those of the maximum-mass star (indicated by
tick marks) of each sequence. Stars beyond the mass peak are unstable
against radial oscillations and thus cannot exist stably (collapse to
black holes) in nature.
\begin{figure}[tb]
\begin{center}
\parbox[t]{5.9 cm}
{\leavevmode
\psfig{figure=msphec.bb,width=6.50cm,angle=90}
{\caption[]{Non-rotating neutron star mass as a function of central 
density.\cite{weber99:book}}\label{fig:ovme1}}}
\ \hskip 0.5 cm   \
\parbox[t]{5.9cm} {\leavevmode
\psfig{figure=msphz.bb,width=6.5cm,angle=90} {\caption[]{Non-rotating
neutron star mass as a function of
redshift.\cite{weber99:book}}\label{fig:ovmz1}}}
\end{center}
\end{figure} One sees that all equations of state are able to support 
non-rotating neutron star models of gravitational masses $M \geq
M({\rm PSR}~1913+16)$.  On the other hand, rather massive stars of say
$M\gsim 2\,M_\odot$ can only be obtained for those models of the
equation of state that exhibit a rather stiff behavior at supernuclear
densities.  The largest maximum mass value, $M=2.2\,\msun$, is
obtained for HFV, which is caused by the stiffening exchange term.
Knowledge of the maximum mass value is of great importance for two
reasons. Firstly, because the largest known neutron star mass imposes
a lower bound on the maximum mass of a theoretical model. The current
lower bound is about $1.60\,\msun$ [neutron star 4U\,0900--40
($\equiv$ Vela X--1)] which does not set too stringent a constraint on
the nuclear equation of state. The situation could easily change if an
accurate future determination of the mass of this neutron star should
result in a value that is close to its present upper bound of
$2.1\,\msun$.  In this case most of the equations of state of our
collection would be ruled out.  The second reason is that the maximum
mass is essential in order to identify black hole
candidates.\cite{brown94:a,bethe95:a}  For example, if the mass of a
compact companion of an optical star is determined to exceed the
maximum mass of a neutron star it must be a black hole. Since the
maximum mass of stable neutron stars studied here is $\sim
2.2\,\msun$, compact companions being more massive than that value are
predicted to be black holes.

The neutron star mass as a function of gravitational redshift, defined
in Eq.\ (\ref{eq:redshift}), is shown in Fig.\
\ref{fig:ovmz1}. Maximum-mass stars have redshifts in the range
$0.4\lsim z \lsim 0.8$, depending on the stiffness of the equation of
state.  Neutron stars of typically $M\approx 1.5 \,\msun$ (e.g., PSR
1913+16) are predicted to have redshifts in the considerably narrower
range $0.2 \leq z \leq 0.32$.  The rectangle covers masses and
redshifts in the ranges of $1.30 \leq M/\msun \leq 1.65$ and $0.25
\leq z \leq 0.35$, respectively. The former range has been determined
from observational data on X-ray burst source MXB~1636--536
\cite{fujimoto86:a}, while the latter is based on the neutron star
redshift data base provided by measurements of gamma-ray burst pair
annihilation lines.\cite{liang86:a} From the redshift value of SNR
N49 one would expect a neutron mass star of mass $1.1 \lsim M/\msun
\lsim 1.6$.
\begin{figure}[tb]
\begin{center}
\parbox[t]{5.9 cm} {\leavevmode
\psfig{figure=rz.bb,width=6.5cm,angle=90} {\caption[]{Radius as a
function of redshift for a collection of equations of
state.\cite{weber99:book}}
\label{fig:ovrm1}}}
\ \hskip 0.5 cm   \
\parbox[t]{5.9 cm} {\leavevmode
\psfig{figure=im.bb,width=6.5cm,angle=90} {\caption[]{Moment of
inertia as a function of mass for a collection of equations of
state.\cite{weber99:book}}\label{fig:ovimg1}}}
\end{center}
\end{figure}
Figure \ref{fig:ovrm1} displays the radius as a function of gravitational 
redshift.  

\begin{figure}[tb]
\begin{center}
\leavevmode
\psfig{figure=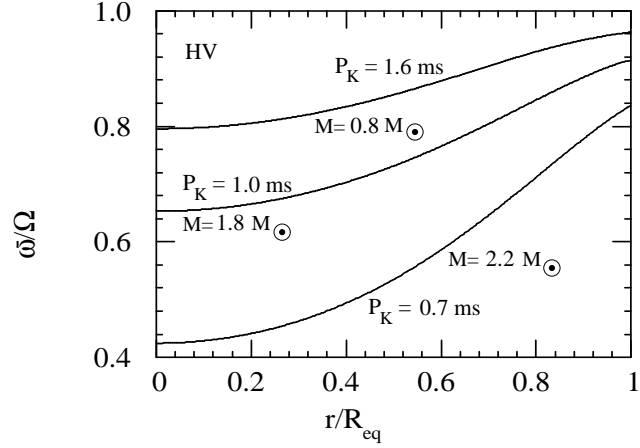,width=6.0cm,angle=-90}
\caption[]{Dragging of the local inertial frames inside rotating
neutron stars, from the center to the equator, calculated for the HV
of Table~\protect{\ref{tab:rel.eos}}.  $\pkgr$ and $M$ denote Kepler
period, defined in (\protect{\ref{eq:okgr}}), and gravitational
mass.\cite{weber99:topr}}
\label{fig:fdragg}
\end{center}
\end{figure}

Figure \ref{fig:ovimg1} shows the moment of inertia of neutron stars,
given by\cite{glen92:crust}
\begin{eqnarray}
 I(\Omega) = 4\,\pi \, \int\limits_0^{\pi/2} d\theta
\int\limits_0^{R(\theta)} dr\, e^{\lambda+\mu+\nu+\psi}\, {{\epsilon +
P(\epsilon)}\over{e^{2\nu-2\psi} - (\Omega-\omega)^2}} \,
{{\Omega-\omega}\over{\Omega}} \; ,
\label{eq:f231} 
\end{eqnarray}
as a function of gravitational mass. In Newtonian mechanics one has
for a sphere of uniform density $I \propto R^2 \, M$. The general
relativistic expression for the moment of inertia, given in Eq.\
(\ref{eq:f231}), is considerably more complicated as it accounts for
the dragging effect of the local inertial frames (see Fig.\
\ref{fig:fdragg}) and the curvature of space-time.\cite{weber99:book}
Nevertheless the qualitative dependence of $I$ on mass and radius as
expressed in the classical expression remains valid.  Estimates for
the upper and lower bounds on the moment of inertia of the Crab pulsar
derived from the pulsar's energy loss rate (labeled Rud72), and the
lower bound on the moment of inertia derived from the luminosity of
the Crab nebula (labeled Crab) are shown in Fig.\
\ref{fig:ovimg1}. The arrows refer only to the value of the moment of
inertial of the Crab pulsar and not to its mass, which is not known.

\goodbreak
\subsection{Rotating Star Models}\label{sec:rotst}

The stellar equations describing rotating compact stars are
considerably more complicated than those of non-rotating compact
stars.\cite{weber99:book} These complications have their cause in
the deformation of rotating stars plus the general relativistic effect
of the dragging of local inertial frames. This reflects itself in a
metric of the form\cite{weber99:book,friedman86:a}
\begin{eqnarray}
  ds^2 = - \, e^{2\,\nu} \, dt^2 + e^{2\,\psi} \, \bigl( d\phi -
  \omega \, dt \bigr)^2 + e^{2\,\mu} \, d\theta^2 + e^{2\,\lambda} \,
  dr^2 \, ,
\label{eq:f220.exact} 
\end{eqnarray} where each metric function $\nu$, $\psi$, $\mu$ and
$\lambda$ depends on the radial coordinate $r$, polar angle $\theta$,
and implicitly on the star's angular velocity $\Omega$.  The quantity
$\omega$ denotes the angular velocity of the local inertial frames,
which are dragged along in the direction of the star's rotation. This
frequency too depends on $r$, $\theta$ and $\Omega$.  Of particular
interest is the relative frame dragging frequency $\bar\omega$ defined
as $ \bar\omega(r,\theta,\Omega) \equiv \Omega -
\omega(r,\theta,\Omega)$, which typically increases from about 15\% at
the surface to about 60\% at the center of a neutron star that rotates
at its Kepler frequency.\cite{weber99:book} 

\subsection{Mass shedding versus Gravity Wave Emission}
\label{ssec:kepler.grr}

The Kepler frequency, $\okgr$, is the maximum frequency a star can
have before mass loss (mass shedding) at the equator sets in. It sets
an absolute upper limit on stable rapid rotation. In classical
mechanics the expression for $\okgr$, determined by the equality
between centrifuge and gravity, is readily obtained as $\okgr =
\sqrt{M/R^3}$. Its general relativistic counterpart is given
by\cite{weber99:book,friedman86:a}
\begin{eqnarray}
  \okgr = \omega +\frac{\omega^\prime}{2\psi^\prime} + e^{\nu -\psi}
    \sqrt{ \frac{\nu^\prime}{\psi^\prime} +
    \Bigl(\frac{\omega^\prime}{2 \psi^\prime}e^{\psi-\nu}\Bigr)^2 } \,
    , \qquad \pkgr \equiv {{2 \pi} \over {\okgr}} \, .
\label{eq:okgr}  
\end{eqnarray} The primes denote derivatives with respect to the
Schwarzschild radial coordinate. In order to construct stellar models
that rotate at their respective Kepler frequencies, equation
(\ref{eq:okgr}) is to be solved self-consistently together with
Einstein's field equations for a given model for the equation of
state.\cite{weber99:book,friedman86:a}

\begin{figure}[tb]
\begin{center}
\leavevmode 
\psfig{figure=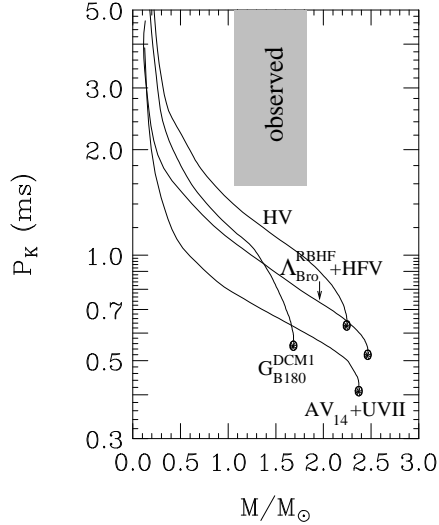,width=7.0cm,angle=90}
\caption[]{Onset of mass shedding from rapidly spinning neutron stars,
computed for a collection of equations of
state.\protect{\cite{weber99:book}} The Kepler period is defined in
Eq.\ (\ref{eq:okgr}).}
\label{fig:1}
\end{center}
\end{figure}

Figure~\ref{fig:1} shows $\okgr$ as a function of rotating star mass. The
rectangle indicates both the approximate range of observed neutron star masses
as well as the observed rotational periods which, currently, are $P \geq 1.6$
ms.  One sees that
 all pulsars so far observed rotate below the
mass shedding frequency and so can be interpreted as rotating neutron
stars. Half-millisecond periods or even smaller ones are excluded for
neutron stars of mass $1.4\,
\msun$.\cite{weber99:book,friedman86:a,friedman89:a} The situation
appears to be very different for stars made up of self-bound strange
quark matter, the so-called strange stars which will be introduced in
Sect.~\ref{sec:strange}. Such stars can withstand stable rotation
against mass shedding down to rotational periods in the
half-millisecond regime or even below.\cite{glen91:a} Consequently,
the possible future discovery of a single sub-millisecond pulsar
spinning at say 0.5~ms could give a strong hint that
strange stars actually exist, and that the deconfined self-bound
phase of 3-flavor strange quark matter is in fact the true ground
state of the strong interaction rather than nuclear matter.

The gravitational radiation-reaction driven instability in compact
stars sets a more stringent limit on rapid stellar rotation than mass
shedding, as can be seen from Figs.\ \ref{fig:3} and \ref{fig:4}. This
instability originates from counter-rotating surface vibrational
modes, which at sufficiently high rotational star frequencies are
dragged forward.  In this case, gravitational radiation which
inevitably accompanies the aspherical transport of matter does not
damp the instability modes but rather drives
them.\cite{chandrasekhar70:a,friedman83:a} Viscosity plays the
important role of damping such gravitational-wave radiation
reaction-driven instabilities at a sufficiently reduced rotational
frequency such that the viscous damping rate and power in gravity
waves are comparable.\cite{lindblom77:a} The instability modes are
taken to have the dependence ${\rm exp}[i\omega_m(\Omega)t + i m\phi -
t/\tau_m(\Omega)]$, where $\omega_m$ is the frequency of the surface
mode which depends on the angular velocity $\Omega$ of the star,
$\phi$ denotes the azimuthal angle, and $\tau_m$ is the time scale for
the mode which determines its growth or damping.  The rotation
frequency $\Omega$ at which it changes sign is the critical frequency
for the particular mode, $m$ (=2,3,4,...).  It is conveniently
expressed as the frequency, denoted by $\Omega_m^\nu$, that
solves\cite{lindblom86:a}
\begin{eqnarray}
\Omega_m^\nu \, =\, {{\omega_m(0)}\over m } \, \left[
\tilde\alpha_m(\Omega_m^\nu) + \tilde\gamma_m(\Omega_m^\nu)\, \Bigl(
{ {\tau_{g,m}}\over{\tau_{\nu,m}} } \Bigr)^{1\over{2\,m + 1}}
\right]\; ,
\label{eq:grr1}
\end{eqnarray}
where
\begin{equation}
\omega_m(0) \, \equiv \,
\sqrt{  { {2\,m\,(m-1)}\over{2\,m\,+\,1} } \, {M \over {R^3} }  }
\label{eq:omegm}
\end{equation}
is the frequency of the vibrational mode in a non-rotating star.  The
time scales for the gravitational radiation reaction to grow,
$\tau_{g,m}$, and for viscous damping to set in, $\tau_{\nu,m}$, are
given by\cite{weber99:book}
\begin{eqnarray}
\tau_{g,m} \, &=& \, {2\over 3} \, { {(m-1)\,
[(2m+1)!!]^2}\over{(m+1)\,(m+2)} } \, \left( { {2m+1}\over{2 m (m-1)}
} \right)^m \, \left( { R \over M } \right)^{m+1} \, R \, ,
\label{eq:grr2} \\
\tau_{\nu,m} \, &=& \,
{ {R^2} \over {(2m+1)\, (m-1)} } ~ {1\over \nu} \, .
\label{eq:grr3}
\end{eqnarray}
The shear viscosity, $\nu$, depends on a star's temperature according
to $\nu(T) \propto T^{-2}$. It is small in very hot ($T\approx
10^{10}$~K) and therefore young stars but considerably larger in cold
ones ($T\approx 10^{6}$~K). Information about the pulsations of the
rotating star is contained in the functions $\tilde\alpha_m$ and
$\tilde\gamma_m$.\cite{lindblom86:a,cutler87:a} A characteristic
feature of equations (\ref{eq:grr1}) through (\ref{eq:grr3}) is that
$\Omega^\nu_m$ merely depends on radius and mass ($R$ and $M$) of the
spherical star model, which renders solving these equations rather
straightforward.

Figure \ref{fig:3} shows the critical rotational periods at which the
emission of gravity waves sets in in hot ($T = 10^{10}$~K) pulsars
newly born in supernova explosions.  Figure \ref{fig:4} is the analog
to Fig.\ \ref{fig:3}, but for old and therefore cold pulsars of
\begin{figure}[tb]
\begin{center}
\parbox[t]{5.9 cm} {\leavevmode
\psfig{figure=grr10.bb,width=6.5cm,angle=90} {\caption[]
{Gravitational radiation-reaction instability period $P^T$ versus mass
for newly born stars of temperature $T =
10^{10}$~K.\protect{\cite{weber99:book}}}\label{fig:3}}} 
\ \hskip 0.5 cm \
\parbox[t]{5.9 cm} {\leavevmode
\psfig{figure=grr06.bb,width=6.5cm,angle=90} {\caption[]
{Gravitational radiation-reaction instability period $P^T$ versus mass
for old stars of temperature $T = 10^6$~K.\protect{\cite{weber99:book}}}
\label{fig:4}}}
\end{center}
\end{figure}
temperature $T=10^6$~K, like neutron stars in binary systems that are
being spun up by mass accretion from a companion.  One sees that the
limiting rotational periods $P^T$ ($\equiv 2\pi/\Omega^\nu_m$) are the
smaller the more massive (the smaller the radius) the star (cf.\ Fig.\
\ref{fig:ovrm1}).  A comparison between Figs.\ \ref{fig:3} and
\ref{fig:4} shows that the instability periods are shifted toward
smaller values the colder the star due to the larger viscosity in
colder objects.  Consequently, the instability modes of neutron stars
in binary systems are excited at smaller rotational periods than for
hot, newly born pulsars in supernovae.
This conclusion hinges on the role of bulk viscosity of neutron star
matter which may become very large at high temperatures. Sawyer has
pointed out that it proportional as the sixth power of the temperature
as compared with a $T^{-2}$ dependence for the shear
viscosity.\cite{sawyer89:b} This means that at temperatures $T \gsim
10^9$~K the bulk viscosity may dominate over the shear viscosity and
regulate the gravitational radiation-reaction driven instability in
rapidly rotating neutron stars, pushing their critical rotational
periods toward smaller values, possibly even as small as the Kepler
period. 

The dependence of $P^T$ on the equation of state is shown too in
Figs.\ \ref{fig:3} and \ref{fig:4} too. One
sees that the lower limits on $P^T$ are set by non-relativistic
\begin{figure}[tb]
\begin{center}
\leavevmode
\psfig{figure=pvse_s.bb,width=7.0cm}
\caption[]{Equation of state of a strange star surrounded by a nuclear
crust.  $\pdrip(\edrip)$ denotes the pressure at the maximum possible
inner crust density determined by neutron drip, $\ecrusti =
0.24~\mevt$ ($4.3\times 10^{11}~\gcmt$). Any inner crust value smaller
than that is possible. As an example, we show the equation of state for
$\ecrusti = 10^{-4}~\mevt$ ($10^8~\gcmt$).\cite{glen92:crust}}
\label{fig:eos.ss}
\end{center}
\end{figure}
equations of state due to the relatively small radii obtained for
stars constructed from them.  The rectangles in Figs.\ \ref{fig:3} and
\ref{fig:4} labeled ``observed'' cover both the range of observed
neutron star masses, $1.1\lsim M/\msun \lsim 1.8$ as well as observed
pulsar periods, $P\geq 1.6$ ms.  One sees that even the most rapidly
rotating pulsars so far observed have rotational periods larger that
those at which the gravitational radiation reaction-driven instability
sets in. Thus all observed pulsars can be understood as rotating
neutron stars. The observation of pulsars possessing masses in the
observed range but rotational periods that are smaller than say $\sim
1$~ms, depending on temperature and thus on the pulsar's history,
would be in contradiction to our collection of equations of state.   

Finally, we mention the $r$-mode instability which has attracted a
great deal of attention over the last several years. Like the
\begin{figure}[tb]
\begin{center}
\leavevmode
\psfig{figure=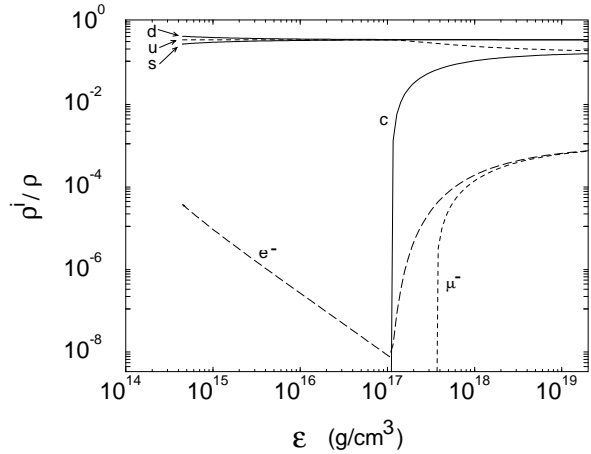,width=6.0cm,angle=-90}
\caption[]{Relative quark and lepton densities in cold quark star
  matter as a function of energy density for a bag constant of
  $\bag=145$~MeV.\protect{\cite{kettner94:b}}}
\label{fig:1.5}
\end{center}
\end{figure}  
$f$-modes discussed above, this instability too is produced by the
Chandrasekhar, Friedman and Schutz (CFS) mechanism and render every
star made of a perfect fluid unstable at all rates of
rotation.\cite{andersson98:a,friedman98:a} The question whether or not
it provides a much more severe constraint on the stable rotation of
viscous stars, however, appears to be still an open issue.

\section{Strange Quark Matter Stars}\label{sec:strange}

\subsection{The Strange Matter Hypothesis}\label{ssec:hyp}

The theoretical possibility that strange quark matter may be
absolutely stable with respect to nuclear matter, that is, the energy
per baryon of such matter is less than 930~MeV, has been pointed out
independently by Bodmer, Terazawa, and
Witten.\cite{bodmer71:a,terazawa89:a,witten84:a} This so-called
strange matter hypothesis constitutes one of the most startling
possibilities regarding the behavior of superdense matter, which, if
true, would have implications of fundamental importance for cosmology,
the early universe, its evolution to the present day, astrophysical
compact objects. Even to the present day there is no sound scientific
basis on which one can either confirm or reject the hypothesis so that
it remains a serious possibility of fundamental significance for
various phenomena.\cite{weber99:book,aarhus91:proc}

\goodbreak
\subsection{Quark-lepton Composition of Strange Matter}\label{sec:qlc}

The relative quark-lepton composition of quark-star matter at zero temperature
is shown in Fig.\ \ref{fig:1.5}.  All quark flavor states that become populated
at the densities shown are taken into account. Strange and charm quark masses
of respectively 0.15 GeV and 1.2 GeV are assumed. Since stars in their
lowest energy state are electrically charge neutral to very high
precision \cite{glen85:a}, any net positive quark charge must be
balanced by leptons. In general, as can be seen in Fig.\
\ref{fig:1.5}, there is only little need for leptons, since charge
neutrality can be achieved essentially among the quarks themselves.
The concentration of electrons is largest at the lower densities of
Fig.\ \ref{fig:1.5} due to the finite $s$-quark mass which leads to a
deficit of net negative quark charge, and at densities beyond which
the $c$-quark state becomes populated which increases the net positive
quark charge.

The presence of electrons in strange quark matter is crucial for the
possible existence of a nuclear crust on such objects, as will be
discussed below.  Recently it has been argued that strange quark
matter is a color superconductor (see Sect.\ \ref{sec:color}) which,
at extremely high densities, is in the Color-Flavor-Locked (CFL)
phase. This phase is rigorously electrically neutral with no electrons
required.\cite{rajagopal01:b} If the CFL phase would extend all the
way to the surface of a strange star, then strange stars would not be
able to carry nuclear crusts because of the missing electric dipole
layer. However, for sufficiently large strange quark masses, the
``low'' density regime of strange matter is rather expected to form a
2-flavor color superconductor (2SC) in which electrons are
present.\cite{rajagopal01:a,alford01:a}

As shown in Refs.\ \cite{kettner94:b,alcock86:a,alcock88:a}, the
electrons, because they are bound to strange matter by the Coulomb
force rather than the strong force, extend several hundred fermi
beyond the surface of the strange star.  Associated with this electron
displacement is a electric dipole layer which can support, out of
contact with the surface of the strange star, a crust of nuclear
material, which it polarizes.\cite{alcock86:a,alcock88:a} The maximal
possible density at the base of the crust (inner crust density) is
determined by neutron drip, which occurs at about $4.3\times
10^{11}~\gcmt$.

\goodbreak
\subsection{Equation of State of Strange Stars with Crust}\label{sec:eos.ss}

The somewhat complicated situation of the structure of a strange star
with crust described above can be represented by a proper choice of
equation of state which consists of two parts.\cite{glen92:crust} At
densities below neutron drip it is represented by the low-density
equation of state of charge-neutral nuclear matter, for which we use
the Baym-Pethick-Sutherland equation of state.\cite{baym71:a,baym71:b}
The star's strange matter core is described by the bag model. The
graphical illustration of such an equation of state is shown in Fig.\
\ref{fig:eos.ss}.  Notice that there is a discontinuity in energy
density between strange quark matter and hadronic matter across the
electron surface (dipole gap) inside the star where the pressure of
the hadronic crust at its base equals the pressure of the strange core
at its surface.\cite{glen92:crust}

\goodbreak
\subsection{Strange and Charm Stars}

\begin{figure}[tb]
\begin{center}
\leavevmode 
\psfig{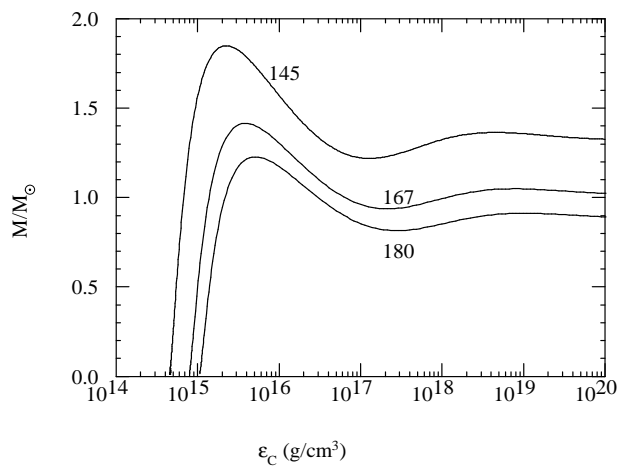}
\caption[] {Gravitational mass of strange and charm stars for
different $\bag$ values.}
\label{fig:mvsec}
\end{center}
\end{figure}

Figure \ref{fig:mvsec} shows the mass of quark matter stars as a
function on central star density for different, representative bag
constants. Stars with central densities smaller than a few times
$10^{15}~\gcmt$ are composed of up, down and strange quarks only and
are referred to as strange stars. Charm quarks are present at central
densities higher than $\sim 10^{17}~\gcmt$. Stars with densities
larger than that are thus denoted charm stars. Only the
sequences computed for $\bag=145$ MeV consist of matter that is
absolutely stable at zero external pressure.  The other two sequences
correspond to strange matter whose energy per baryon number is about
$960$~MeV ($\bag=167$~MeV) and $1030$~MeV ($\bag=180$~MeV).  A
stability analysis shows that charm stars are unstable against radial
oscillations and thus cannot exist stably in the
universe.\cite{kettner94:crete1} So the only type of quark stars that
could possess physical significance are strange stars.

\goodbreak
\subsection{Complete Sequences of Strange Matter Stars}

\begin{figure}[tb]
\begin{center}
\leavevmode
\psfig{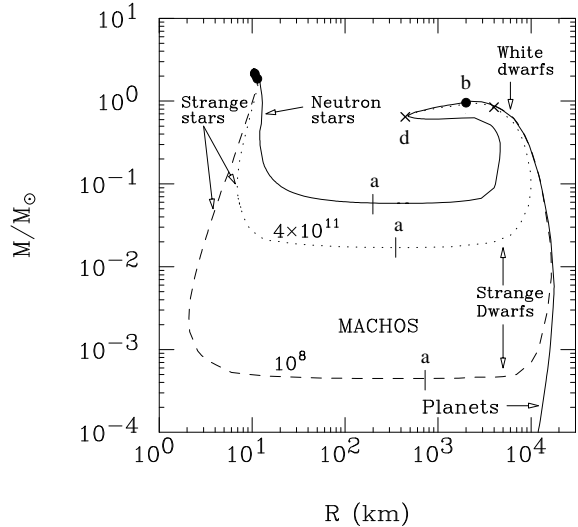}
\caption[]{Mass versus radius of strange stars with nuclear crusts
(dashed and dotted curves) and neutron stars and white dwarfs (solid
curve). The strange stars carry nuclear crusts with chosen inner
densities of $\ecrusti = 4\times 10^{11}~\gcmt$ and $10^{8}~\gcmt$.
The two crosses denote the termination points of the strange star
sequences where the quark matter cores have shrunk to zero. Dots refer
to maximum mass stars, minimum mass stars are located at the vertical
bars labeled ``a''.\protect{\cite{weber99:book}}}
\label{fig:mr145}
\end{center}
\end{figure}

Since the nuclear crusts surrounding the cores of strange stars are
bound by the gravitational force rather than confinement, the
mass-radius relationship of strange matter stars with crusts is
qualitatively similar to the one of purely gravitationally bound
compact stars, that is, neutron stars and white dwarfs. This is
illustrated in Fig.\ \ref{fig:mr145}.  The strange star sequence is
computed for the maximal possible inner crust density,
$\ecrusti=\edrip$.  Of course there are other possible sequences of
strange stars with any smaller value of
inner crust density, which will be discussed below.  From the maximum
mass star, the central density decreases monotonically through the
sequence in each case.  The neutron star sequence is computed for HFV
of Table \ref{tab:rel.eos}, a representative model for the equation of
state of neutron star matter, which has been combined at subnuclear
densities with the Baym-Pethick-Sutherland equation of state.  Hence
the white dwarfs shown in Fig.\ \ref{fig:mr145} are computed for the
latter equation of state.  Those gravitationally bound stars with
radii $\lsim 200$~km and $\gsim 3000$~km represent stable neutron
stars and white dwarfs, respectively. The fact that strange stars with
crust possess smaller radii than neutron stars leads to smaller mass
shedding periods, as already indicated by the classical expression
${P_{\rm K}}=2\pi\sqrt{R^3/M}$. Of course the general relativistic
expression (\ref{eq:okgr}), which is to be applied to neutron and
strange stars, is considerably more complicated. However the
qualitative dependence of ${P_{\rm K}}$ on mass and radius remains
valid.\cite{glen93:drag}  Due to the smaller radii of strange stars,
the complete sequence of such objects (and not just those close to the
mass peak, as is the case for neutron stars) can sustain extremely
rapid rotation.\cite{weber93:b}  In particular, a strange star with a
typical pulsar mass of $\sim 1.45\,\msun$ can rotate at Kepler periods
as small as $0.55\lsim{P_{\rm K}}/{\rm msec}\lsim 0.8$, depending on
crust thickness and bag constant.\cite{glen92:crust,weber93:b} This
range is to be compared with ${P_{\rm K}} \sim 1~{\rm msec}$ obtained
for neutron stars of the same mass.\cite{weber91:d}

The minimum-mass configuration of the strange star sequence, labeled
``a'' in Fig.\ \ref{fig:1}, has a mass of about $\sim 0.017\, \msun$
(about 17 Jupiter masses). Depending on the chosen value of inner crust
density, strange matter stars can be even by orders of magnitude
lighter than this value.\cite{weber93:b} If abundant enough in our
Galaxy, such low-mass strange stars could be seen by the gravitational
microlensing searches.  Strange stars located to the right of ``a''
consist of small strange cores ($\lsim 3$ km) surrounded by a thick
nuclear crust made up of white dwarf material. Such objects are thus
called strange dwarfs.  Their cores have shrunk to zero at the points
labeled ``X''. What is left is an ordinary white dwarf with a central
density equal to the inner crust density of the former strange
dwarf.\cite{weber93:b,romanelli86:a} A detailed stability analysis of
strange stars against radial oscillations shows that the strange
dwarfs between ``b'' and ``d'' are unstable against the fundamental
eigenmode.\cite{weber93:b} Hence such objects cannot exist stably in
nature.  However all other stars of this sequence are stable against
oscillations.  So, in contrast to neutron stars and white dwarfs, the
branches of strange stars and strange dwarfs are stably connected with
each other.\cite{weber93:b,glen94:a}

\begin{figure}[tb] 
\begin{center}
\leavevmode
\psfig{figure=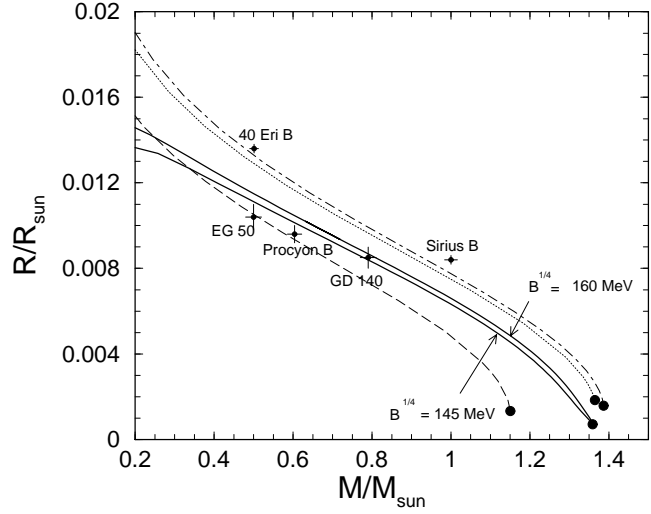,width=7.0cm,angle=-90}
\caption[]{Comparison of theoretical mass-radius relationships of
strange dwarfs ($\ecrusti=\edrip$) and several ordinary white dwarf
stars ($^{12}{\rm C}$ dwarfs: dot-dashed, $^{24}{\rm Mg}$: dotted,
$^{56}{\rm Fe}$: dashed double-dotted) with data from the Hipparcos
project.\cite{mathews02:a}}
\label{fig:mvsr}
\end{center}
\end{figure}

Until recently, only rather vague tests of the theoretical mass-radius
relation of white dwarfs were possible. This has changed dramatically
because of the availability of new data emerging from the Hipparcos
project.\cite{provencal98:a} These data allow the first accurate
measurements of white dwarf distances and, as a result, establishing
the mass-radius relation of such objects empirically. Figure
\ref{fig:mvsr} shows a
comparison of several data from the Hipparcos project with the
mass-radius relationships of strange dwarfs (solid lines) and ordinary
white dwarfs computed for different compositions. The outcome could
suggests that there is evidence for a bimodality in the white-dwarf
populations one of which may contain strange-matter cores.

\goodbreak
\subsection{Hadronic Crust on Strange Stars and Pulsar Glitches}
\label{ssec:crust}

Of considerable relevance for the viability of the strange matter
hypothesis is the question of whether strange stars can exhibit
glitches in rotation frequency. From the study performed in
Refs.~\cite{glen92:crust,zdunik01:a} it is known that the ratio of the
crustal moment of inertia to the total moment of inertia,
$\icrust/\itotal$, varies between $10^{-3}$ and $\sim 10^{-5}$.  If
the angular momentum of the pulsar is conserved in a stellar quake
then the relative frequency change and moment of inertia change are
equal, and one arrives for the change of the star's frequency
at\cite{glen92:crust}
\begin{equation}
       {{\Delta \Omega}\over{\Omega}} \; = \; {{|\Delta I|}\over
       {I_0}} \; > \; {{|\Delta I|}\over {I}} \; \equiv \; f \;
       {\icrust\over I}\; \sim \; (10^{-5} - 10^{-3})\, f \; , ~{\rm
       with} \quad 0 < f < 1\, .
\label{eq:delomeg}
\end{equation}
Here $I_0$ denotes the moment of inertia of that part of the star
whose frequency is changed in the quake. It might be that of the crust
only, or some fraction, or all of the star. The factor $f$ in Eq.\
(\ref{eq:delomeg}) represents the fraction of the crustal moment of
inertia that is altered in the quake, i.e., $f \equiv |\Delta I|/
\icrust$.  Since the observed glitches have relative frequency changes
$\Delta \Omega/\Omega = (10^{-9} - 10^{-6})$, a change in the crustal
moment of inertia of $f\lsim 0.1$ would cause a giant glitch even in
the least favorable case.\cite{glen92:crust} Moreover it turns out
that the observed range of the fractional change in the spin-down
rate, $\dot \Omega$, is consistent with the crust having the small
moment of inertia calculated and the quake involving only a small
fraction $f$ of that, just as in Eq.\ (\ref{eq:delomeg}).  For this
purpose we write\cite{glen92:crust}
\begin{equation} 
        { {\Delta \dot\Omega}\over{\dot\Omega } } \; = \;
        { {\Delta \dot\Omega /  \dot\Omega} \over  
          {\Delta    \Omega  /     \Omega }  } \,
        { {|\Delta I |}\over{I_0} } \; = \; 
        { {\Delta \dot\Omega /  \dot\Omega} \over  
          {\Delta    \Omega  /     \Omega }  } \; f \;
         {\icrust\over {I_0} } \; > \; 
       (10^{-1}\; {\rm to} \; 10) \; f \, , 
\label{eq:omdot}
\end{equation} 
where use of Eq.\ (\ref{eq:delomeg}) has been made. Equation
(\ref{eq:omdot}) yields a small $f$ value, i.e., $f < (10^{-4} \; {\rm
to} \; 10^{-1})$, in agreement with $f\lsim 10^{-1}$ established just
above. Here measured values of the ratio $(\Delta
\Omega/\Omega)/(\Delta\dot\Omega/\dot\Omega) \sim 10^{-6}$ to
$10^{-4}$ for the Crab and Vela pulsars, respectively, have been used.

\goodbreak
\section{Thermal evolution of Neutron and Strange Stars}
\label{sec:thermal}

The predominant cooling mechanism of hot (temperatures of several
$\sim 10^{10}$~K) newly formed neutron stars immediately after
formation is neutrino emission, with an initial cooling time scale of
seconds. Already a few minutes after birth, the internal neutron star
temperature drops to $\sim 10^9$~K. Photon emission overtakes
neutrinos only when the internal temperature has fallen to $\sim
10^8$~K, with a corresponding surface temperature roughly two orders
of magnitude smaller. Neutrino cooling dominates for at least the
first $10^3$ years, and typically for much longer in standard
\begin{figure}[tb]
\begin{center}
\leavevmode 
\psfig{figure=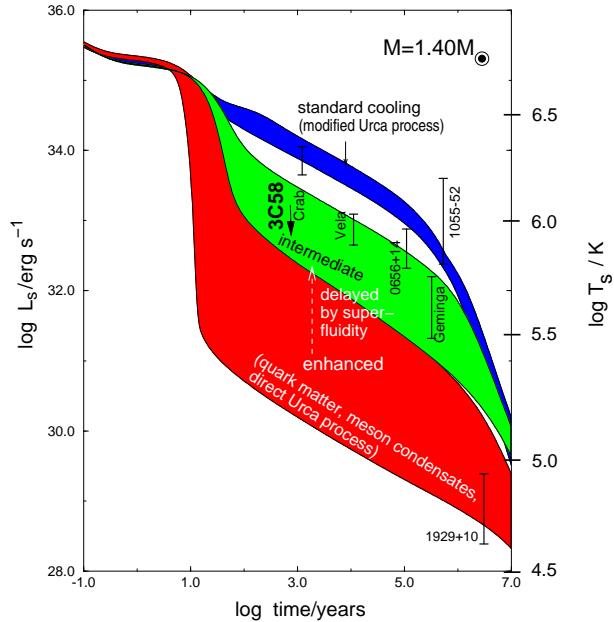,width=8.0cm}
\caption[]{Cooling behavior of a $1.4\,\msun$ neutron star based on
competing assumptions about the behavior of superdense matter. Three
distinct cooling scenarios, referred to as ``standard'',
``intermediate'', and ``enhanced'' can be distinguished (cf.\ Sect.\
\ref{ssec:cool}). The band-like structures reflects the uncertainties
inherent in the equation of state of superdense
matter.\cite{weber99:book}}
\label{fig:cool} 
\end{center}
\end{figure}
cooling (modified Urca) calculations.  Being sensitive to the adopted
nuclear equation of state, the neutron star mass, the assumed magnetic
field strength, the possible existence of superfluidity, meson
condensates and quark matter, theoretical cooling calculations, as
summarized in Fig.\ \ref{fig:cool}, provide most valuable information
about the interior matter and neutron star structure. The thermal
evolution of a neutron star also yields information about such
temperature sensitive properties as transport coefficients, transition
to superfluid states, crust solidification, and internal pulsar
heating mechanisms such as frictional dissipation at the
crust-superfluid interfaces.\cite{schaab99:b} The stellar cooling
tracks in Fig.\ \ref{fig:cool} are computed for the broad collection
of equations of state introduced in Sect.\ \ref{sec:eos}. Recent
Chandra observations have identified pulsar J0205+6449 at the center
of the young Crab-like supernova remnant 3C58. Historical evidence
suggests an association of the remnant with supernova SN~1181, which
makes 3C58 younger (see Table \ref{tab:observations1}) than Crab. The
temperature of J0205+6449 was recently determined to be
$\logtem=6.0334$~\cite{slane02:a} which, as can be seen from Fig.\
\ref{fig:cool}, falls well below predictions of standard cooling
calculations.\cite{slane02:a} Processes which speed up cooling
considerably range from the presence of meson
condensates~\cite{brown96:a,page98:a}, to quark
matter~\cite{blaschke01:a,page02:a}, to the direct Urca
process~\cite{lattimer91:a} (see Table~\ref{tab:emis.over} for an
overview of neutron star cooling mechanisms).

\goodbreak
\section{Color Superconductivity of Quark Matter}\label{sec:color}

There has been much recent progress in our understanding of quark
matter, culminating in the discovery that if quark matter exists it
will be in a color superconducting
state.\cite{rajagopal01:a,alford01:a,alford98:a,rapp98:a} The phase
diagram of such matter is very complex.\cite{rajagopal01:a,alford01:a}
At asymptotic densities the ground state of QCD with a vanishing
strange quark mass is the color-flavor locked (CFL) phase. This phase
is electrically neutral in bulk for a significant range of chemical
potentials and strange quark masses.\cite{rajagopal01:b} If the
strange quark mass is heavy enough to be ignored, then up and down
quarks may pair in the two-flavor superconducting (2SC) phase.  Other
possible condensation patters are the CFL--$K^0$ phase
\cite{bedaque01:a} and the color-spin locked (2SC+s)
phase.\cite{schaefer00:a} The magnitude of the gap energy lies between
$\sim 50$ and $100$~MeV. Color superconductivity thus modifies the
equation of state at the order $(\Delta / \mu)^2$ level, which is only
a few percent. Such small effects can be safely neglected in present
determinations of models for the equation of state of neutron stars
and strange quark matter stars, as is the case here. There has been
much recent work on how color superconductivity in neutron stars could
affect their properties. (See Refs.\
\cite{rajagopal01:a,alford01:a,rajagopal00:a,alford00:a,alford00:b,%
blaschke99:a} and references therein.)  These studies reveal that
possible signatures include the cooling by neutrino emission, the
pattern of the arrival times of supernova neutrinos, the evolution of
neutron star magnetic fields, rotational (r-mode) instabilities, and
glitches in rotation frequencies.

Aside from neutron star properties, an additional test of color
superconductivity may be provided by upcoming cosmic ray space
experiments such as AMS~\cite{ams01:homepage} and
ECCO.\cite{ecco01:homepage} As shown in Ref.\ \cite{madsen01:a},
finite lumps of color-flavor locked strange quark matter
(strangelets), which should be present in cosmic rays if strange
matter is the ground state of the strong interaction, turn out to be
significantly more stable than strangelets without color-flavor
locking for wide ranges of parameters. In addition, strangelets made
of CFL strange matter obey a charge-mass relation of $Z/A \propto
A^{-1/3}$, which differs significantly from the charge-mass relation
of strangelets made of ``ordinary'' strange quark matter. In the
latter case, $Z/A$ would be constant for small baryon numbers $A$ and
$Z/A \propto A^{-2/3}$ for large
$A$.\cite{aarhus91:proc,madsen01:a,madsen98:b} This difference may
allow an experimental test of CFL locking in strange quark
matter.\cite{madsen01:a}

\goodbreak
\section{Particle Thresholds and Quark Deconfinement in Rotating Neutron 
Stars}\label{ssec:fdepqd}

\begin{figure}[tb]
\parbox[t]{6.2cm} {\leavevmode
\psfig{figure=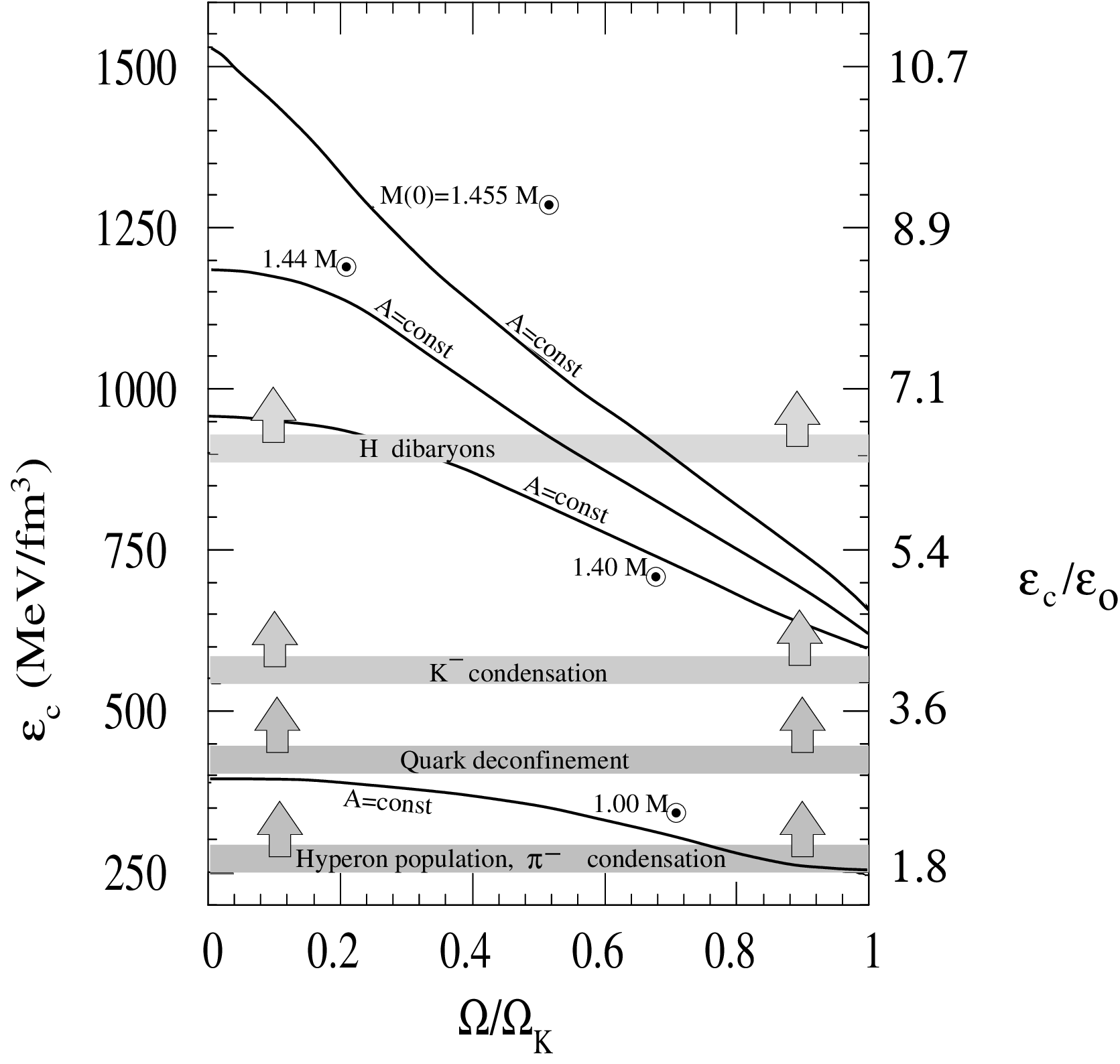,width=6.0cm,angle=-90}
\caption[]{Central density versus rotational frequency of neutron
  stars of constant baryon number, $A$, computed for $\KBt$.
  Threshold densities of various possible phases of superdense matter
  are indicated. $M(0)$ is the non-rotating star mass, $\okgr$ stands
  for the Kepler frequency.\protect{\cite{weber99:topr}}}
\label{fig:ec1445fig}}
\ \hskip0.25cm   \
\parbox[t]{6.2cm} {\leavevmode
\psfig{figure=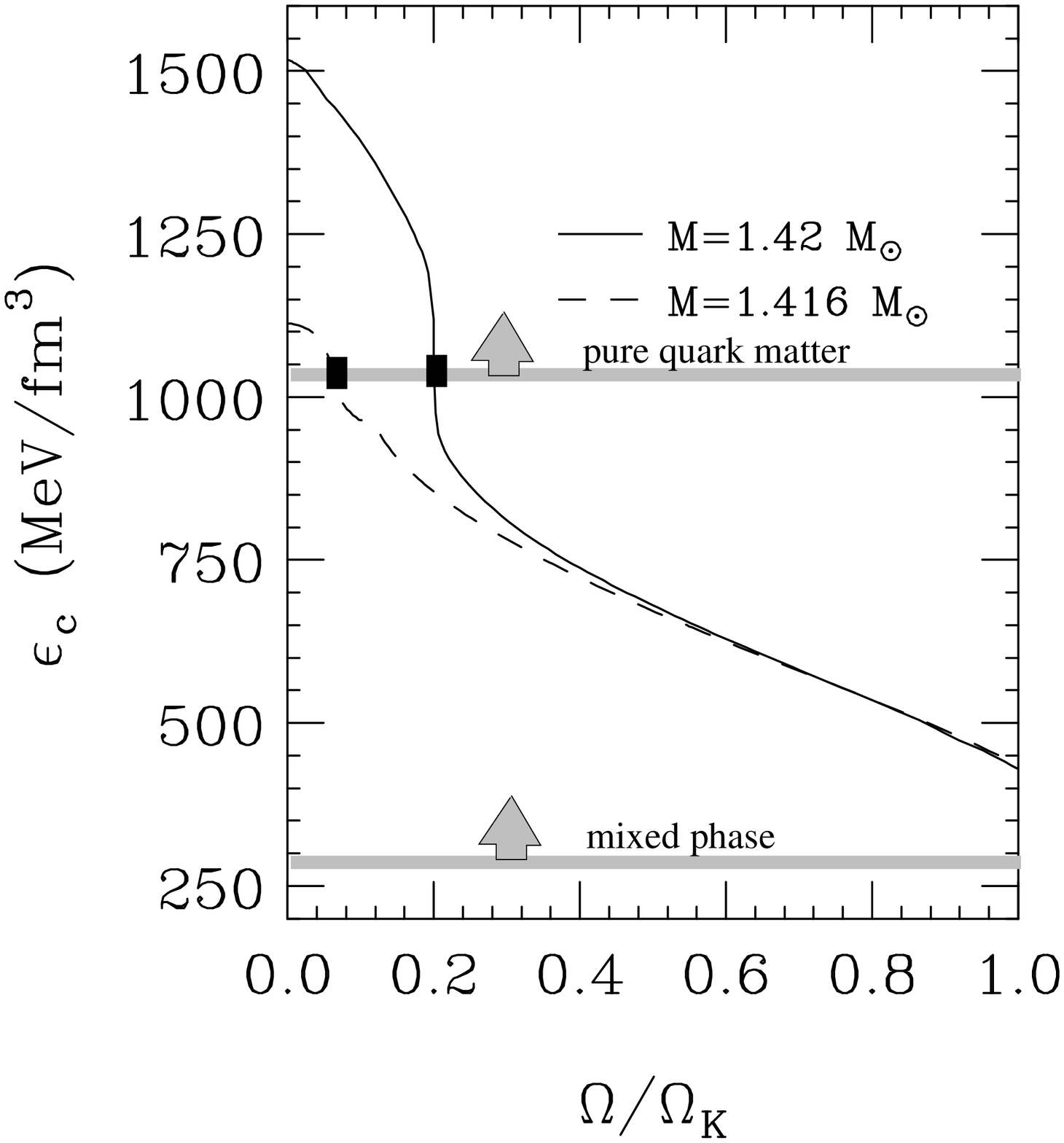,width=6.0cm,angle=-90}
{\caption[]{Central star density versus rotational frequency for two
sample stars computed for equation of state $\KBt$.}
\label{fig:ec3B18}}}
\end{figure}

\begin{figure}[tb] 
\parbox[t]{6.2cm} {\leavevmode
\psfig{figure=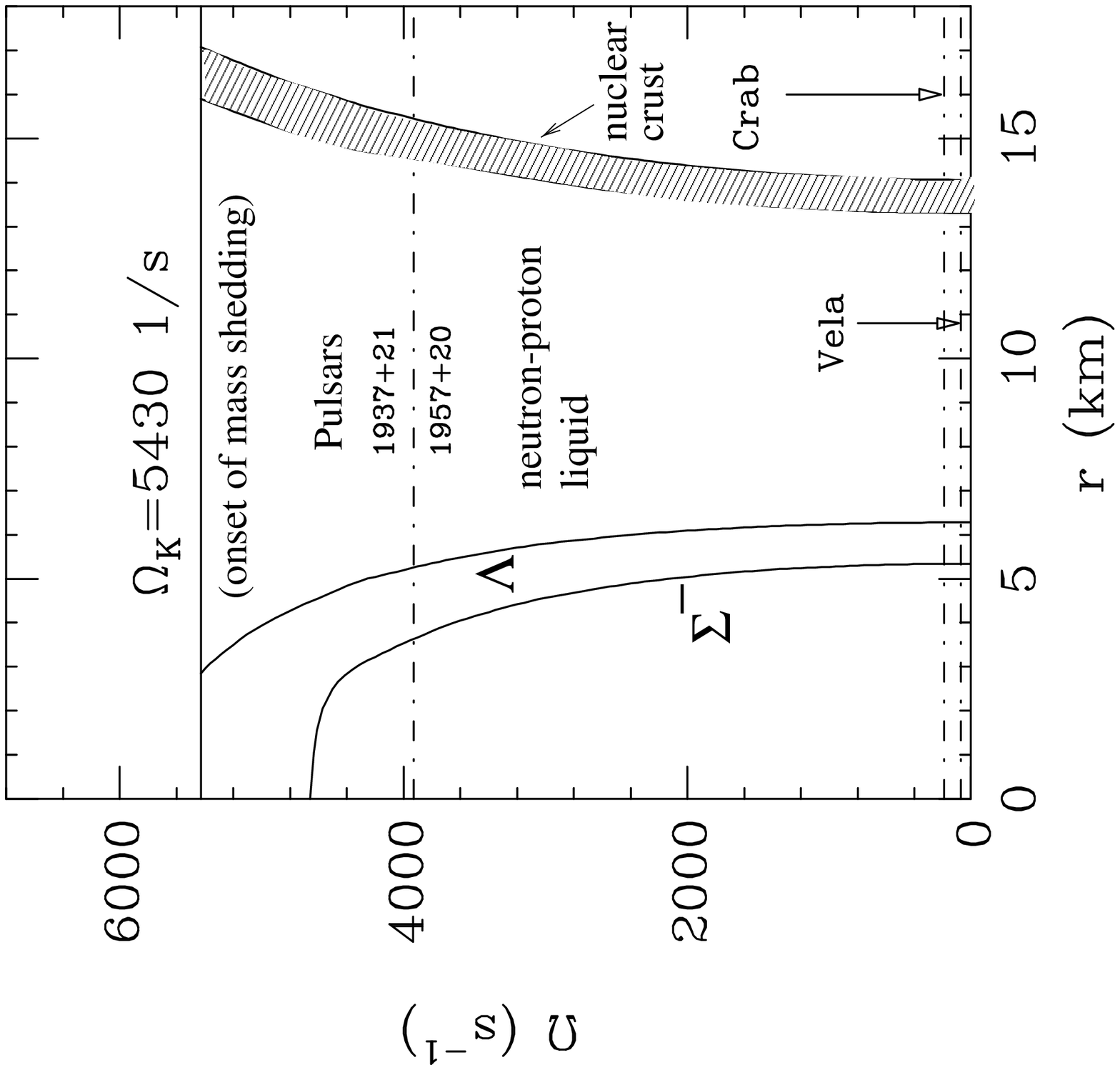,width=6.0cm,angle=-90} {\caption[]{Frequency
dependence of hyperon thresholds in equatorial neutron star direction
computed for HV. The star's non-rotating mass is
$1.40\,\msun$.\protect{\cite{weber99:book}}}
\label{fig:freqhp1}}}
\ \hskip0.25cm   \
\parbox[t]{6.2cm} {\leavevmode
\psfig{figure=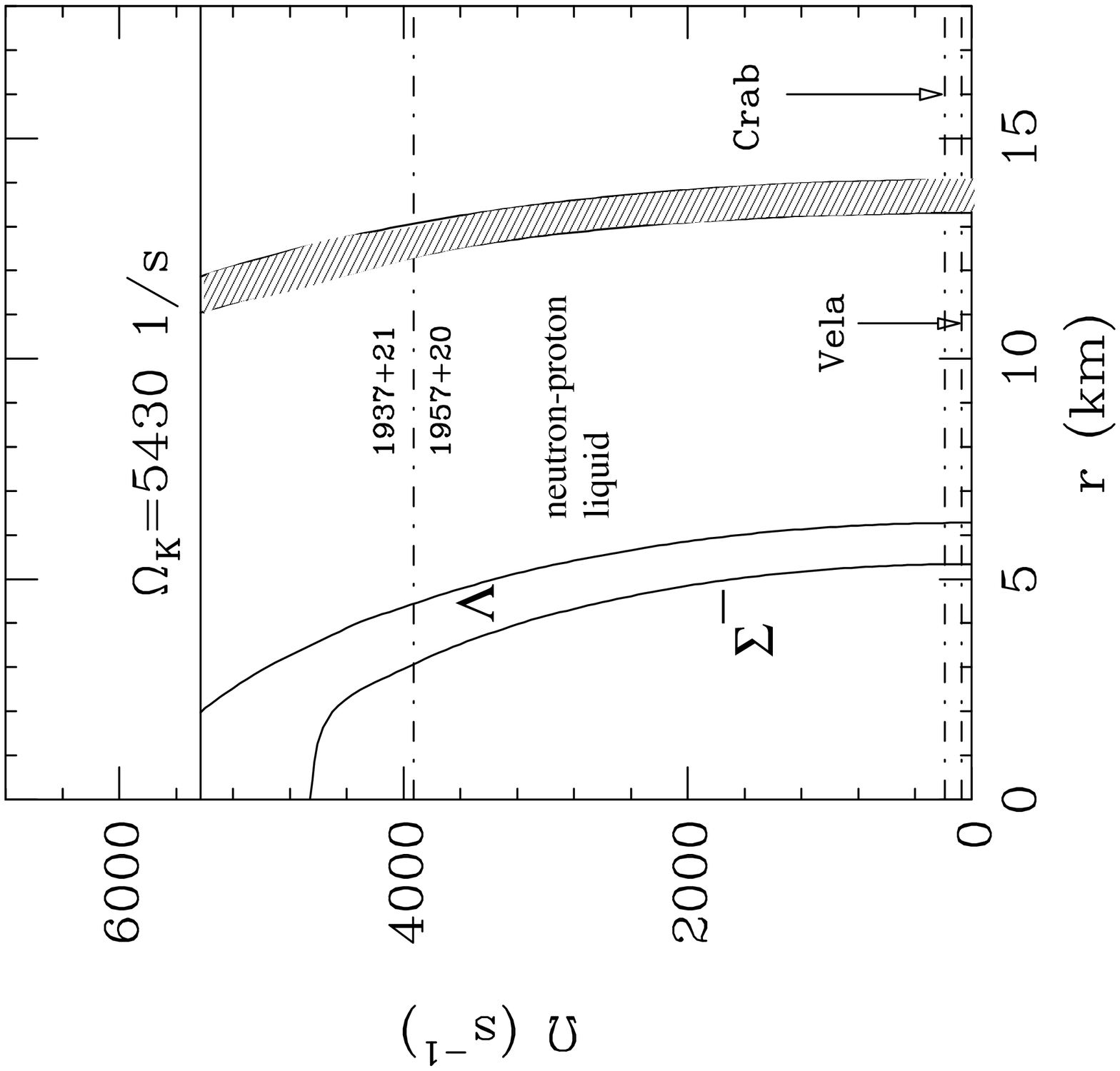,width=6.0cm,angle=-90} {\caption[]{Same as
Fig.\ \protect{\ref{fig:freqhp1}}, but in polar
direction.\protect{\cite{weber99:book}}}
\label{fig:freqhp2}}}
\end{figure}

\begin{figure}[tb] 
\parbox[t]{6.2cm} {\leavevmode
\psfig{figure=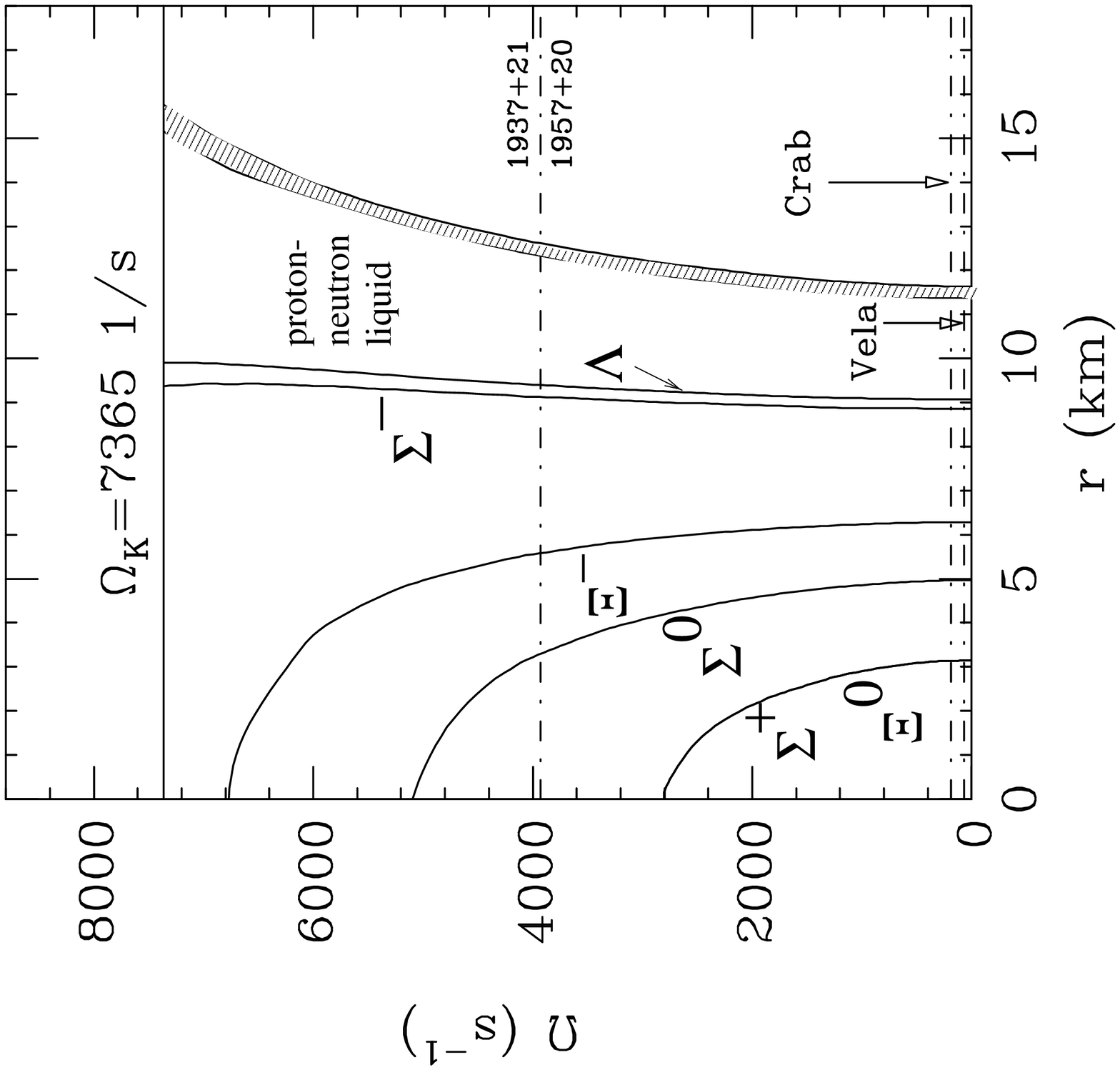,width=6.0cm,angle=-90} {\caption[]{Frequency
dependence of hyperon thresholds in equatorial neutron star direction
computed for HV. The star's non-rotating mass is
$1.978\,\msun$.\protect{\cite{weber99:book}}}
\label{fig:freqhp3}}}
\ \hskip0.25cm   \
\parbox[t]{6.2cm} {\leavevmode
\psfig{figure=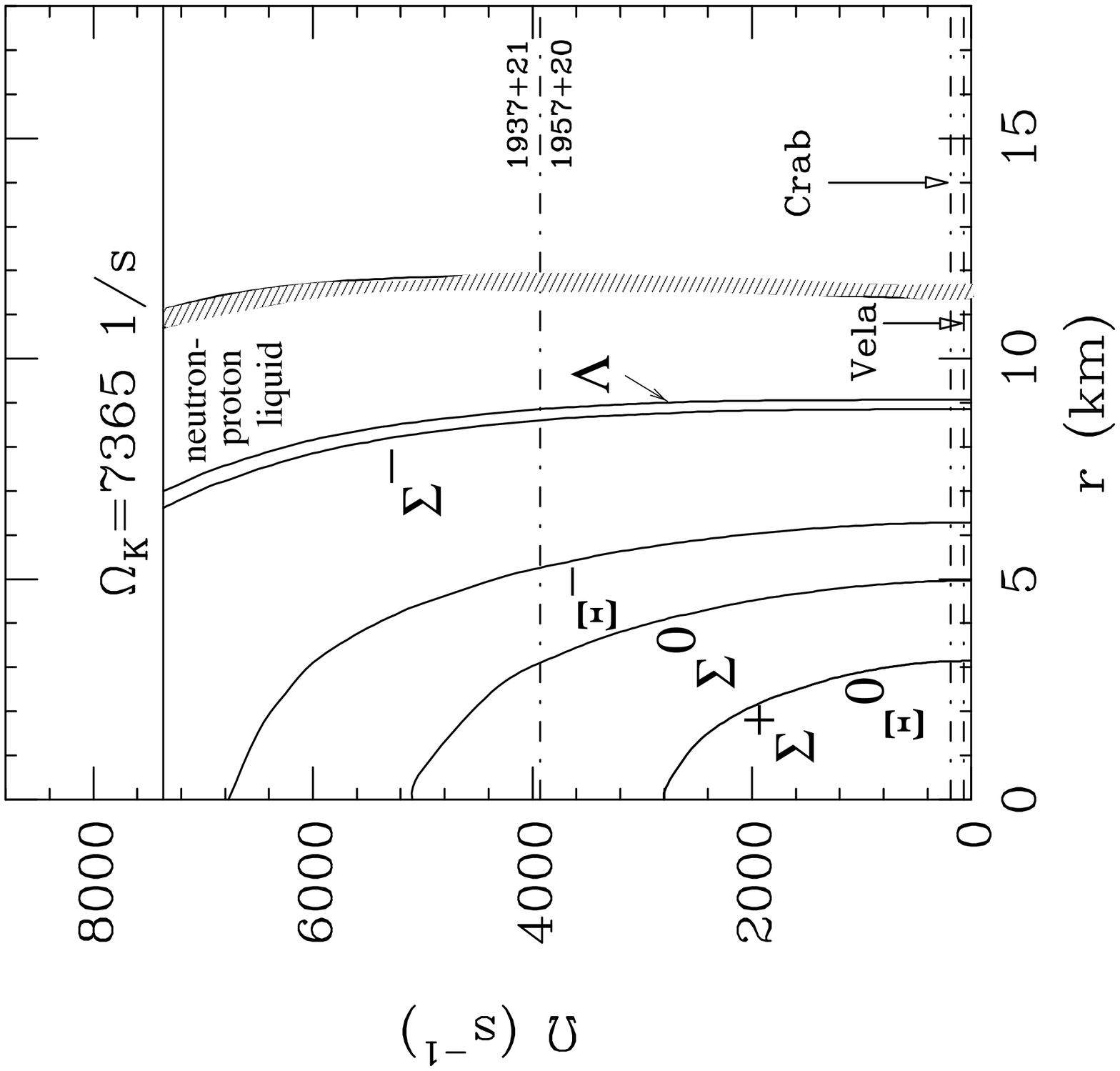,width=6.0cm,angle=-90} {\caption[]{Same as
Fig.\ \protect{\ref{fig:freqhp3}}, but in polar
direction.\protect{\cite{weber99:book}}}
\label{fig:freqhp4}}}
\end{figure}

\begin{figure}[tb] 
\parbox[t]{6.2cm} {\leavevmode
\psfig{figure=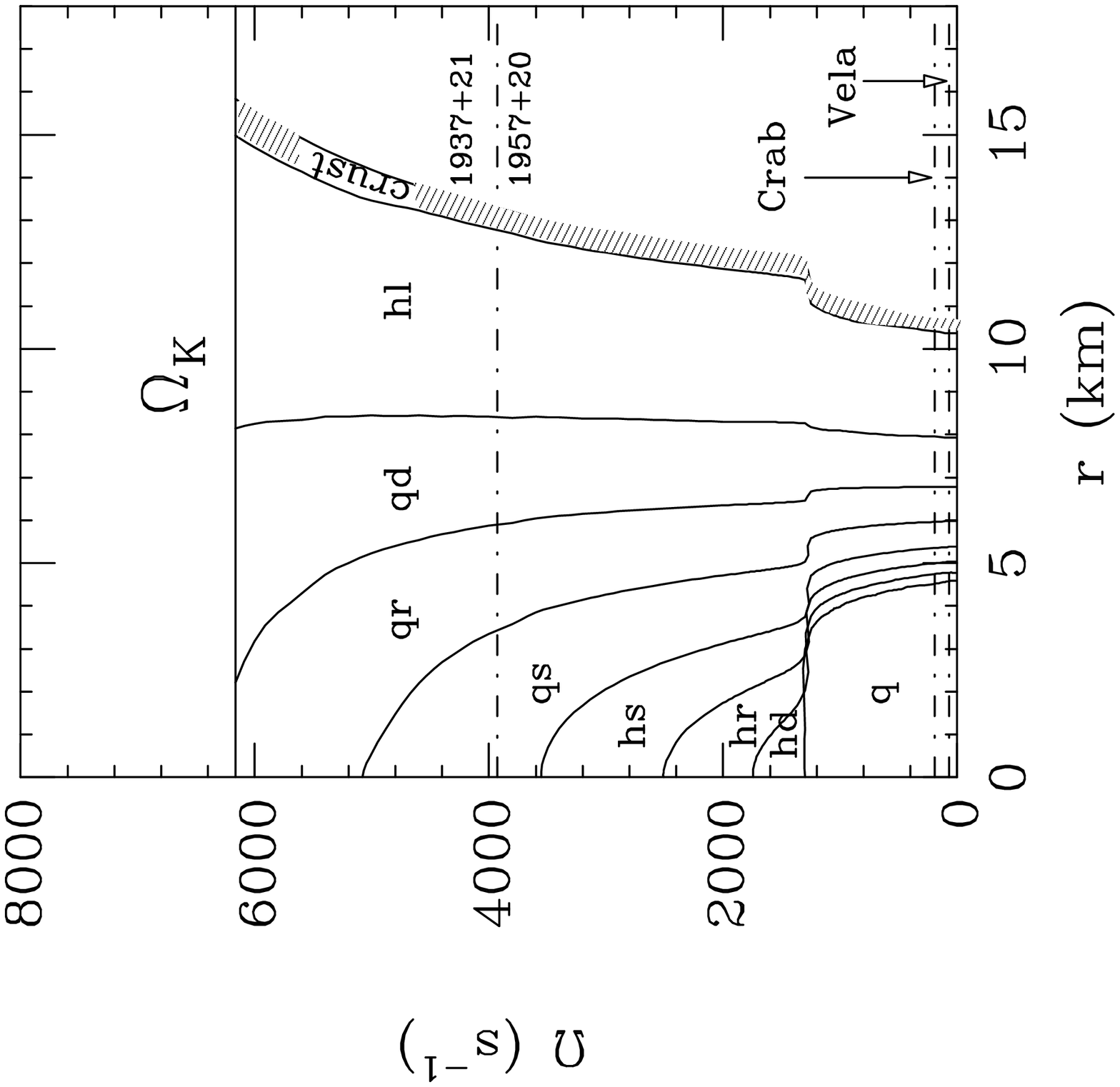,width=6.0cm,angle=-90}
{\caption[]{Frequency dependence of quark structure in equatorial
neutron star direction computed for $\KBt$ and a non-rotating star
mass of $1.42\,\msun$.\protect{\cite{weber99:book}}}
\label{fig:OkEq_1.42_G3B18}}}
\ \hskip0.25cm   \
\parbox[t]{6.2cm} {\leavevmode
\psfig{figure=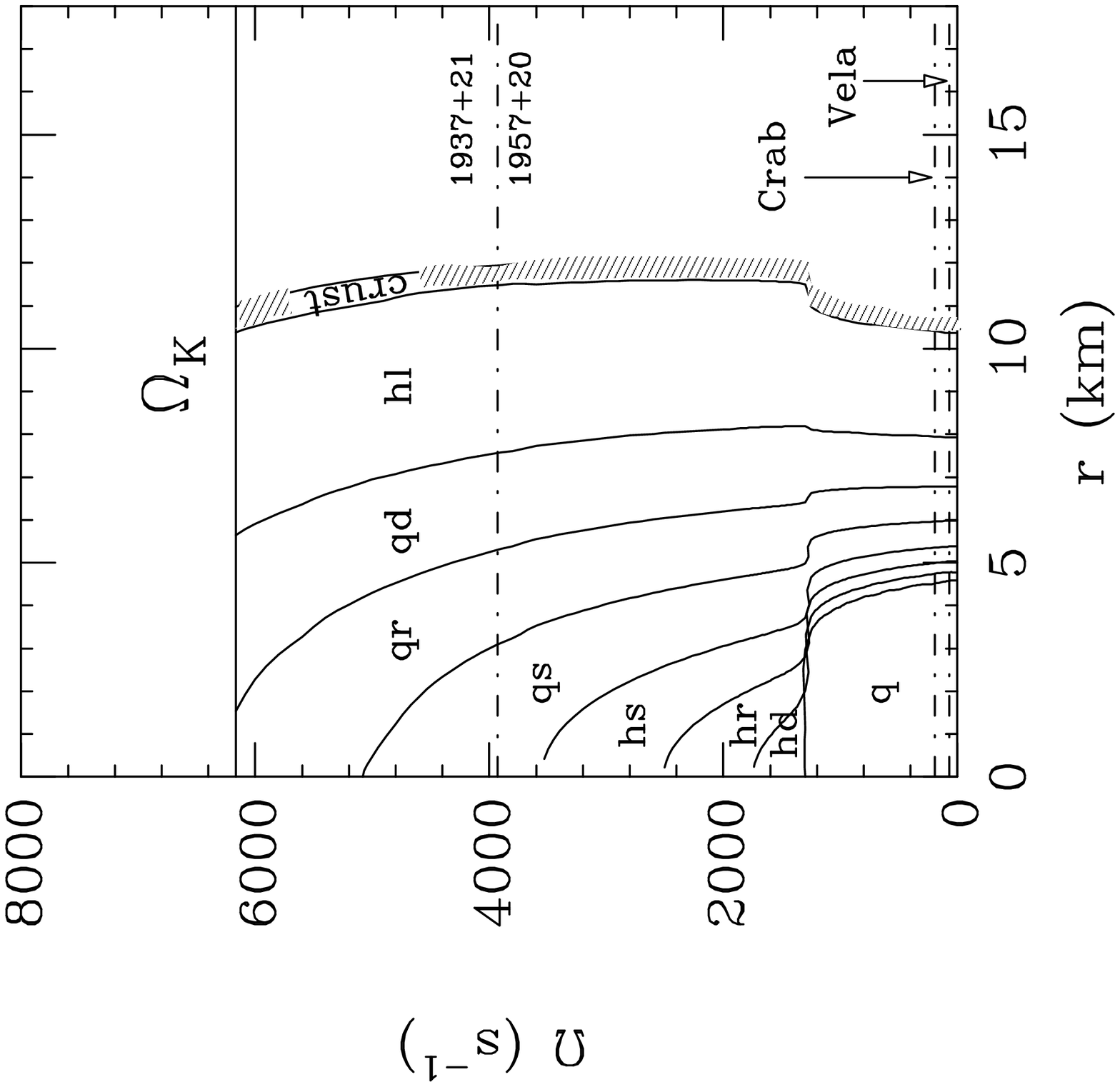,width=6.0cm,angle=-90} {\caption[]{Same
as Fig.\ \protect{\ref{fig:OkEq_1.42_G3B18}}, but in polar
direction.\protect{\cite{weber99:book}}}
\label{fig:OkPo_1.42_G3B18}}}
\end{figure}

\begin{figure}[tb] 
\parbox[t]{6.2cm} {\leavevmode
\psfig{figure=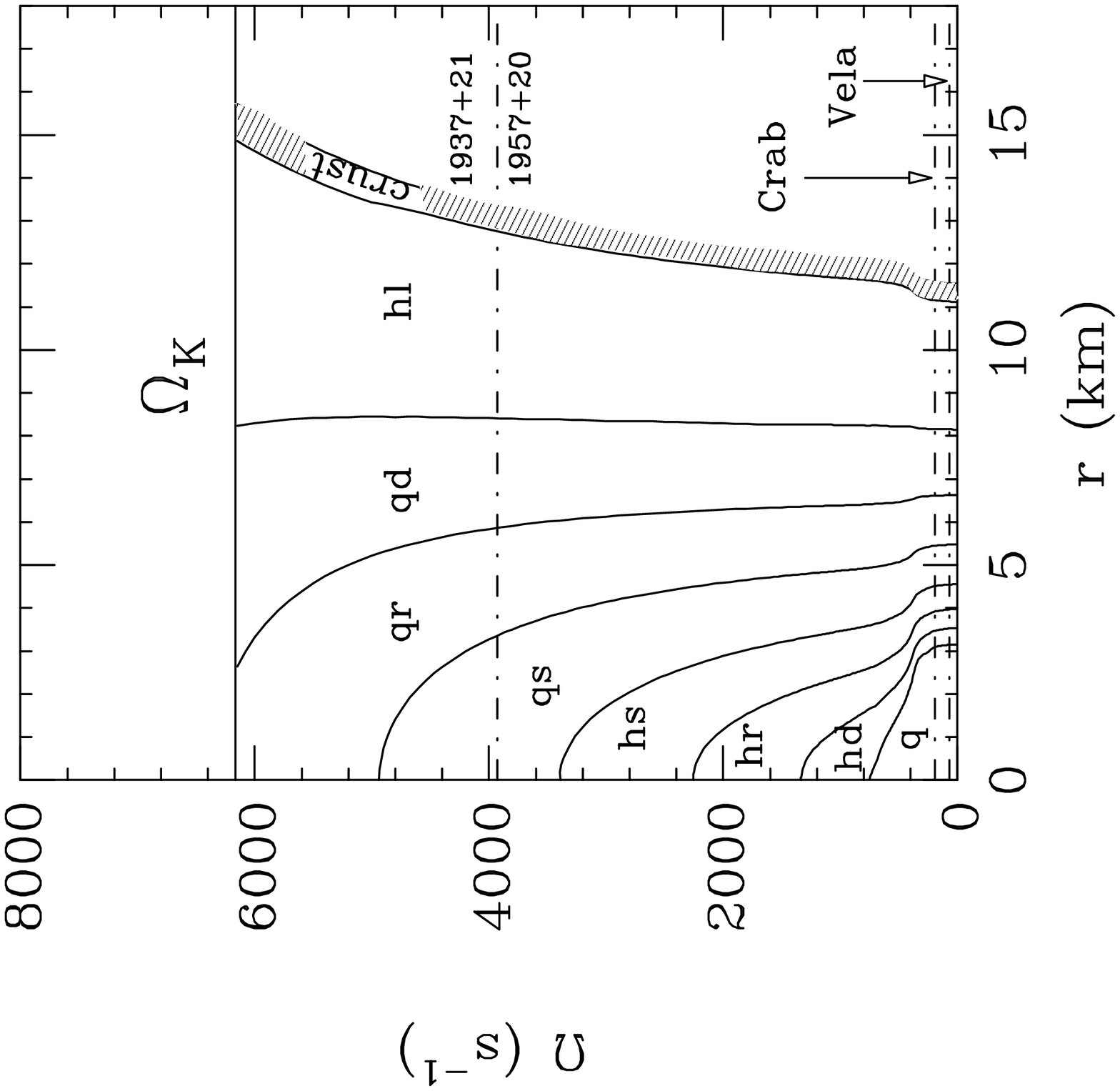,width=6.0cm,angle=-90}
{\caption[]{Frequency dependence of quark structure in equatorial
direction computed for $\KBt$ and a non-rotating star mass of
$1.416\,\msun$.\protect{\cite{weber99:book}}}
\label{fig:OkEq_1.416_G3B18}}}
\ \hskip0.25cm   \
\parbox[t]{6.2cm} {\leavevmode
\psfig{figure=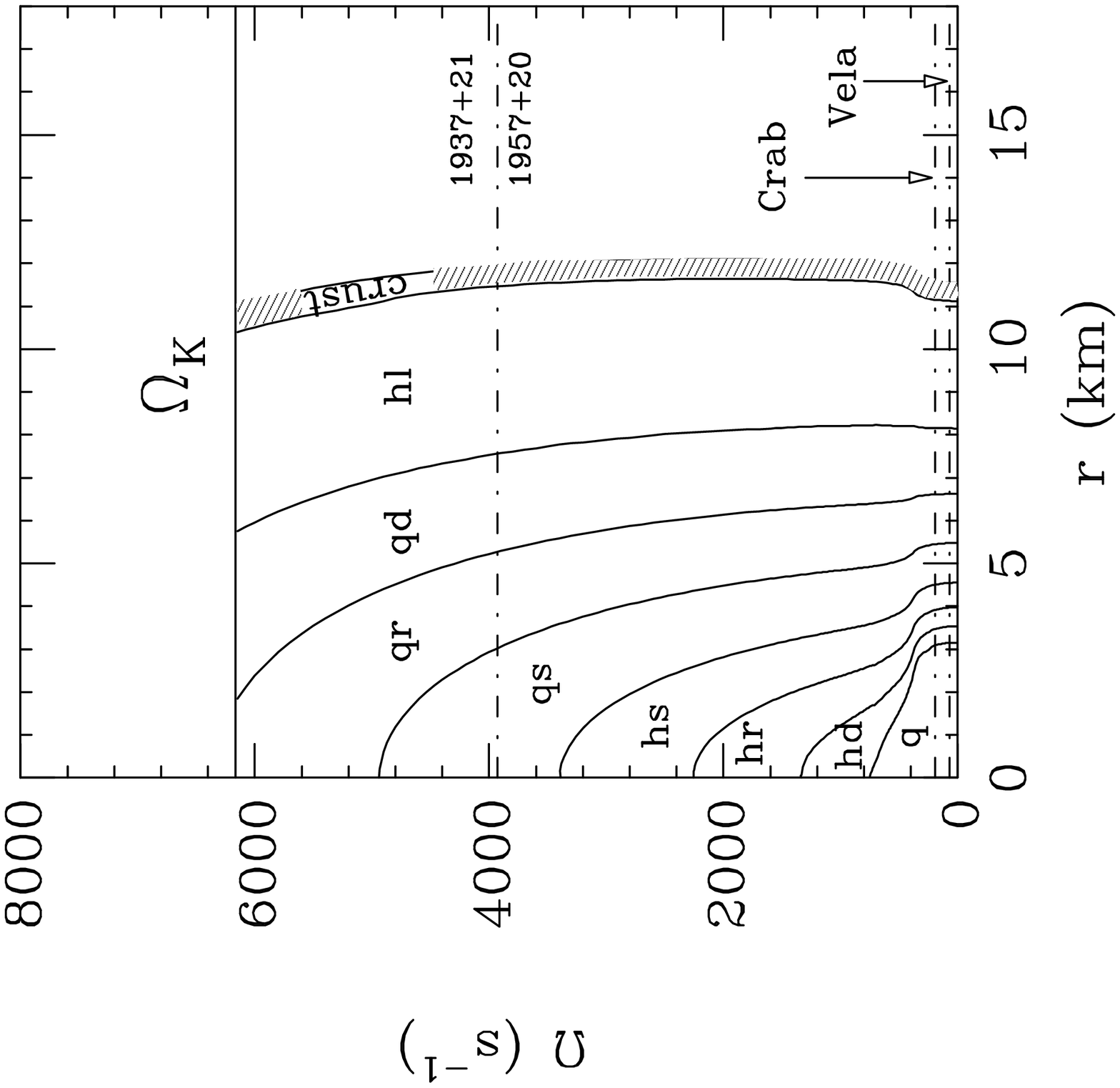,width=6.0cm,angle=-90} {\caption[]{Same
as Fig.\ \protect{\ref{fig:OkEq_1.416_G3B18}}, but in polar
direction.\protect{\cite{weber99:book}}}
\label{fig:OkPo_1.416_G3B18}}}
\end{figure}

Figures~\ref{fig:ec1445fig} and \ref{fig:ec3B18} reveal that the
weakening of centrifuge accompanied by the slowing-down of a rotating
neutron star causes a significant increase of its central density,
which is accompanied by a rearrangement of the particle population and
the creation of new phases of matter. From Fig.\ \ref{fig:ec3B18}, for
instance, one reads off that the central density of a neutron star of
mass $M=1.42\, \msun$ increases from about $450~\mevt$ for rotation at
$\okgr$ to more than $1500~\mevt$ for $\Omega = 0$, which is a $\sim
66$\% effect.  Of course, this effect is smaller for the lighter
stars, since for them the changes of the gravitational pull are
weaker.
Such dramatic changes in the interior density of slowing-down neutron
stars imply profound changes of their interior composition, because of
the changing particle composition with density. Figures
\ref{fig:freqhp1} through \ref{fig:freqhp4} illustrate how these
changes carry over to the internal structure of conventional neutron
stars. In each case the star's rotational frequency covers the maximal
possible range $0 \leq \Omega \leq \okgr$.
 It is evident that as rotating stars spin down they become
significantly less deformed, which leads to closer equality between
the polar and equatorial radii.  At the same time the central density
rises from below to above the threshold densities of heavier baryons
(i.e., $\Sigma, \Lambda, \Xi$).  For some pulsars the mass and initial
rotational frequency may be such that the central density rises from
below the critical density for dissolution of baryons into their quark
constituents.  Examples of which are shown in Figs.\
\ref{fig:OkEq_1.42_G3B18} through \ref{fig:OkPo_1.416_G3B18}.
 The evolution of the central mass densities of these stellar
models can be inferred in reference with Fig.\ \ref{fig:ec3B18}.  For
rotational frequencies below $\sim 1250~\secm$ the quark-hybrid star
models consist of an inner region of pure quark matter (q)
 surrounded by a few kilometer thick shell of mixed phase
of hadronic and quark matter. This phase consists of structures like
hadronic drops (hd), hadronic rods (hr), hadronic slabs (hs), quark
slabs (qs), quark rods (qr), and quark drops (qd), in each case
arranged in a lattice structure.\cite{glen91:pt} This shell is
surrounded by a sphere of hadronic liquid (hl), itself enveloped in a
thin nuclear crust of heavy ions. These quark-hadron structures are a
consequence of the competition of the Coulomb and surface energies of
the hadronic and quark matter phase.\cite{glen91:pt} They may have
dramatic effects on pulsar observables including transport properties
and the theory of glitches.

If the mass and initial rotational frequency of a pulsar is such that
during its slowing-down phase the interior density rises from below to
above the critical density for the quark-hadron phase transition,
first at the center where the density is highest (Figs.\
\ref{fig:ec1445fig} and \ref{fig:ec3B18}) and then in a region
expanding in the radial outward direction away from the star's center,
matter will be converted from the relatively incompressible nuclear
matter phase to the highly compressible quark matter phase.  The
tremendous weight of the overlaying layers of nuclear matter tends to
compress the quark matter core, which causes the entire star to shrink
on a macroscopic length scale, as shown in Figs.\
\ref{fig:OkEq_1.42_G3B18} through \ref{fig:OkPo_1.416_G3B18}. The mass
concentration in the core will be further enhanced by the increasing
gravitational attraction of the quark core on the overlaying nuclear
matter.  The moment of inertia thus decreases anomalously with
decreasing rotational frequency as the new phase slowly engulfs a
growing fraction of the star \cite{glen97:a}, as can be seem from
Figs.\ \ref{fig:moi1} and \ref{fig:moi2}.  Figure~\ref{fig:moi1} shows
the moment of inertia, $I$, computed self-consistently from Eq.\
(\ref{eq:f231}) for several sample stars having the same baryon number
but different internal constitutions.\cite{weber97:jaipur} The curve
labeled $M=1.420\, \msun$, computed for $\KBt$, shows the moment of
inertia of the quark-hybrid star of Figs.\ \ref{fig:OkEq_1.42_G3B18}
and \ref{fig:OkPo_1.42_G3B18}.  The other curves correspond to a
standard hyperon star constructed for $\KM$ and a standard neutron
star where hyperons were ignored. In accordance with what has been
said just above, the shrinkage of quark-hybrid stars driven by the
development of quark matter cores is the less pronounced the smaller
the quark matter cores which are being built up in their centers
during spin-down.  Correspondingly the reduction of $I$ weakens with
decreasing star mass, as shown in Fig.\ \ref{fig:moi1} for several
sample star masses.

The decrease of the moment of inertia caused by the quark-hadron phase
transition shown in Fig.\ \ref{fig:moi1} is superimposed on the
response of the stellar shape to a decreasing centrifugal force as the
star spins down due to the loss of rotational energy.  In order to
conserve angular momentum not carried off by particle radiation from
the star, the deceleration rate $\dot\Omega$~($< 0$) must respond
correspondingly by decreasing in absolute magnitude. More than that,
$\dot\Omega$ may even change sign, as shown in Figs.\ \ref{fig:moi2}
\cite{glen97:a}, which carries the important astrophysical information
that an isolated pulsar may spin up during a certain period of its
stellar evolution.  Such an anomalous decrease of $I$ is analogous to
the backbending phenomenon known from nuclear physics, in which case
the moment of inertia of an atomic nucleus changes anomalously because
of a change in phase from a nucleon spin aligned state at high angular
momentum to a pair correlated superfluid phase at low angular
momentum.\cite{mottelson82:a,stephens72:a,johnson72:a}  The stellar
backbending shown in Fig.\ \ref{fig:moi2} shows that stars evolving
from $b \rightarrow a$ are rotationally accelerated ($\dot\Omega >
0$), while stars evolving from $a \rightarrow b$ are rotationally
decelerated ($\dot\Omega < 0$).  The structure in the moment of
inertia and, specifically, the backbending phenomenon may dramatically
modify the timing structure of pulsar spin-down, rendering the
observation of quark matter in neutron stars accessible to radio
astronomy.

\goodbreak
\section{Evolution of a Pulsar's Braking Index}\label{sec:braking}

Pulsars are identified by their periodic signal believed to be due to
a strong magnetic field fixed in the star and oriented at an angle
from the rotation axis.  The star's angular velocity of rotation
decreases slowly but measurably over time, and usually the first and
occasionally second time derivative can also be measured. Various
energy loss mechanisms could be at play such as the magnetic dipole
radiation, part of which is detected on each revolution, as well as
\begin{figure}[tb] 
\parbox[t]{6.2cm} {\leavevmode
\psfig{figure=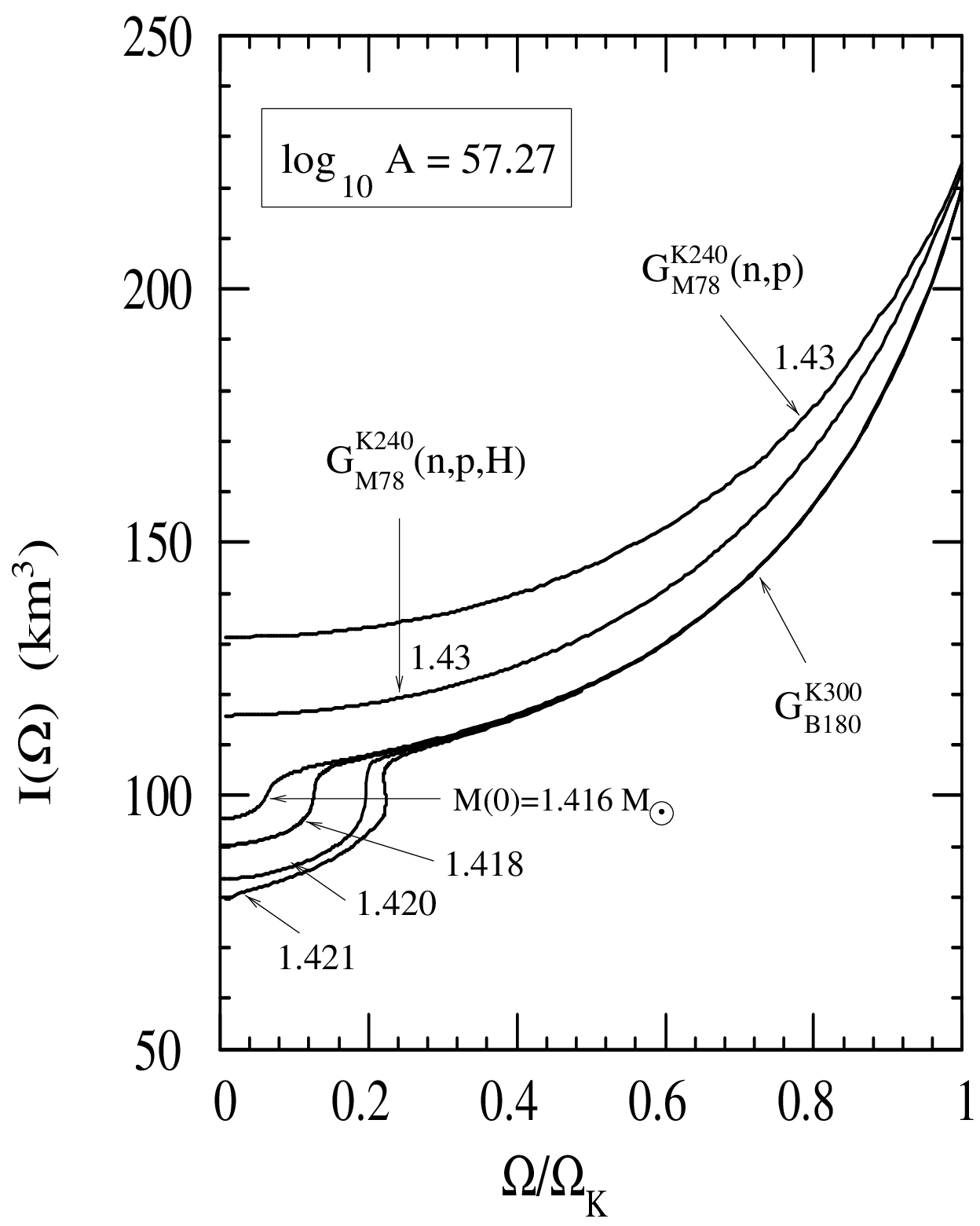,width=5.8cm,angle=-90} {\caption[]{Moment of
inertia versus frequency of neutron stars with the same baryon number,
$A$, but different constitutions. The dips at low $\Omega$'s are
caused by quark deconfinement.\cite{weber99:topr}}
\label{fig:moi1}}}
\ \hskip0.25cm \ 
\parbox[t]{6.2cm} {\leavevmode
\psfig{figure=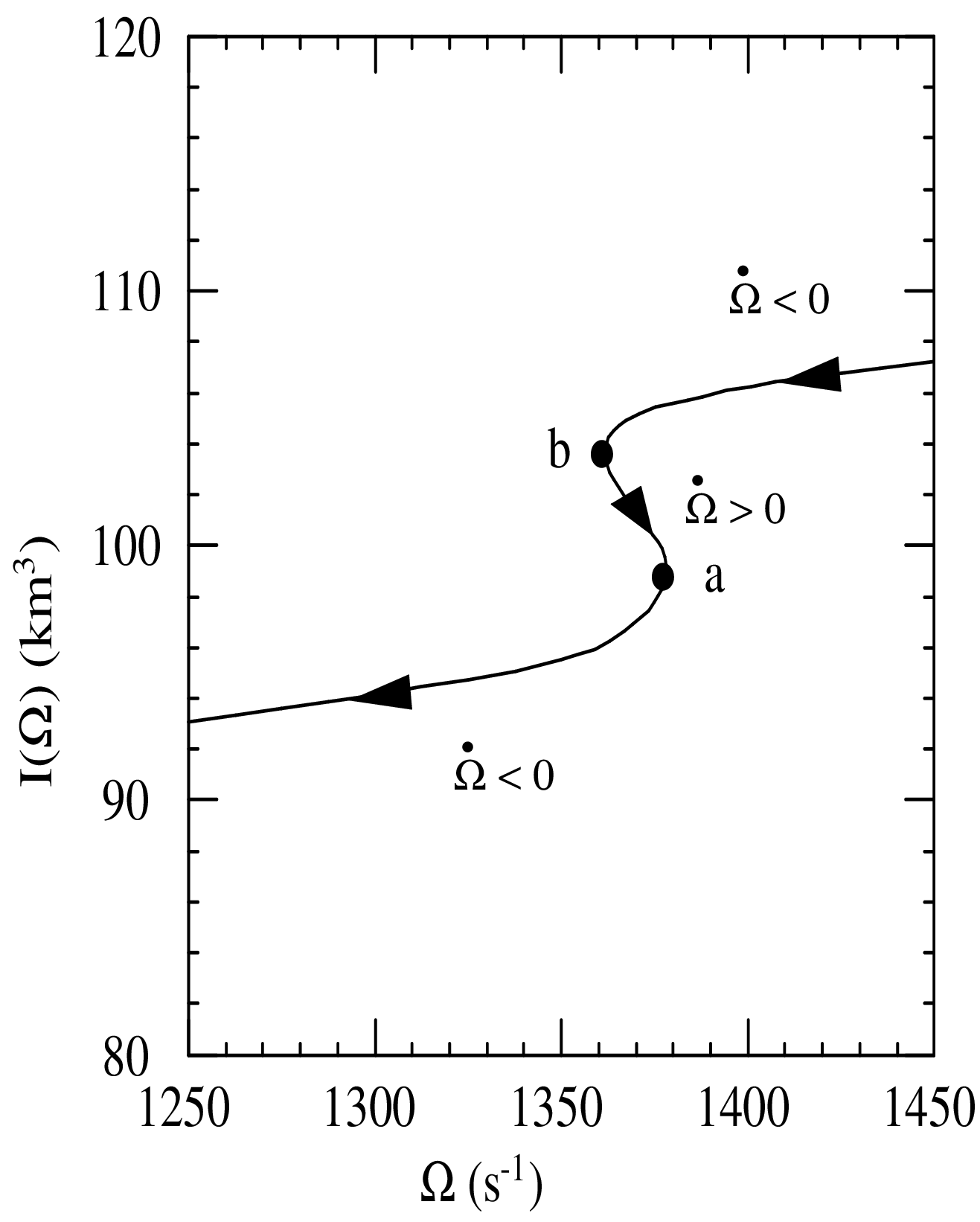,width=5.8cm,angle=-90} {\caption[]{Enlargement
of the lower-left portion of Fig.\ \protect{\ref{fig:moi1}} for the
star of mass $M=1.421\, \msun$, which is characterized by a
backbending of $I$ for frequencies between $a$ and
$b$.\protect{\cite{weber99:book,glen97:a}}}
\label{fig:moi2}}}
\end{figure}
other losses such as ejection of charged particles.\cite{ruderman87:a}
Without loss of generality, one can assume that pulsar slow-down is
governed by a single mechanism, or several mechanisms having the same
power law.  The energy balance equation can then be written in the
form
\begin{equation}
  \frac{dE}{dt} = \frac{d}{dt} \, \left( \frac{1}{2}\,
  I(\Omega) \, \Omega^2 \right) = - \, C \ \Omega^{n+1} \, .
  \label{eq:engloss}
\end{equation} In the case of magnetic dipole radiation, the constant $C$
is equal to $C= \frac{2}{3} \mu^2 \sin^2 \alpha$ ($\mu$ denotes the
star's magnetic moment), and $n=3$ if $I$ is kept constant during
spin-up (down).  If, as is customary, the star's angular velocity
$\Omega$ is regarded as the only time dependent quantity, one obtains
the usual formula for the rate of change of the pulsar frequency, given by
\begin{equation}
\dot{\Omega} = - \ K \ \Omega^n \, ,
\label{eq:braking}
\end{equation} with $K$ a constant and $n$ the braking index.  
As known from Figs.\ \ref{fig:moi1} and \ref{fig:moi2}, the moment of
inertia is not constant in time but responds to changes in rotational
frequency.  This response changes the value of the braking index in a
frequency dependent manner, that is $n = n(\Omega)$. Thus during any
epoch of observation, the braking index will be measured to be
different from it canonical value $n=3$ by a certain amount. How
different depends, for any given pulsar, on its rotational frequency
and for different pulsars of the same frequency, on their mass and on
their internal constitution.  When the frequency response of the
moment of inertia is taken into account, Eq.\ (\ref{eq:braking}) is to
be replaced with\cite{weber99:book,glen97:a}
\begin{equation}
  \dot{\Omega}= - \, 2 \, I \, K \, \frac{\Omega^n}{ 2 \, I + { { d
 I}\over{d\Omega} } \, \Omega } = - \, K \, \Omega^n \Biggl( 1 -
 \frac{ { {dI}\over{d\Omega} } }{2 \, I} \, \Omega + \Biggl(\frac{
 { {dI}\over{d\Omega} } }{2 \, I} \, \Omega \Biggr)^2 - \cdots
 \Biggr) \, ,
\label{eq:braking2}
\end{equation} where $K=C/I$.  This explicitly shows that the frequency 
dependence of $\dot{\Omega}$ corresponding to any mechanism that
absorbs (or deposits) rotational energy cannot be a simple power law
as given in equation~(\ref{eq:braking}) (with $K$ a constant). It must
depend on the mass and internal constitution of the star through the
response of the moment of inertia to rotation as expressed in
(\ref{eq:braking2}).  It is interesting to note that Eq.\
(\ref{eq:braking2}) can be represented in the form of
(\ref{eq:braking}) but now with a frequency dependent prefactor, by
evaluating
\begin{equation}  
  n(\Omega) \equiv \frac{\Omega\, \ddot{\Omega} }{\dot{\Omega}^2} = 3
- \frac{ 3 \, { {dI}\over{d\Omega} } \, \Omega + { {d^2
I}\over{d\Omega^2} } \, \Omega^2 } {2\, I + { {dI}\over{d\Omega}
} \, \Omega} \, .
\label{eq:index}
\end{equation} 
One sees that this effective braking index depends explicitly and
implicitly on $\Omega$ and reduces to the canonical expression $n=3$
only if $I$ is independent of frequency.

\begin{figure}[tb]
\begin{center}
\leavevmode
\psfig{figure=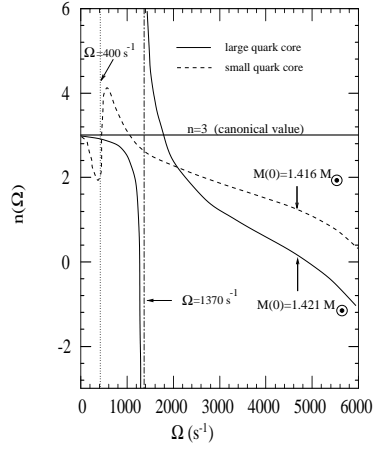,width=6.0cm,angle=-90}
\caption[]{Braking index, $n$, of quark-hybrid stars of Figs.\
\protect{\ref{fig:moi1}} and \protect{\ref{fig:moi2}}.  The anomalies
in $n$ at $\Omega \sim 400~\secm$ and $\Omega \sim 1370~\secm$ are
caused by quark deconfinement. The overall reduction of $n$ below 3 is
caused by the frequency dependence of the moment of
inertia.\cite{weber99:topr}}
\label{fig:n}
\end{center}
\end{figure}

As an example, we show in Fig.\ \ref{fig:n} the change of the
braking index with frequency for two selected quark-hybrid stars.
 For illustration purposes we assume dipole radiation.  
Because of the structure in the moment of inertia, driven by the phase
transition of baryonic matter into deconfined quark matter, the
braking index deviates dramatically from $n=3$ at those rotational
frequencies where quark deconfinement leads to the built-up of pure
quark matter cores in the centers of these stars.  Such anomalies in
$n(\Omega)$ are not obtained for conventional neutron stars or hyperon
stars because their moments of inertia increase smoothly with
$\Omega$, as known from Fig.\ \ref{fig:moi1}.  The observation of
such an anomaly in the timing structure of pulsar spin-down could thus
be interpreted as a possible signal of quark deconfinement in the
center of a pulsar. Of course, because of the extremely small temporal
change of a pulsar's rotational period, one cannot measure the shape
of the curve which is in fact not necessary. Just a single anomalous
value of $n$ that differed significantly from the canonical value of
$n=3$ would suffice.\cite{glen97:a,glen97:b}
 
\begin{figure}[tb]
\begin{center}
\leavevmode
\psfig{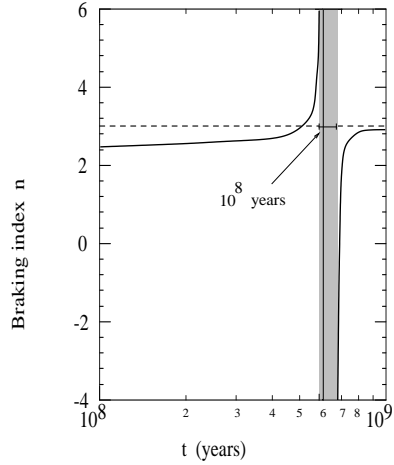}
\caption[]{Braking index versus time for the quark-hybrid star of mass
$M=1.421\, \msun$ of Fig.\ \protect{\ref{fig:n}}. The epoch over which
$n$ is anomalous because of quark deconfinement, $\sim 10^8$~years, is
indicated by the shaded area.\protect{\cite{weber99:book}}}
\label{fig:nvst_1}
\end{center}
\end{figure}

As a final but very important point on this subject, we estimate the
typical duration over which the braking index is anomalous if
deconfinement is well pronounced. The time span can be estimated from
\begin{equation}
  \Delta T \simeq - \, {{\Delta\Omega}\over{\dot\Omega}} = {{\Delta
      P}\over{\dot{P}}} \, ,
\label{eq:estim1}
\end{equation} where $\Delta \Omega$ is the frequency interval of the
anomaly.  The range over which $n(\Omega)$ is smaller than zero and
larger than six (Fig.\ \ref{fig:n}) is $\Delta \Omega \approx - \,
100~\secm$, or $\Delta P \approx - 2 \pi \Delta \Omega / \Omega^2
\approx 3 \times 10^{-4}$~s at $\Omega=1370~\secm$.  So, for a
millisecond pulsar whose period derivative is typically $\dot{P}\simeq
10^{-19}$ one has $\Delta T \simeq 10^8$~years, as illustrated in
Fig.\ \ref{fig:nvst_1}. The dipole age of such pulsars is about
$10^9$~years. So, as a rough estimate one may expect about 10\% of the
25 presently known solitary millisecond pulsars to be in the
transition epoch during which pure quark matter cores are gradually
being built up in their centers. These pulsars could be signaling the
ongoing process of quark deconfinement in their cores.  Last but not
least we note that the spin-up time (region $b$--$a$ in Fig.\
\ref{fig:moi2}) is about 1/5 of the time span $\Delta T$, or about
1/50 of the dipole age $\tau = - \, \Omega / (\dot{\Omega} (n-1))$.

It is probably needless to say that the observation of a pulsar with
an anomalously large braking index, signaling the existence of quark
matter in the centers of pulsars, would be a momentous discovery with
most far-reaching consequences for nuclear and particle physics.  It
would help to clarify how quark matter behaves, and give a boost to
theories about the early Universe as well as laboratory searches for
the production of quark matter in heavy-ion collisions, which has
become a forefront area of modern physics. On an even more fundamental
level, such a discovery would prove that the essentially free quark
state predicted for matter at very high energy densities actually
exists, and give us a picture of an early phase of the Universe that
is based on radio pulsar observation, which may be coined quark
astronomy.

\section{Quark Matter in Neutron Stars in Low-Mass X-ray Binaries}
\label{sec:spin}

The signal of quark deconfinement described in Sect.\
\ref{sec:braking} applies only to isolated neutron stars, where
deconfinement is driven by the gradual stellar contraction as the star
spins down.  The situation is reversed in neutron stars in binary
systems, which experience a spin-up torque because of the transfer of
angular momentum carried by the matter picked up by the star's
magnetic field from the surrounding accretion disk.  The spin-up
torque causes a change in the stars' angular momentum according to the
relation\cite{glen01:a}
\begin{eqnarray}
{{dJ} \over {d t}} = {\dot M} {\tilde l}(r_{\rm m}) - N(r_{\rm c}) \,
,
\label{eq:dJdt}
\end{eqnarray}
where $\dot{M}$ denotes the accretion rate and
\begin{eqnarray}
{\tilde l}(r_{\rm m}) = \sqrt{M r_{\rm m}} 
\label{eq:l}
\end{eqnarray}
is the angular momentum added to the star per unit mass of accreted
matter. The quantity $N$ stands for the magnetic plus viscous torque
term,
\begin{equation}
N(r_{\rm c}) = \kappa \, \mu^2 \, r_{\rm c}^{-3} \, ,
\label{eq:N}
\end{equation}
with $\mu \equiv R^3 B$ the star's magnetic moment.  The quantities
\begin{figure}[tb]
\begin{center} 
\leavevmode 
\psfig{figure=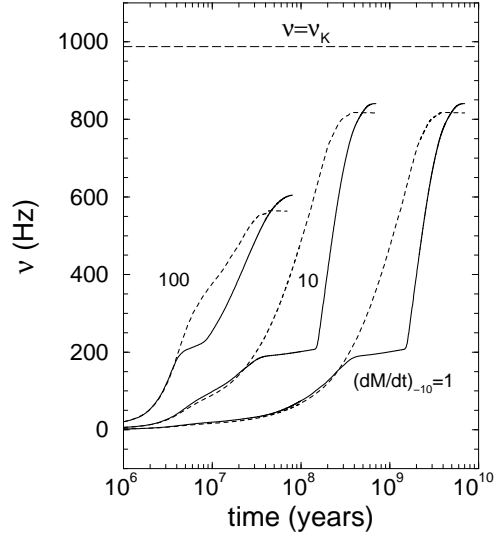,width=6.5cm}
\caption[]{Evolution of spin frequencies of accreting X-ray neutron
stars with (solid curves) and without (dashed curves) quark
deconfinement.\protect{\cite{glen01:a}}  The spin plateau around
200~Hz signals the ongoing process of quark re-confinement in the
stellar centers}
\label{fig:nue_t}
\end{center}
\end{figure}
$r_{\rm m}$ and $r_{\rm c}$ denote the radius of the inner edge of the
accretion disk and the co-rotating radius, respectively, and are given
by $(\xi \sim 1)$
\begin{equation}
  r_{\rm m} = \xi \, r_{\rm A} \, ,
\label{eq:r_m}
\end{equation}
and 
\begin{eqnarray}
r_{\rm c} = \left( M \Omega^{-2} \right)^{1/3} \, .
\label{eq:r_c}
\end{eqnarray}
Accretion will be inhibited by a centrifugal barrier if the neutron
star's magnetosphere rotates faster than the Kepler frequency at the
magnetosphere. Hence $r_{\rm m} < r_{\rm c}$, otherwise accretion onto
the star will cease.  The Alf\'en radius $r_{\rm A}$ in
Eq.\ (\ref{eq:r_m}) is defined by
\begin{equation}
r_{\rm A} = \left( { {\mu^4} \over {2 M \dot{M}^2} } \right)^{1/7} \, .
\label{eq:r_A}
\end{equation}
The rate of change of a star's angular frequency $\Omega$ $(\equiv
J/I)$ then follows from Eq.\ (\ref{eq:dJdt}) as
\begin{equation}
  I(t) {{d\Omega(t)} \over {d t}} = {\dot M} {\tilde l}(t) - \Omega(t)
  {{dI(t)}\over{dt}} - \kappa \mu(t)^2 r_{\rm c}(t)^{-3} \, ,
\label{eq:dOdt.1}
\end{equation} with the explicit time dependences as indicated.
There are two terms on the right-hand-side of Eq.\ (\ref{eq:dOdt.1})
that grow linearly respectively quadratically with $\Omega$.  Ignoring
the linear term shows that mass transfer can spin up a neutron star to
an equilibrium period ($P = 2\pi/ \Omega$) of\cite{heuvel91:a}
\begin{equation}
P_{\rm eq} = 2.4~{\rm ms} \left({{\dot M}\over{\dot M_{\rm
Edd}}}\right)^{-3/7} \left({{M}\over{M_{\odot}}}\right)^{-5/7} ~
R_6^{15/7} ~ B_9^{6/7} \, ,
\label{eq:equil.1}
\end{equation}
where $R_6$ and $B_9$ are the star's radius and its magnetic field in
units of $10^6$~cm and $10^9$~G, respectively.  $\dot{M}_{\rm Edd}$ in
Eq.\ (\ref{eq:equil.1}) denotes the maximum possible accretion rate,
known as the Eddington limit, at which the accretion luminosity equals
the luminosity at which the radiation pressure force on ionized
\begin{figure}[tb] 
\begin{center}
\leavevmode
\psfig{figure=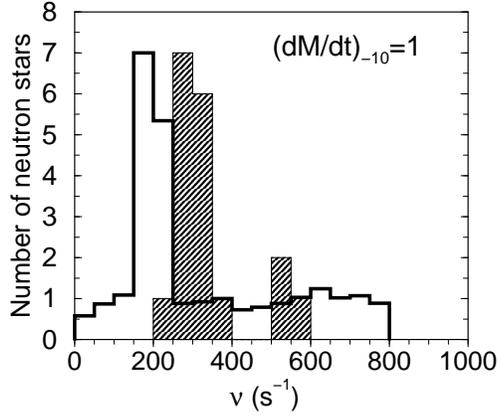,width=6.5cm}
\caption[]{Calculated spin distribution of X-ray neutron stars.  The
spike in the calculated distribution (unshaded diagram) corresponds to
the spinout of the quark matter phase. Otherwise the spike would be
absent. The shaded histogram displays the observed data.\cite{glen01:a}}
\label{fig:histo}
\end{center}
\end{figure}
hydrogen plasma near the star balances the gravitational acceleration
force exerted by the star on the plasma.  This condition leads to an
Eddington accretion rate of $\dot{M}_{\rm Edd} = 1.5\times 10^{-8} R_6
M_\odot/{\rm yr}$.  For a typical accretion rate of ${\dot M}_{-10}
\equiv {\dot M}/(10^{-10} M_{\odot} / {\rm yr})$, the Eddington rate
can be expressed as ${\dot M}_{\rm Edd} = 150 \, R_6 \, {\dot
M}_{-10}^{-1} \, {\dot M}$.  The low-mass X-ray binaries (LMXBs)
observed with the Rossi X-ray Timing Explorer are divided into Z
sources and A(toll) sources, and appear to accrete at rates of
$\dot{M}_{-10} \sim 200$ and $\dot{M}_{-10} \sim 2$, respectively.

The temporal change of the moment of inertia of accreting neutron
stars which undergo phase transitions is crucial for their spin
evolution.\cite{glen01:a} This temporal change, on the other hand,
renders the computation of the moment of inertia, defined in Eq.\
(\ref{eq:f231}), very cumbersome since each quantity on the
right-hand-side of (\ref{eq:f231}) varies accordingly during stellar
spin-up.  The solution of Eq.\ (\ref{eq:dOdt.1}) in combination with
Eq.\ (\ref{eq:f231}) is shown in Fig.\ \ref{fig:nue_t}. The result is
most striking. One sees that quark matter remains relatively dormant
in the stellar core until the star has been spun up to frequencies at
which the central density is about to drop below the threshold density
at which quark matter exists. As known from Fig.\ \ref{fig:moi2}, this
manifests itself in a significant increase of the star's moment of
inertia. The angular momentum added to a neutron star during this
phase of evolution is therefore consumed by the star's expansion,
inhibiting a further spin-up until the quark matter has been converted
into a mixed phase of matter made up of hadrons and quarks.  Such
accreters, therefore, tend to spend a greater length of time in the
critical frequencies than otherwise. There will be an anomalous number
of accreters that appear at or near the same frequency, as shown in
Fig.\ \ref{fig:histo}. This is what was actually found recently with
the Rossi X-ray Timing Explorer (shown by the shaded area in Fig.\
\ref{fig:histo}).  Quark deconfinement constitutes a most striking
explanation for this anomaly \cite{glen01:a,glen00:b,poghosyan01:a}
though alternative explanations were suggested
too.\cite{bildsten98:a,andersson00:a}

\section*{Acknowledgments}
I wish to thank the organizers of Hadron 2002, Prof.\ Cesar A.\ Z.\
Vasconcellos and Prof.\ Victoria E.\ Herscovitz, for creating a very
stimulating and enjoyable atmosphere at this meeting as well for their
outstanding hospitality.

\end{document}